\documentclass[preprint,12pt]{elsarticle}

\include{packages}

\usepackage{tikz}
\usepackage{tikz-3dplot}
\usepackage{xifthen}
\usetikzlibrary{spy}
\usetikzlibrary{arrows}
\usetikzlibrary{arrows.meta}
\usetikzlibrary{shapes}
\usetikzlibrary{intersections}
\usetikzlibrary{positioning}
\usetikzlibrary{backgrounds}
\usepackage{pgfplots}
\usepackage{pgfplotstable}
\pgfplotsset{compat=1.13}
\usetikzlibrary{calc}
\tikzstyle{rounded}  = [rounded corners]     
\tikzstyle{function} = [rectangle, text centered, draw=black,minimum width=6.0cm,minimum height=1.0cm]
\tikzstyle{mpi}      = [anchor=south west, xshift=0.2cm]

\usetikzlibrary{3d}

\usetikzlibrary{fillbetween}
\usetikzlibrary{decorations,decorations.text,backgrounds}

\def\Dbb{{\mathbb D}}

\def\Fbb{{\mathbb F}}

\def\Ibb{{\mathbb I}}
\def\Nbb{{\mathbb N}}
\def\Rbb{{\mathbb R}}

\def\Tbb{{\mathbb T}}

\def\Ecal{{\mathcal E}}

\def\Lcal{{\mathcal L}}

\def\Ocal{{\mathcal O}}
\def\Rcal{{\mathcal R}}
\def\Scal{{\mathcal S}}

\newcommand{\liq}{_{\mathrm{liq}}}
\newcommand{\vap}{_{\mathrm{vap}}}
\newcommand{\spin}{{\mathrm{spin}}}
\newcommand{\sat}{{\mathrm{sat}}}
\newcommand{\surf}[1]{{#1}_{\Gamma}}
\newcommand{\surftang}[1]{{#1}_{\Gamma,\tau}}
\newcommand{\surfnorm}[1]{{#1}_{\Gamma,\mathrm{n}}}
\newcommand{\tang}[1]{{#1}_{\tau}}

\newcommand{\overbar}[1]{\mkern 1.0mu\overline{\mkern-1.0mu#1\mkern-1.0mu}\mkern 1.0mu}
\newcommand{\overhat}[1]{%
    \savestack{\tmpbox}{\stretchto{%
            \scaleto{%
                \scalerel*[\widthof{\ensuremath{#1}}]                         {\kern-.6pt\bigwedge\kern-.6pt}%
                    {\rule[-\textheight/2]{1ex}{\textheight}}%
            }{\textheight}%
        }{0.5ex}}%
    \ensurestackMath{\stackon[1pt]{#1}{\tmpbox}}%
}

\newcolumntype{?}{!{\vrule width 1pt}}

\newcommand{\eos}{EoS}
\newcommand{\cit}{CIT}
\newcommand{\nsk}{NSK}
\newcommand{\flexi}{\textit{FLEXI}}
\newcommand{\dgsem}{DGSEM}
\newcommand{\dg}{DG}
\newcommand{\fv}{FV}
\newcommand{\rk}{RK}

\newcommand{\brone}{BR1}
\newcommand{\bc}{BC}
\newcommand{\dof}{DOF}
\newcommand{\originalmodel}{\textit{original model}}
\newcommand{\relaxationmodelone}{\textit{relaxation model 1}}
\newcommand{\relaxationmodeltwo}{\textit{relaxation model 2}}

\newcommand{\gradient}{\nabla}
\newcommand{\laplace}{\mathop{}\!\mathcal{4}}

\newcommand{\fracp}[2]{\frac{\partial #1}{\partial #2}}
\newcommand{\fracpp}[2]{\frac{\partial^2 #1}{\partial^2 #2}}

\newcommand{\material}[1]{\dot{#1}}
\newcommand{\transpose}[1]{{#1}^{\mathrm{T}}}

\newcommand{\jump}[1]{\ensuremath{ [\![ #1 ]\!]}}
\newcommand{\jumpbig}[1]{\ensuremath{ \big[\!\big[ #1 \big]\!\big]}}
\newcommand{\jumpBig}[1]{\ensuremath{ \Big[\!\!\Big[ #1 \Big]\!\!\Big]}}

\newcommand{\domain}{\Omega}
\newcommand{\domainsurf}{\partial \Omega}
\newcommand{\normvec}{\mathbf{n}}
\newcommand{\surfnormvec}{\surf{\normvec}}
\newcommand{\dimvec}{\mathbf{x}}
\newcommand{\dimensions}{\operatorname{d}}
\newcommand{\timeinterval}{\l( 0 , T \r)}
\newcommand{\spacetimeinterval}{\timeinterval \times \domain}
\newcommand{\intdomain}{\int_\Omega}
\newcommand{\dx}{\mathrm{d} \dimvec}
\newcommand{\tend}{t}
\newcommand{\wallsurface}{\Gamma}
\newcommand{\xdir}{x}
\newcommand{\ydir}{y}
\newcommand{\zdir}{z}
\newcommand{\dist}{\mathrm{d}}
\newcommand{\height}{h}

\newcommand{\Rgzero}{\Rbb_{> 0}}
\newcommand{\Ngzero}{\Nbb_{> 0}}

\newcommand{\alphakorteweg}{\ensuremath{\alpha}}
\newcommand{\betakorteweg}{\ensuremath{\beta}}
\newcommand{\upsilonkorteweg}{\ensuremath{\Upsilon}}
\newcommand{\capillarycoef}{\ensuremath{\gamma}}
\newcommand{\epsilonkorteweg}{\ensuremath{\varepsilon}}
\newcommand{\gammaepsilonkorteweg}{\capillarycoef \epsilonkorteweg^2}
\newcommand{\gammakorteweg}{\ensuremath{\capillarycoef_\mathrm{K}}}

\newcommand{\dens}{\rho}

\newcommand{\dense}{\dens \ener}

\newcommand{\vel}{\mathbf{u}}
\newcommand{\pres}{p}
\newcommand{\temp}{\ensuremath{\vartheta}}
\newcommand{\ckorteweg}{\ensuremath{c}}

\newcommand{\velx}{u}

\newcommand{\eintmass}{\epsilon}
\newcommand{\ener}{e}
\newcommand{\entropymass}{\eta}

\newcommand{\helmfreevol}{\psi}
\newcommand{\helmfreevolsurftot}{\surf{\overhat{\helmfreevol}}}
\newcommand{\enervolsurf}{\surf{\overhat{\ener}}}
\newcommand{\entropyvolsurf}{\surf{\overhat{\entropymass}}}

\newcommand{\gradvel}{\Dbb}

\newcommand{\chempotc}{\chi_{\ckorteweg}}

\newcommand{\curvature}{\kappa_\ensuremath{M}}
\newcommand{\surftenscoeff}{\sigma}
\newcommand{\contactangle}{\theta}
\newcommand{\bitangent}{\helmfreevol_{\mathrm{M}}}

\newcommand{\Dbbdevfree}{\Dbb^{\mathrm{d}}}
\newcommand{\Dbbdevfreealpha}{\Dbb^{\mathrm{d},\alphakorteweg}}

\newcommand{\speedofsound}{c_s}

\newcommand{\dummyorderparameter}{\phi}

\newcommand{\helmfreefunc}{\Ecal_\helmfreevol \l( \dens , \temp \r)}
\newcommand{\surfacefuncvdw}{\Ecal_{\mathrm{vdW}} \l( \gradient \dens \r)}
\newcommand{\Ekin}{\Ecal_\mathrm{kin} \l( \dens,\vel \r)}
\newcommand{\helmfreefunctot}{\Ecal_\mathrm{tot} \l( \dens , \temp , \gradient \rho , \vel \r)}
\newcommand{\helmfreefuncalphatot}{\Ecal^{\alphakorteweg}_\mathrm{tot} \l( \dens^\alphakorteweg , \temp^\alphakorteweg , \ckorteweg^\alphakorteweg , \gradient \ckorteweg^\alphakorteweg , \vel^\alphakorteweg \r)}

\newcommand{\arbvolquant}{\Psi}

\newcommand{\force}{\mathbf{b}}
\newcommand{\statevariable}{y}
\newcommand{\statevector}{\mathbf{y}}
\newcommand{\mstatevariables}{M}

\newcommand{\affinitiesvec}{\mathbf{A}}
\newcommand{\affinitycoef}[1]{\gamma_{#1}}

\newcommand{\rhocrit}{\tilde{\dens}_{\mathrm{c}}}
\newcommand{\Tcrit}{\tilde{\temp}_{\mathrm{c}}}
\newcommand{\pcrit}{\tilde{\pres}_{\mathrm{c}}}
\newcommand{\avdw}{\ensuremath{a}}
\newcommand{\bvdw}{\ensuremath{b}}
\newcommand{\Rvdw}{\ensuremath{R}}
\newcommand{\cvdw}{\ensuremath{c_v}}
\newcommand{\specheatcapvol}{\ensuremath{\tilde{c}_v}}
\newcommand{\gasconst}{\ensuremath{\Rcal}}
\newcommand{\Lref}{\ensuremath{\tilde{L}}}
\newcommand{\Uref}{\tilde{\vel}_0}
\newcommand{\Tref}{\tilde{t}_0}
\newcommand{\Eref}{\tilde{e}_0}

\newcommand{\physflux}{\mathbf{j}}
\newcommand{\je}{\physflux_{\ensuremath{e}}}
\newcommand{\jc}{\physflux_{\ensuremath{c}}}
\newcommand{\js}{\frac{\mathbf{q}}{\temp}}
\newcommand{\jfourier}{\physflux_{\ensuremath{F}}}
\newcommand{\jnsk}{\physflux_{\ensuremath{int}}}

\newcommand{\jkappa}{\ensuremath{J}_{\ensuremath{\kappa}}}
\newcommand{\jkappaone}{\ensuremath{J}_{\ensuremath{1}}}
\newcommand{\jkappatwo}{\ensuremath{J}_{\ensuremath{2}}}
\newcommand{\jkappavec}{\mathbf{J}}
\newcommand{\viscousstress}{\Tbb^\mathrm{V}}
\newcommand{\kortewegstress}{\Tbb^\mathrm{K}}
\newcommand{\viscousstressalpha}{\Tbb^{\mathrm{V},\alphakorteweg}}
\newcommand{\kortewegstressalpha}{\Tbb^{\mathrm{K},\alphakorteweg}}

\newcommand{\jesurf}{\physflux_{\ensuremath{e},\Gamma}}
\newcommand{\jssurf}{\frac{\surf{\mathbf{q}}}{\surf{\temp}}}

\newcommand{\arbvolflux}{\Phi}

\newcommand{\src}{\zeta_{\ensuremath{c}}}
\newcommand{\srstemp}{\frac{\xi}{\temp}}
\newcommand{\srs}{\Pi^\entropymass}

\newcommand{\srdummy}{\zeta_{\dummyorderparameter}}

\newcommand{\srx}[1]{\zeta_{#1}}

\newcommand{\srarb}{\Pi}

\newcommand{\srssurf}{\frac{\surf{\xi}}{\surf{\temp}}}
\newcommand{\srcsurf}{\zeta_{\ensuremath{c},\Gamma}}

\newcommand{\munewton}{\mu}
\newcommand{\lambdastokes}{\lambda}
\newcommand{\heatcoef}{k}

\newcommand{\fkappa}[1]{f_\kappa \l({#1}\r)}
\newcommand{\fkappavec}{\mathbf{f}}
\newcommand{\function}[1]{f \l( {#1} \r)}

\newcommand{\Reynolds}{\ensuremath{\mathrm{Re}}}
\newcommand{\Prandtl}{\ensuremath{\mathrm{Pr}}}
\newcommand{\Weber}{\ensuremath{\mathrm{We}}}

\newcommand{\eigenvalue}{\lambda}
\newcommand{\UPDE}{\mathbf{U}}
\newcommand{\FPDE}{\Fbb}
\newcommand{\SPDE}{\mathbf{S}}
\newcommand{\FcPDE}{\FPDE^\mathrm{C}}
\newcommand{\FvPDE}{\FPDE^\mathrm{V}}
\newcommand{\refelem}{E}
\newcommand{\refspace}{\tilde{\boldsymbol{\xi}}}
\newcommand{\xiref}{\tilde{\xi}_1}
\newcommand{\etaref}{\tilde{\xi}_2}
\newcommand{\zetaref}{\tilde{\xi}_3}
\newcommand{\numflux}[1]{{#1}^{*}}
\newcommand{\idir}{i}
\newcommand{\jdir}{j}
\newcommand{\kdir}{k}
\newcommand{\ijk}{i,j,k}
\newcommand{\odir}{o}

\newcommand{\qdir}{q}
\newcommand{\opq}{o,p,q}
\newcommand{\oq}{o,q}

\newcommand{\Npoly}{N}
\newcommand{\Ngeo}{N_{geo}}

\newcommand{\radius}{r}
\newcommand{\interfacewidth}{L_i}

\renewcommand{\l}{\left}
\renewcommand{\r}{\right}

\journal{Journal of Computational Physics}

\begin{document}

\begin{frontmatter}

\title{A Relaxation Model for the Non-Isothermal Navier-Stokes-Korteweg Equations in Confined Domains}

\author[label1]{Jens Keim}
\ead{keim@iag.uni-stuttgart.de}
\author[label1]{Claus-Dieter Munz}
\author[label2]{Christian Rohde}

\address[label1]{Institute of Aerodynamics and Gas Dynamics, University of Stuttgart, 70569 Stuttgart, Germany}
\address[label2]{Institute of Applied Analysis and Numerical Simulation, University of Stuttgart, 70569 Stuttgart, Germany}

\begin{abstract}
The Navier-Stokes-Korteweg (\nsk) ~system is a classical diffuse interface model which is based on van der Waals' theory of capillarity.
Diffuse interface methods have gained much interest to model two-phase flow in porous media.
However, for the numerical solution of the \nsk ~equations two major challenges have to be faced.
First, an extended numerical stencil is required due to a third-order term in the linear momentum and the total energy equations.
In addition, the dispersive contribution in the linear momentum equations prevents the straightforward use of contact angle boundary conditions.
Secondly, any real gas equation of state is based on a non-convex Helmholtz free energy potential which may cause the eigenvalues of the Jacobian of the first-order fluxes to become imaginary numbers inside the spinodal region.

In this work, a thermodynamically consistent relaxation model is presented which is used to approximate the \nsk ~equations.
The model is complimented by thermodynamically consistent non-equilibrium boundary conditions which take contact angle effects into account.
Due to the relaxation approach, the contribution of the Korteweg tensor in the linear momentum and total energy equations can be reduced to second-order terms which enables a straightforward implementation of contact angle boundary conditions in a numerical scheme.
Moreover, the definition of a modified pressure function enables to formulate first-order fluxes which remain strictly hyperbolic in the entire spinodal region.
The present work is a generalization of a previously presented parabolic relaxation model for the isothermal \nsk ~equations.

A high-order discontinuous Galerkin spectral element method which supports curved elements and hanging nodes is employed to discretize the system.
The relaxation model and its corresponding boundary conditions are validated using solutions of the original \nsk ~model and analytical results for one-, two- and three-dimensional test cases.
The simulation of a spinodal decomposition in a three-dimensional porous structure underlines the capability of the presented approach.

\end{abstract}


\end{frontmatter}

\section{Introduction}
\label{sec:introduction}
Multiphase phenomena are present in a variety of natural and technical processes related to porous media domains.
Applications include wetting properties of plant leaves \cite{guo2007}, underground $CO_2$ storage \cite{class2009}, coating processes \cite{weinstein2004}, inkjet printing \cite{degans2004} and natural processes investigated in geosciences \cite{blunt2013}.

Numerous continuum based models have been developed for the direct numerical simulation of multiphase flow.
Two popular but fundamentally different approaches are the sharp \cite{ishii2011} and the diffuse \cite{anderson1998} interface methods.
In the sharp interface approach, the computational domain is partitioned into distinct subdomains, called bulk domains, which contain either the liquid or the vapor phase.
Inside the bulk domains, the flow dynamics are governed by the well-known and widely accepted Navier-Stokes equations. 
Across the phase boundary, discontinuities in the solution, e.g. in the density or the pressure, are allowed.
Hence, suitable jump conditions have to be applied at the interface to provide a physically sound coupling of the subdomains.
This is a challenging task, especially if a non-isothermal compressible flow with phase transition is considered, see e.g. \cite{hitz2021,joens2022}.
In addition, an explicit tracking of the phase interface is required, e.g. via the volume-of-fluid \cite{hirt1981} or the level-set method \cite{fedkiw1999}.

A promising alternative for the simulation of multiphase phenomena with phase transition emerges from models based on diffuse interface or phase field approaches \cite{cueto2017}.
They originate from the seminal works of van der Waals and Korteweg \cite{vanderwaals1894,korteweg1901} and provided the first thermodynamic insight into the physics of capillarity.
In diffuse interface methods, the phase interface is assumed to be an interfacial zone of finite thickness, where the density changes continuously but with a strong gradient.
The properties of this interfacial zone are based on a Helmholtz free energy, which is composed of two parts; a double well potential with two minima that account for the two coexisting bulk phases and a gradient term with respect to an order parameter \cite{vanderwaals1894,anderson1998}.
The latter is directly related to the surface energy and models the surface tension effects.
A thermodynamically consistent coupling of this van der Waals-Korteweg-type fluid with the governing equations of fluid dynamics, the Navier-Stokes equations, was achieved by Dunn and Serrin \cite{dunn1985}, see also \cite{anderson1998,soucek2020}.
This system of partial differential equations, namely the Navier-Stokes-Korteweg (\nsk) equations, inherently accounts for surface tension and phase change.
Since in diffuse interface methods the interfacial zone has to be well resolved, they are restricted to applications where the characteristic length scale of the problem is of the order of the interface thickness.
Thus, van der Waals-Korteweg-type models became attrative in the modeling of porous media flow due to their inherent ability to deal with two phase flow and complex domains \cite{cueto2017,rohde2018}.

Several researchers investigated the \nsk ~model from an analytical, see e.g. \cite{hattori1994,bresch2003,rohde2005,kotschote2008}, and a numerical, see e.g. \cite{jamet2001,coquel2005,diehl2007,haink2008,gomez2010,braack2013,giesselmann2015,tian2015,diehl2016,gelissen2018,martinez2019,gelissen2020}, point of view.
If the numerical discretization of the \nsk ~system is targeted, two issues have to be faced.
First, the dependence of the Helmholtz free energy on the density gradient results in a third-order term in the linear momentum equations, which contributes to the surface tension effects.
This requires, on the one hand, an additional discretization effort compared with the Navier-Stokes equations and, on the other hand, a special numerical treatment to allow wall boundary conditions with contact angle phenomena.
Secondly, the non-convex bulk potential induces a non-monotonous pressure function, which in turn results in a mixed hyperbolic-elliptic type of the first-order fluxes.
Hence, the straightforward use of classical upwind finite volume (\fv) schemes, which rely on the solution of a Riemann problem, is not possible.
Moreover, due to the mixed hyperbolic-elliptic structure of the first order operator, the system is hardly accessible for asymptotic analysis methods, see e.g. \cite{rohde2020}.
This prevents the derivation of homogenized  models on the macro-scale.

Different solution approaches have been presented to overcome these disadvantages.
In \cite{rohde2010}, the author proposed a relaxation system for the isothermal \nsk ~equations that handles the capillarity effects by a local and low-order differential operator.
The system is of second-order with an additional relaxation parameter as an unknown which has to fulfill a linear elliptic equation.
\citeauthor{neusser2015} \cite{neusser2015} utilized the structure of this elliptic relaxation system to develop a numerical scheme which avoids the mentioned numerical issues of the original \nsk ~system.
\citeauthor{chertock2017} \cite{chertock2017} constructed an asymptotic preserving method for the elliptic relaxation system.
Their approach relies on an implicit-explicit operator splitting combined with a first-order \fv ~method.
However, the elliptic constraint renders the overall system to be of mixed type, which aggravates a consistent high-order and locally mesh-resolved numerical approach.
Motivated by the work in \cite{corli2014,rohde2018} on scalar model problems, \citeauthor{hitz2020} \cite{hitz2020} suggested a parabolic evolution equation instead of the elliptic constraint for the additional relaxation parameter.
This modified system enabled them to solve much more complex two-phase problems.
The system was discretized with a high-order discontinuous Galerkin spectral element method (DGSEM) in combination with a second-order total variation diminishing \fv~ subcell scheme which was used as a local limiter.
Their numerical approach enabled stable and thermodynamically consistent computations for high Korteweg parameters.
As far as the authors are aware, no attempts have been made to extend any relaxation scheme of the \nsk ~equations to the full non-isothermal system.

Moreover, there is only a small amount of publications which address the application of the \nsk ~system in confined and porous media domains.
\citeauthor{tian2015} \cite{tian2015} used a local discontinuous Galerkin scheme for the discretization of the non-isothermal \nsk ~system in a confined domain, but restricted themselves to $90 \degree$ contact angles and adiabatic walls.
In \cite{jamet2001}, the authors investigated water in the vicinity of the critical point.
For this, they used a modified van der Waals equation of state which is only valid in the vicinity of the critical point.
They considered contact angles prescribed by a spatially fixed density gradient normal to the wall.
However, they did not provide a thermodynamic motivation for their choice of the contact angle.
\citeauthor{desmarais2016} \cite{desmarais2016} considered a static contact angle boundary condition which is consistent with the Second Law of Thermodynamics.
For this, based on the works of \cite{jacqmin2000,sibley2013a,sibley2013b}, the author introduced a polynomial function at the solid surface as an additional wall interaction energy.
This boundary condition was then used to investigate two-dimensional bubble nucleation processes in water on a heated wall.
The same numerical framework was employed by \citeauthor{gelissen2020} \cite{gelissen2020} for simulations of three-dimensional droplets which impinge on a heated solid wall.
However, due to their conservative second-order \fv ~discretization of the \nsk ~system, they have to use an iterative scheme at the solid boundaries to guarantee the consistency of the density and its first and second gradients at the solid surface.
This procedure enabled them for a consistent surface flux computation.
\citeauthor{soucek2020} \cite{soucek2020} presented a general mathematical framework for the derivation of thermodynamically consistent boundary conditions for Korteweg-type fluids.
Based on different, physically motivated, surface Helmholtz free energies, they derived a variety of non-equilibrium boundary conditions, which are a generalization of the non-equilibrium boundary conditions previously presented in \cite{jacqmin2000,heida2013}.
In particular, they identified extensions of the classical Navier slip condition and the static contact angle boundary condition.
In both boundary conditions, the additional contributions arise from dynamic contact angle effects induced by velocity gradients.
In addition to their theoretical derivations, they presented numerical results based on a finite element discretization of the isothermal \nsk ~equations.
They used a mixed formulation for the discretization of the third-order Korteweg tensor with a non-conservative first-order contribution in the momentum equations and an additional elliptic constraint for the density.

Turning to porous media flow, in \cite{cueto2017} the authors used a reduced formulation of the \nsk ~equations for the investigation of spinodal decompositions and evaporation processes in different complex two-dimensional pore structures.
However, the model considered is based on several assumptions: isothermal flow, negligence of inertia and the viscous forces are assumed to be proportional to the gap-integrated fluid velocity.
Therefore, the linear momentum equations reduced to a Darcy-type volumetric flux which limits the applicability of the model to Hele-Shaw cell flow.
Moreover, the boundary conditions imposed during their investigations do not fit into the thermodynamic framework of \citeauthor{soucek2020}.
The reason for this is that they would require a functional dependence of the surface Helmholtz free energy on the gradient of the density which contradicts with the statistical physics description of the surface free energy \cite{soucek2020}.
To the authors best knowledge, there was no application of the complete \nsk ~model in porous media flow.

The present paper can be seen in line with the works \cite{rohde2010,neusser2015,chertock2017,rohde2018,dhaouadi2019,hitz2020} towards a robust, efficient and thermodynamically consistent approximation of non-isothermal liquid-vapor flow with phase transition, capable to handle flow in complex pore structures.
The derivation of the non-isothermal relaxation system is based on the free energy potential initially proposed in \cite{rohde2010} as well as the principle of entropy maximization, see e.g. \cite{rajagopal2004} and follows the thermodynamic framework of \citeauthor{heida2012a} \cite{heida2012a,heida2012b}.
In accordance to the works \cite{rohde2010,neusser2015,rohde2018,hitz2020}, the relaxation model is parameterized by a Korteweg parameter such that if it tends to infinity, the original \nsk ~model is formally recovered.
Moreover, the relaxation model enables to introduce a modified pressure that guarantees hyperbolicity of the convective fluxes if the Korteweg parameter is kept fixed and is large enough.
This allows the straightforward use of upwind based numerical schemes and further asymptotic analysis.
Furthermore, the proposed relaxation model enables the direct use of contact angle boundary conditions without the requirement of a mixed discretization \cite{soucek2020} or an iterative scheme at the solid boundaries \cite{desmarais2016,gelissen2020}.

The remainder of the paper is structured as follows.
A review of the thermodynamic setting and the principles of classical irreversible thermodynamics (\cit) is provided in the \cref{sec:thermodynamic_setting,sec:cit}, respectively.
We revisit the original Navier-Stokes-Korteweg model in \cref{sec:navier_stokes_korteweg} and finally present the new non-isothermal relaxation formulation in \cref{sec:relaxation_model}.
In a next step, thermodynamically consistent boundary conditions for both models are given in \cref{sec:boundary_conditions}.

With the physical models fixed, in \cref{sec:numerics_dgsem} we continue with the numerical scheme used for the discretization of the bulk flow which is based on an extension of the open-source framework \flexi \footnote{https://www.flexi-project.org}.
In addition, it is shown that the numerical method can easily cope with the contact angle boundary conditions for both model formulations.

Turning to numerical experiments, in \cref{sec:results} we first show that the solution of the original NSK model can be recovered by the relaxation model for different one- and two-dimensional problems.
Moreover, we demonstrate for all test cases that the presented relaxation formulation is consistent with the Second Law of Thermodynamics.
The numerical experiments in the bulk are concluded in \cref{sec:3D}, where three-dimensional simulations of head-on droplet collisions are presented.

Then, the contact angle boundary conditions from \cref{sec:boundary_conditions} are validated.
First, a static setup is investigated, where the results are compared to the Young-Laplace law, and in a second test case dynamic contact angle effects are analyzed.
All simulations are performed in three dimensions.
The section concludes with the three-dimensional simulation of a spinodal decomposition in a porous media, where we exploited the high-order boundary approximation techniques of the \flexi ~framwork.
A short conclusion is given in \cref{sec:conclusion}.
\section{Diffuse Interface Models}
\label{sec:diffuseinterfacemodels}
\subsection{Thermodynamic Settings}
\label{sec:thermodynamic_setting}
We consider a time interval $\timeinterval$ and a bounded domain $\domain \subset \Rbb^{\dimensions}$, $\dimensions \in \l\lbrace 1, 2 , 3 \r\rbrace$ with the boundary $\domainsurf$.
The domain is occupied by a homogeneous fluid which can occur in a liquid and a vapor state.
If the van der Waals equation of state (\eos) is assumed as the material law of the fluid, the non-dimensional Helmholtz free energy per unit volume is given by
\begin{align}
\label{eq:helmholtzfreeenergyvol}
\helmfreevol \l( \dens , \temp \r) = \dens \Rvdw \temp \cvdw \l( 1 - \ln \l( \temp \r) \r) + \dens \Rvdw \temp \ln \l( \frac{\bvdw \dens}{1 - \bvdw \dens} \r) - \avdw \dens^2.
\end{align}
In \cref{eq:helmholtzfreeenergyvol}, the symbols $\dens: \spacetimeinterval \rightarrow \l[ 0 , \frac{1}{b} \r)$ and $\temp: \spacetimeinterval \rightarrow \Rgzero$ indicate the non-dimensional mass density and the temperature, respectively.
The reference states used to de-dimensionalize the van der Waals {\eos} are the density $\rhocrit$, the temperature $\Tcrit$ and the pressure $\pcrit$ at the critical point, a reference length $\Lref$ and the gas constant $\gasconst$ of the fluid.
The critical states are given by
\begin{align}
\label{eq:criticalstates}
\rhocrit = \frac{1}{3 \tilde{\bvdw}}, \quad \pcrit = \frac{\tilde{\avdw}}{27 {\tilde{\bvdw}}^2}, \quad \Tcrit = \frac{8 \pcrit}{3 \gasconst \rhocrit}
\end{align}
with the cohesion pressure $\tilde{\avdw}$ and the the co-volume $\tilde{\bvdw}$.
Based on the critical states, \cref{eq:criticalstates}, and the reference length $\Lref$, reference quantities for the velocity, the time and the energy
\begin{align}
\label{eq:referencestates}
\Uref = \sqrt{\frac{\pcrit}{\rhocrit}}, \quad \Tref = \frac{\Lref}{\Uref}, \quad \Eref = \Uref^2
\end{align}
can be defined.
Consequently, the constant material parameters $\Rvdw , \avdw , \bvdw , \cvdw$ in \cref{eq:helmholtzfreeenergyvol} are given as
\begin{align*}
\Rvdw = \frac{8}{3} , ~\avdw = 3 , ~\bvdw = \frac{1}{3}, ~\cvdw = \frac{\specheatcapvol}{\gasconst}
\end{align*}
with the specific heat capacity ratio at a constant volume $\specheatcapvol > 0$.

\begin{figure}
\begin{center}
	\includegraphics[width=0.9\linewidth,trim = 0.5cm 0 0 0, clip]{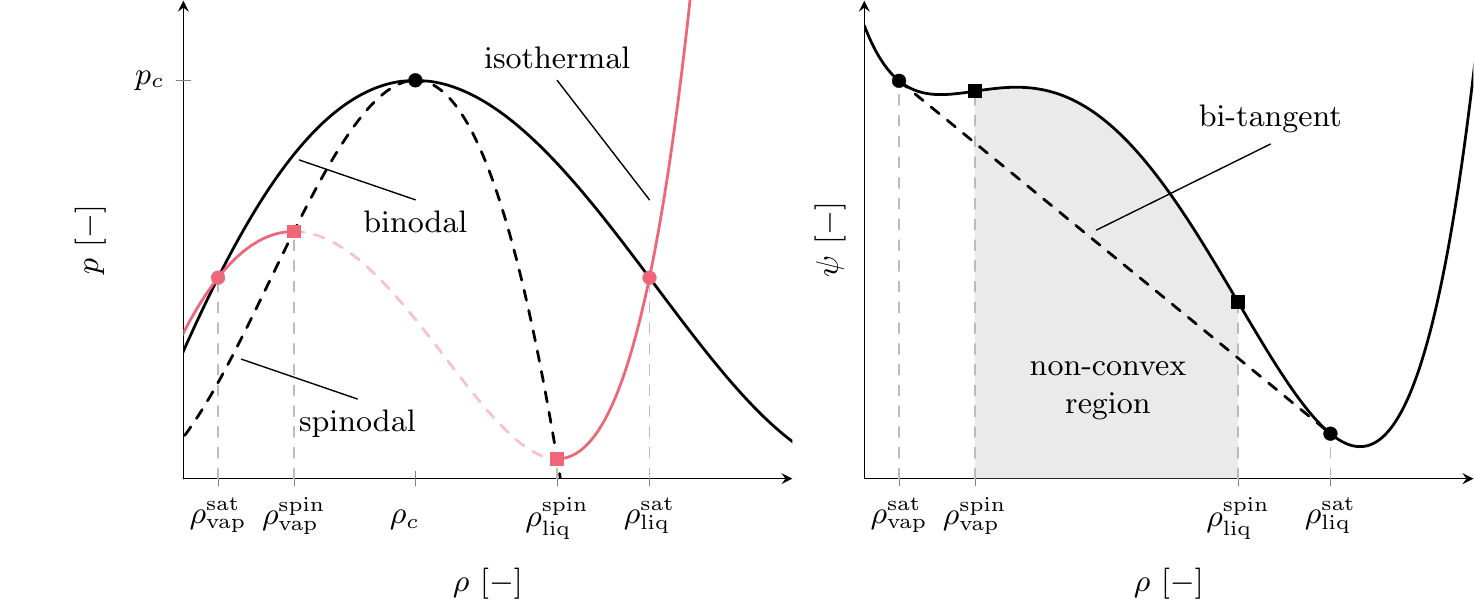}
\end{center}
\caption{\label{fig:thermodynamics}Graphs of the pressure $\pres$ (left) and the Helmholtz free energy per unit volume $\helmfreevol$ (right).}
\end{figure}

\cref{eq:helmholtzfreeenergyvol} is a thermodynamic potential, which allows the derivation of further thermodynamic state properties, e.g. via derivatives of \cref{eq:helmholtzfreeenergyvol}.
E.g., the pressure of the van der Waals fluid is given by
\begin{align}
\label{eq:pressure}
\pres = \dens \l( \fracp{\helmfreevol}{\dens} \r)_{\temp} - \helmfreevol = \frac{\dens \Rvdw \temp}{1 - \bvdw \dens} - \avdw \dens^2
\end{align}
and its first derivative as
\begin{align}
\label{eq:firstpressurederivative}
\l( \fracp{\pres}{\dens} \r)_\temp = \dens \l( \fracpp{\helmfreevol}{\dens} \r)_{\temp}.
\end{align}
The Helmholtz free energy per unit volume, \cref{eq:helmholtzfreeenergyvol}, and the pressure law, \cref{eq:pressure}, are depicted in \cref{fig:thermodynamics}.
The saturation dome with the saturation curves (or binodal) are illustrated on the left (black solid line).
On this curve, the fluid may exist in a saturated vapor or a saturated liquid state, denoted by the Maxwellian densities $\dens^{\sat}_{\vap}$ and $\dens^{\sat}_{\liq}$, respectively.
At a fixed temperature, the saturation states obtain equivalent values for the pressure as depicted by the isothermal (red line).
The spinodal curves are indicated by the black dashed line.
These curves coincide with the local extrema of the isothermals, which are located at the spinodal densities $\dens^{\spin}_{\vap}$ and $\dens^{\spin}_{\liq}$.
Due to the so called van der Waals loop which connects a liquid to a vapor state, the first derivative of the pressure exhibits a negative sign  inside the spinodal region (red dashed line), i.e. for $\dens \in ( \dens^{\spin}_{\vap} , \dens^{\spin}_{\liq} )$.
Hence, according to \cref{eq:firstpressurederivative}, the spinodal region coincides with a non-convex region in the Helmholtz free energy per unit volume, as depicted on the right in \cref{fig:thermodynamics}.
Thus, as long as $\temp < \Tcrit$, the van der Waals fluid is able to model a homogeneous fluid in a vapor and a liquid state.
Assuming thermodynamic equilibrium, the phase change will take place at a constant pressure, as indicated by the black dashed line on the right in \cref{fig:thermodynamics}.
In the context of the $\pres$-$\rho$ diagram this line is known as the Maxwell line and in the context of the $\helmfreevol$-$\rho$ diagram (right panel) as the so called bi-tangent.

Neglecting capillary effects, equilibrium states of the fluid in $\domain$ have to be minimizers of the Helmholtz free energy functional
\begin{align}
\label{eq:helmholtzfreefunctional}
\helmfreefunc = \intdomain \helmfreevol \l( \dens , \temp \r) \dx.
\end{align}
However, minimizers for \cref{eq:helmholtzfreefunctional} are not unique, even for an isothermal fluid and a fixed total mass.
Nevertheless, at a constant temperature, minimizers in the set of piecewise continuous functions take exactly the Maxwellian densities $\dens^\sat_\vap$ and $\dens^\sat_\liq$ \cite{carr1984}.
To consider capillary effects, van der Waals \cite{vanderwaals1894} proposed in his square gradient theory for liquid-vapor phase interfaces to extend the Helmholtz free energy functional by an additional contribution
\begin{align}
\label{eq:surfacefunctionalvdw}
\surfacefuncvdw = \intdomain \frac{\gammakorteweg}{2} \l| \gradient \dens \r|^2 \dx
\end{align}
with some coefficient $\gammakorteweg > 0$.
This additional energy functional penalizes the occurrence of spatial gradients in the density and can be understood as a resistance to the formation of an interface \cite{heinen2022}.
Consequently, under consideration of \cref{eq:helmholtzfreefunctional,eq:surfacefunctionalvdw}, the total Helmholtz free energy functional is given by
\begin{align}
\label{eq:totalhelmholtzfreefunctional}
\helmfreefunctot &= \Ekin + \helmfreefunc + \surfacefuncvdw
\end{align}
with the kinetic energy defined as
\begin{align*}
\Ekin = \intdomain \frac{1}{2} \dens \l| \vel \r|^2 \dx,
\end{align*}
where $\vel: \spacetimeinterval \rightarrow \Rbb^\dimensions$ indicates the velocity vector.
\subsection{Principles from Classical Irreversible Thermodynamics}
\label{sec:cit}
Given the total Helmholtz free energy functional in \cref{eq:totalhelmholtzfreefunctional}, the non-dimensional total energy per unit mass is defined as
\begin{align}
\label{eq:totalenergymass}
\ener \l( \entropymass, \vel, \dens, \gradient \dens \r) = \frac{1}{2} \vel \cdot \vel + \eintmass \l( \dens , \entropymass \r) + \frac{\gammakorteweg}{2 \dens} \gradient \dens \cdot \gradient \dens,
\end{align}
where the symbol $\eintmass \l( \dens , \entropymass \r)$ indicates the non-dimensional internal energy per unit mass as a function of the density and the entropy per unit mass $\entropymass$.
Once the total energy per unit mass is specified, an elegant derivation of the balance equations of mass, linear momentum and energy can be achieved by the First Law of Thermodynamics and the imposition of the principle of Galilean invariance \cite{jou2010}.
Neglecting external energy supply, this yields
\begin{subequations}
\label{eq:balancelaws}
	\begin{align}
	\label{eq:lagrange_mass}
	\material{\dens} + \dens \gradient \cdot \vel &= 0, \\
	\label{eq:lagrange_linearmomentum}
	\dens \material{\vel} - \gradient \cdot \Tbb &= \dens \force, \\
	\label{eq:lagrange_energy}
	\dens \material{\ener} - \gradient \cdot \l( \Tbb \cdot \vel + \je \r) &= \dens \force \cdot \vel
	\end{align}
\end{subequations}
for the non-dimensional balance of mass, linear momentum and total energy in the Lagrangian frame, respectively.
In \cref{eq:balancelaws}, the Cauchy stress $\Tbb: \spacetimeinterval \rightarrow \Rbb^{\dimensions \times \dimensions}$ indicates the conductive momentum transport and the vectors $\je: \spacetimeinterval \rightarrow \Rbb^{\dimensions}$ and $\force: \spacetimeinterval \rightarrow \Rbb^{\dimensions}$ the diffusive energy flux and an external body force, respectively.
Moreover, the conservation of the angular momentum implies the symmetry of the Cauchy stress $\Tbb = \transpose{\Tbb}$ in the absence of intrinsic rotational motion and external couples \cite{jou2010}.
The balance laws in \cref{eq:balancelaws} combined with the EoS in \cref{eq:helmholtzfreeenergyvol} only consider the microscopic interactions in the fluid.
The mesoscopic or macroscopic structure of the fluid enter the balance equations via the thermodynamic fluxes $\Tbb$ and $\je$ \cite{heida2012a}.
However, these constitutive relations have to be determined on the basis of the underlying total Helmholtz free energy functional and the corresponding total energy per unit mass.

Following the works of \citeauthor{heida2012a} \cite{heida2012a,heida2012b}, the total energy per unit mass, \cref{eq:totalenergymass}, can be written in the general form
\begin{align}
\label{eq:totalenergygeneric}
\ener \l( \entropymass, \statevector \r) \quad \text{with} \quad \statevector = \transpose{\l( \statevariable_1 , ... , \statevariable_\mstatevariables \r)} = \transpose{\l( \vel, \dens, \gradient \dens\r)},
\end{align}
where $\statevector$ indicates the vector of the state variables.
Hence, multiplication of the material derivative of \cref{eq:totalenergygeneric} by the density yields
\begin{align}
\label{eq:totalenergybalancegeneric}
\dens \material{\ener} = \dens \material{\entropymass} \temp + \sum_{i = 1}^\mstatevariables \dens \material{\statevariable_i} \l( \fracp{\ener}{\statevariable_i} \r)_{\entropymass , \lbrace \statevariable_{j \neq i} \rbrace}
\end{align}
for the temporal evolution equation of the total energy per unit mass.
In \cref{eq:totalenergybalancegeneric}, the identity $\temp = \l( \fracp{\ener}{\entropymass} \r)_{\lbrace \statevariable_i \rbrace}$ has been used.
By substituting the \cref{eq:lagrange_mass,eq:lagrange_linearmomentum,eq:lagrange_energy} and an additional evolution equation for the gradient of the density $\gradient \dens$ which can be derived from \cref{eq:lagrange_mass} into \cref{eq:totalenergybalancegeneric}, one obtains a balance equation for the non-dimensional entropy per unit mass of the form
\begin{align}
\label{eq:entropybalance}
\dens \material{\entropymass} - \gradient \cdot \l( \js \r) = \srs.
\end{align}
The expressions $\js$ and $\srs \coloneqq \srstemp$ indicate the entropy flux due to diffusion and the rate of entropy production, respectively.
If $\srs$ can be shown to be positive, the Second Law of Thermodynamics is automatically fulfilled.
As in \cite{heida2012a,soucek2020}, we restrict ourselves to the concept of linear irreversible thermodynamics.
Thus, $\srs$ is required to fulfill the bilinear structure
\begin{align}
\label{eq:rateofentropyproduction}
\srs \l( \jkappavec , \affinitiesvec \r) = \sum_\kappa \jkappa \cdot \fkappa{\affinitiesvec}, \qquad \kappa \in \Ngzero.
\end{align}
In \cref{eq:rateofentropyproduction}, $\jkappavec = \l( \jkappaone , \jkappatwo \r) = \l( \Tbb , \je \r)$ and $\affinitiesvec$ are the vector of the thermodynamic fluxes and the vector of the thermodynamic affinities (forces), respectively, and $\fkappa{\affinitiesvec}$ indicates a function of $\affinitiesvec$.
According to \cite{rajagopal2004,heida2012a}, the rate of entropy production should not only fulfill the Second Law of Thermodynamics but also be maximized with respect to the thermodynamic fluxes.
Since we only consider the quadratic relation
\begin{align*}
\srs \l( \jkappavec \r) = \sum_\kappa \frac{1}{\affinitycoef{\kappa}} \l| \jkappa \r|^2 \geq 0,
\end{align*}
the bilinear structure for the rate of entropy production in \cref{eq:rateofentropyproduction} motivates to enforce the non-negativity by choosing $\jkappa$ according to
\begin{align}
\label{eq:linearfluxforcerelation}
\jkappa = \affinitycoef{\kappa} \fkappa{\affinitiesvec}.
\end{align}
The symbols $\affinitycoef{\kappa}$ in \cref{eq:linearfluxforcerelation} have to be positive parameters to ensure the positivity of the entropy production and may depend on the density and the temperature $\affinitycoef{\kappa} = \affinitycoef{\kappa} \l( \dens, \temp \r) > 0$.
Additionally, the linear flux-force relation in \cref{eq:linearfluxforcerelation} suggests to formulate the rate of entropy production $\srs$ in terms of the function vector of the thermodynamic affinities
\begin{align*}
\srs \l( \affinitiesvec \r) = \sum_\kappa \affinitycoef{\kappa} \l| \fkappa{\affinitiesvec} \r|^2 \geq 0.
\end{align*}

Thus, given a total Helmholtz free energy functional, \cref{eq:totalhelmholtzfreefunctional}, temporal evolution equations for the state variables $\statevector$, \cref{eq:balancelaws}, the balance equation for the entropy, \cref{eq:entropybalance}, the constitutive equations for the rate of entropy production and the thermodynamic fluxes, \cref{eq:rateofentropyproduction,eq:linearfluxforcerelation} as well as an educated choice for the thermodynamic affinities $\affinitiesvec$ and the function $\fkappa{\affinitiesvec}$, constitutive relations for the thermodynamic fluxes $\Tbb$ and $\je$ can be identified.
Based on these constitutive relations and \cref{eq:balancelaws}, a set of balance equations for the Korteweg-van der Waals fluid can be specified.

The interested reader is referred to \cite{truesdell1984,rajagopal2004,jou2010} for more details on \cit ~and the derivation of constitutive relations and to \cite{heida2012a,heida2012b,soucek2020} for the specific application to Korteweg-type fluids.
\subsection{The Equations in the Bulk}
In this section, the framework presented in \cref{sec:cit} will be used to derive thermodynamically consistent constitutive relations for the thermodynamic fluxes of the original and the relaxation formulation of the \nsk ~model.
\subsubsection{Navier-Stokes-Korteweg Equations}
\label{sec:navier_stokes_korteweg}
The Helmholtz free energy functional, \cref{eq:helmholtzfreefunctional}, in combination with the balance laws in \cref{eq:lagrange_mass,eq:lagrange_linearmomentum,eq:lagrange_energy} and the vector valued function of the affinities
\begin{align*}
\fkappavec \l( \affinitiesvec \r) = \l( \gradient \cdot \vel , \Dbbdevfree, \frac{\gradient \temp}{\temp} \r)
\end{align*}
yields the constitutive relations
\begin{subequations}
	\begin{align}
	\label{eq:cauchystressnsk}
	\Tbb \coloneqq &- \pres \Ibb + \viscousstress + \kortewegstress \nonumber \\
	= &- \pres \Ibb + 2 \munewton \Dbb + \l[ \lambdastokes \gradient \cdot \vel \r] \Ibb + \gammakorteweg \l[ \frac{1}{2} \l| \gradient \dens \r|^2 + \dens \laplace \dens \r] \Ibb - \gammakorteweg \gradient \dens \otimes \gradient \dens, \\
	\label{eq:energyfluxnsk}
	\je \coloneqq &~\jfourier + \jnsk = \heatcoef \gradient \temp - \gammakorteweg \dens \gradient \dens \gradient \cdot \vel
	\end{align}
\end{subequations}
for the thermodynamic fluxes.
Here, $\Ibb$ indicates the identity matrix, $\Dbb$ the rate of deformation tensor $\Dbb = \frac{1}{2} ( \gradient \vel + \transpose{\l( \gradient \vel \r)} )$, $\Dbbdevfree$ the deviatoric free rate of deformation tensor $\Dbbdevfree = \Dbb - \frac{1}{3} \l( \gradient \cdot \vel \r) \Ibb$ and the symbols $\munewton$, $\lambdastokes$, $\heatcoef$, $\gammakorteweg$ are non-dimensional parameters which correspond to the parameters $\affinitycoef{\kappa}$ in \cref{eq:linearfluxforcerelation}.
With the reference quantities defined in \cref{eq:criticalstates,eq:referencestates} they are given as
\begin{align}
\label{eq:non-dimensionals}
\munewton = \frac{\tilde{\munewton} \epsilonkorteweg}{\rhocrit \Uref \Lref}, \quad
\lambdastokes = \frac{\tilde{\lambdastokes} \epsilonkorteweg}{\rhocrit \Uref \Lref}, \quad
\heatcoef = \frac{\tilde{\heatcoef} \Tcrit \epsilonkorteweg}{\rhocrit \Uref^3 \Lref} =
\frac{\tilde{\munewton} \epsilonkorteweg}{\rhocrit \Uref \Lref} ~ \frac{\tilde{\heatcoef}}{\tilde{\munewton} \specheatcapvol} ~ \frac{8 \specheatcapvol}{3 \gasconst}, \quad
\gammakorteweg = \frac{\gammaepsilonkorteweg \rhocrit}{\Uref^2 \Lref^2}
\end{align}
with the dynamic viscosity $\tilde{\munewton} > 0$, the bulk viscosity $\tilde{\lambdastokes} > 0$, the thermal conductivity $\tilde{\heatcoef} > 0$, the capillary coefficient $\capillarycoef >0$ and some scaling parameter $\epsilonkorteweg > 0$, which is assumed as a constant parameter during this work.
The first term in \cref{eq:cauchystressnsk} is related to the thermodynamic pressure, the second and third terms can be identified as the viscous stress tensor
\begin{align}
\viscousstress = 2 \munewton \Dbb + \l[ \lambdastokes \gradient \cdot \vel \r] \Ibb = \munewton \l[ \gradient \vel + \transpose{\l( \gradient \vel \r)} - \frac{2}{3} \gradient \cdot \vel \Ibb \r],
\end{align}
where we applied only for simplicity the Stokes hypothesis
\begin{align*}
\tilde{\lambdastokes} = \tilde{\munewton}_2 - \frac{2}{3} \tilde{\munewton}
\end{align*}
with the volumetric viscosity $\tilde{\munewton}_2 = 0$.
The last two terms in \cref{eq:cauchystressnsk} contribute to the so called Korteweg tensor $\kortewegstress$, and in \cref{eq:energyfluxnsk}, the first term represents the diffusive energy flux due to a temperature gradient modeled by Fourier's law.
Finally, the second term is the interstitial work flux
\begin{align*}
\jnsk =  -\gammakorteweg \dens \gradient \dens \gradient \cdot \vel
\end{align*}
first introduced by \citeauthor{dunn1985} to guarantee a positive rate of entropy production.
Thus, in terms of the function vector of the affinities, the rate of entropy production for the original \nsk ~model is given by
\begin{align*}
\srs \l( \affinitiesvec \r) = \frac{1}{\temp} \l\lbrace \frac{4}{3} \munewton \l| \gradient \cdot \vel \r|^2 + \munewton \l| \Dbbdevfree \r|^2 + \heatcoef \l| \frac{\gradient \temp}{\temp} \r|^2 \r\rbrace \geq 0,
\end{align*}
which underlines the consistence with the Second Law of Thermodynamics.

\begin{remark}[Interstitial work flux]
In the literature, an alternative formulation of the interstitial work flux has been proposed, where a contribution in the linear momentum equations instead of the total energy equation is considered \cite{heida2010,freistuehler2017,soucek2020}.
However, \citeauthor{giovangigli2020} \cite{giovangigli2020} argued that this alternative formulation of the interstitial work flux, while mathematically correct, is nonphysical.
He outlined that \cit ~and rational thermodynamics are well suited for linear relations between the fluxes and the affinities.
Nevertheless, for nonlinear systems like capillary fluids, where multiple affinities can be related to exactly the same contribution in the entropy production rate \cref{eq:rateofentropyproduction}, \cit ~is unable to distinguish between the physical and the nonphysical flux and additional finer physical theories are required \cite{giovangigli2020}.
\citeauthor{giovangigli2020} favoured the interstitial work flux in the form of \citeauthor{dunn1985} which has been obtained, e.g. from Hamiltonian's principle in \cite{gavrilyuk1996} and in \cite{giovangigli2020} from the kinetic theory of dense gases.
\end{remark}

Hence, the Navier-Stokes-Korteweg equations in the Eulerian frame are given by
\begin{subequations}
	\label{eq:navierstokeskorteweg}
	\begin{align}
	\label{eq:euler_mass}
	\dens_t = \gradient \cdot \l( \dens \vel \r) &= 0, \\
	\label{eq:euler_linearmomentum}
	\l( \dens \vel \r)_t + \gradient \cdot \l( \dens \vel \otimes \vel + \pres \Ibb \r) & = \gradient \cdot \l[ \viscousstress +\kortewegstress \r], \\
	\label{eq:euler_energy}
	\l( \dens \ener \r)_t + \gradient \cdot \l( \l( \dens \ener + \pres \r) \vel \r) &= \gradient \cdot \l[ \l( \viscousstress + \kortewegstress \r) \cdot \vel \r] + \gradient \cdot \je.
	\end{align}
\end{subequations}
In the following, we refer to the formulation in \cref{eq:navierstokeskorteweg} as the \originalmodel.
According to \citeauthor{dreyer2012} \cite{dreyer2012} and \citeauthor{neusser2015} \cite{neusser2015}, in the sharp interface limit $\epsilonkorteweg \rightarrow 0$ of \cref{eq:navierstokeskorteweg}, the Reynolds numbers as well as the Prandtl and the Weber number
\begin{align*}
\Reynolds = \frac{\rhocrit \Uref \Lref}{\tilde{\munewton}}, \quad
\Reynolds_\lambdastokes = \frac{\rhocrit \Uref \Lref}{\tilde{\lambdastokes}}, \quad
\Prandtl = \frac{\tilde{\munewton} \specheatcapvol}{\tilde{\heatcoef}}, \quad
\Weber = \frac{\Uref^2 \Lref^2}{\gamma \rhocrit}
\end{align*}
can be identified in \cref{eq:non-dimensionals}.
Similar to \cite{gelissen2020}, the specific heat capacity at a constant volume $\specheatcapvol$ is used for the definition of the Prandtl number, since the specific heat capacity at a constant pressure takes negative values inside the spinodal region.

The eigenvalues of the Jacobian of the first-order fluxes are given by
\begin{align*}
\eigenvalue_1 = \vel \cdot \normvec - \sqrt{ \l( \fracp{\pres}{\dens} \r)_\entropymass }, \quad
\eigenvalue_{2,...,\dimensions+1} = \vel \cdot \normvec, \quad
\eigenvalue_{\dimensions+2} = \vel \cdot \normvec + \sqrt{ \l( \fracp{\pres}{\dens} \r)_\entropymass },
\end{align*}
for an arbitrary normal vector $\normvec \in {\Scal}^{\dimensions-1}$ with the square of the speed of sound
\begin{align}
\label{eq:speedofsound}
\speedofsound^2 = \l( \fracp{\pres}{\dens} \r)_\entropymass = \l( \fracp{\pres}{\dens} \r)_\temp - \l( \fracp{\pres}{\temp} \r)_\dens \l( \fracp{\entropymass}{\dens} \r)_\temp \l( \fracp{\entropymass}{\temp} \r)^{-1}_\dens.
\end{align}
Due to the non-convexity of the Helmholtz free energy per unit volume in the spinodal region, the speed of sound might become an imaginary number, which causes the first-order fluxes in \cref{eq:navierstokeskorteweg} to be of mixed hyperbolic-elliptic type.
In the case of an isothermal process $\cvdw \rightarrow \infty$, this is true in the entire spinodal region as \cref{eq:speedofsound} reduces to the first term on the right hand side and this term is negative in the interval $( \dens_\vap^\spin , \dens_\liq^\spin )$ for $\temp < \Tcrit$, see also \cref{fig:thermodynamics}.
However, in a non-isothermal process the loss of hyperbolicity in the spinodal region additionally depends on the specific choice of $\cvdw$, see also \cite{desmarais2014}.
From a numerical point of view, the loss of hyperbolicity prevents the straightforward use of upwind schemes which are based on a Riemann solver.
Thus, numerical methods are commonly based on a simple numerical flux function which is independent of the local wave speeds, e.g. the global Lax-Friedrichs flux \cite{tian2015} or the central flux \cite{diehl2016,gelissen2018}.
Furthermore, any explicit time-stepping scheme requires an estimate of the expected maximum wave speed, which is impossible if $\eigenvalue_{\dimensions+2}$ is a complex number.
From an analytical point of view, due to the mixed hyperbolic-elliptic structure of the first-order operator, the system is hardly accessible for asymptotic analysis methods, see e.g. \cite{rohde2020} and thus prevents the derivation of homogenized models on the macro-scale.

Moreover, due to the Korteweg stress, the linear momentum equation, \eqref{eq:euler_linearmomentum}, is a third-order convection-diffusion-dispersion equation and therefore requires the calculation of higher-order gradients.
This is especially aggravating in the isothermal case \cite{rohde2010}.
In the non-isothermal case, two consecutive gradient computations are required anyway.
First the density gradient is needed to determine the internal energy per unit mass and thus the temperature, see also \cref{eq:totalenergymass}, and only in a second step the temperature gradient required for the Fourier energy fluxes can be calculated.
\subsubsection{The Relaxation Model}
\label{sec:relaxation_model}
To overcome the numerical difficulties mentioned in the previous section, a relaxation system for the isothermal \nsk ~equations has been proposed in \cite{rohde2010}.
It handles the capillarity effects by a local and low-order differential operator.
The key idea is to introduce for a Korteweg parameter $\alphakorteweg>0$ an additional scalar field $\ckorteweg^\alphakorteweg$ which acts as a new order parameter and converges for $\alphakorteweg \rightarrow \infty$ to $\dens$.
For this, the total Helmholtz free energy functional, \cref{eq:totalhelmholtzfreefunctional}, is modified in such a way that the gradient of the density $\gradient \dens$ is substituted by the gradient of the order parameter $\gradient \ckorteweg^\alphakorteweg$ and a penalty term is added which vanishes in the limit $\alphakorteweg \rightarrow \infty$.
Hence, the total Helmholtz free energy functional of the relaxation system reads as
\begin{align}
\label{eq:totalhelmfreefunctionalalpha}
\helmfreefuncalphatot = \intdomain \l( \frac{1}{2} \dens^\alphakorteweg \l| \vel^\alphakorteweg \r|^2 + \helmfreevol \l( \dens^\alphakorteweg , \temp^\alphakorteweg \r) + \frac{\alphakorteweg}{2} \l( \dens^\alphakorteweg - \ckorteweg^\alphakorteweg \r)^2 + \frac{\gammakorteweg}{2} \l| \gradient \ckorteweg^\alphakorteweg \r|^2 \r) \dx
\end{align}
with the Korteweg parameter $\alphakorteweg > 0$, which guarantees asymptotic convergence towards the original \nsk ~equations for $\alpha \rightarrow \infty$.
Mass-constrained minimizers of $\Ecal^\alphakorteweg_{\mathrm{tot}}$ at thermal equilibrium ($\temp = \mathrm{const.}$) then satisfy the Euler-Lagrange equations
\begin{align}
\label{eq:eulerlagrange}
\l( \fracp{\helmfreevol}{\dens^\alphakorteweg} \r)_{\temp^\alphakorteweg} - \alphakorteweg \l( \ckorteweg^\alphakorteweg - \dens^\alphakorteweg \r) = \Lcal, \quad \alphakorteweg \l( \ckorteweg^\alphakorteweg - \dens^\alphakorteweg \r) - \gammaepsilonkorteweg \laplace \ckorteweg^\alphakorteweg = 0,
\end{align}
where $\Lcal$ is a Lagrange multiplier.
\citeauthor{rohde2010} \cite{rohde2010} suggested to use the second Euler-Lagrange equation in \cref{eq:eulerlagrange} as an elliptic constraint for the order parameter.
\citeauthor{neusser2015} \cite{neusser2015} proved the model to be consistent with the First and Second Law of Thermodynamics and performed numerical investigations in the Korteweg and the sharp interface limit.
Based on \cref{eq:totalhelmfreefunctionalalpha}, Noether's theorem \cite{noether1918} and the scalar model problem in \cite{corli2014}, \citeauthor{hitz2020} proposed an alternative parabolic relaxation instead of the elliptic constraint to avoid the mixed discretization of the governing equations.

In the present work, we propose an alternative formulation based on the total Helmholtz free energy functional, \cref{eq:totalhelmfreefunctionalalpha}, its corresponding total energy per mass
\begin{align}
\label{eq:totalenergyalpha}
\ener \l( \dens^\alphakorteweg , \entropymass^\alphakorteweg , \ckorteweg^\alphakorteweg , \gradient \ckorteweg^\alphakorteweg , \vel^\alphakorteweg \r) = \frac{1}{2} \l| \vel^\alphakorteweg \r|^2 + \eintmass \l( \dens^\alphakorteweg , \entropymass^\alphakorteweg \r) + \frac{\alphakorteweg}{2 \dens^\alphakorteweg} \l( \dens^\alphakorteweg - \ckorteweg^\alphakorteweg \r)^2 + \frac{\gammaepsilonkorteweg}{2 \dens^\alphakorteweg} \l| \gradient \ckorteweg^\alphakorteweg \r|^2
\end{align}
and the thermodynamic framework presented in \cref{sec:cit}, see also \cite{heida2012a,heida2012b}.
In addition to the balance laws for the mass, the linear momentum and the total energy in \cref{eq:balancelaws}, a relaxation equation for the order parameter is required.
Therefore, we postulate an evolution equation of the form
\begin{align}
\label{eq:lagrange_orderparameter}
\material{\ckorteweg}^\alphakorteweg + \ckorteweg^\alphakorteweg \gradient \cdot \vel^\alphakorteweg + \gradient \cdot \jc^\alphakorteweg = \src^\alphakorteweg.
\end{align}
This specific choice is based on the constraint that the order parameter $\ckorteweg^\alphakorteweg$ should asymptotically converge to $\dens$ in some limit.
Hence, \cref{eq:lagrange_orderparameter} should asymptotically converge to the conservation equation of mass \eqref{eq:lagrange_mass} in the limit $\ckorteweg^\alphakorteweg \rightarrow \dens$, which corresponds to a vanishing diffusive flux $\jc$ and source term $\src$, respectively.
Following \cite{heida2012b} and similar to the procedure presented in the \cref{sec:cit,sec:navier_stokes_korteweg}, the vector valued function of the affinities
\begin{align*}
\fkappavec \l( \affinitiesvec^\alphakorteweg \r) = \l( \gradient \cdot \vel^\alphakorteweg , \Dbbdevfreealpha , \chempotc^\alphakorteweg , \gradient \chempotc^\alphakorteweg , \frac{\gradient \temp^\alphakorteweg}{\temp^\alphakorteweg} \r)
\end{align*}
with
\begin{align}
\chempotc^\alphakorteweg &= \dens^\alphakorteweg \l( \fracp{\ener}{\ckorteweg} \r)^\alphakorteweg_{\dens,\entropymass,\gradient \ckorteweg} - \gradient \cdot \l[ \dens^\alphakorteweg \l( \fracp{\ener}{\gradient \ckorteweg} \r)^\alphakorteweg_{\dens,\entropymass,\ckorteweg} \r] \nonumber \\
\label{eq:chemicalpotentialalpha}
&= -\gammakorteweg \laplace \ckorteweg^\alphakorteweg - \alphakorteweg \l( \dens^\alphakorteweg - \ckorteweg^\alphakorteweg \r)
\end{align}
yields the constitutive relations
\begin{subequations}
	\begin{align}
	\label{eq:cauchystressnskalpha}
	\Tbb^\alphakorteweg &= - \pres^\alphakorteweg \Ibb + \Tbb^{\mathrm{V},\alphakorteweg} + \l[ \frac{\gammakorteweg}{2} \l| \gradient \ckorteweg^\alphakorteweg \r|^2 + \frac{\alphakorteweg}{2} \l( \l(\ckorteweg^\alphakorteweg \r)^2 - \l( \dens^\alphakorteweg \r)^2 \r) \r] \Ibb - \gammakorteweg \gradient \ckorteweg^\alphakorteweg \otimes \gradient \ckorteweg^\alphakorteweg, \\
	\label{eq:energyfluxnskalpha}
	\je^\alphakorteweg &= \jfourier^\alphakorteweg -\chempotc^\alphakorteweg \jc^\alphakorteweg - \gammakorteweg \gradient \ckorteweg^\alphakorteweg \l( \ckorteweg^\alphakorteweg \gradient \cdot \vel^\alphakorteweg + \gradient \cdot \jc^\alphakorteweg \r) + \gammakorteweg \gradient \ckorteweg^\alphakorteweg \src^\alphakorteweg, \\
	\label{eq:orderparametersourcenskalpha}
	\src^\alphakorteweg &= - \betakorteweg \chempotc^\alphakorteweg, \\
	\label{eq:orderparameterfluxnskalpha}
	\jc^\alphakorteweg &= - \upsilonkorteweg \gradient \chempotc^\alphakorteweg
	\end{align}
\end{subequations}
for the thermodynamic fluxes, with the parameters $\betakorteweg > 0$ and $\upsilonkorteweg > 0$.
Similar to \cref{eq:cauchystressnsk}, the Cauchy stress of the relaxation model in \cref{eq:cauchystressnskalpha} is composed of the thermodynamic pressure $\pres^\alphakorteweg$, the viscous stress tensor $\viscousstressalpha$ and the Korteweg tensor
\begin{align}
\label{eq:kortewegstressalpha}
\kortewegstressalpha = \l[ \frac{\gammakorteweg}{2} \l| \gradient \ckorteweg^\alphakorteweg \r|^2 + \frac{\alphakorteweg}{2} \l( \l(\ckorteweg^\alphakorteweg \r)^2 - \l( \dens^\alphakorteweg \r)^2 \r) \r] \Ibb - \gammakorteweg \gradient \ckorteweg^\alphakorteweg \otimes \gradient \ckorteweg^\alphakorteweg.
\end{align}
The modified energy potential, \cref{eq:totalenergyalpha}, enables the formulation of the Korteweg stress in terms of the order parameter $\ckorteweg^\alphakorteweg$ and avoids any second-order contribution.
Moreover, the energy flux, \cref{eq:energyfluxnskalpha}, has contributions due to Fourier's law $\jfourier^\alphakorteweg$, the interstitial work flux in terms of the order parameter
\begin{align}
\jnsk^\alphakorteweg = - \gammakorteweg \ckorteweg^\alphakorteweg \gradient \ckorteweg^\alphakorteweg \gradient \cdot \vel^\alphakorteweg,
\end{align}
and fluxes due to the diffusion of the order parameter known from Cahn-Hillard equations, where \cref{eq:orderparameterfluxnskalpha} is related to Fick's law, see also \cite{heida2012a}.
The last term in \cref{eq:energyfluxnskalpha} is an energy flux due to the additional source term in \cref{eq:lagrange_orderparameter}, where \cref{eq:orderparametersourcenskalpha} is known from Allen-Cahn equations to model phase transition effects between different components, see also \cite{heida2012b}.
Substitution of \cref{eq:chemicalpotentialalpha,eq:orderparametersourcenskalpha,eq:orderparameterfluxnskalpha} in the balance equation of the order parameter, \cref{eq:lagrange_orderparameter}, yields
\begin{align}
\label{eq:euler_orderparameterprototype}
\ckorteweg^\alphakorteweg_t + \gradient \cdot \l( \ckorteweg^\alphakorteweg \vel^\alphakorteweg \r) = \gradient \cdot \l[ \upsilonkorteweg \gradient \l( - \gammakorteweg \laplace \ckorteweg^\alphakorteweg - \alphakorteweg \l( \dens^\alphakorteweg - \ckorteweg^\alphakorteweg \r) \r) \r] + \betakorteweg \l[ \gammakorteweg \laplace \ckorteweg^\alphakorteweg + \alphakorteweg \l( \dens^\alphakorteweg - \ckorteweg^\alphakorteweg \r)  \r]
\end{align}
with a fourth-order Cahn-Hillard-type diffusion term and a second-order Allen-Cahn-type source term.
The evolution equation for the order parameter has a structure similar to the species equations of Cahn-Hillard and Allen-Cahn equations, cf. \cite{heida2012a,heida2012b}.
However, the order parameter $\ckorteweg^\alphakorteweg$ is only a relaxation parameter which should asymptotically converge to the density in the limit $\alphakorteweg \rightarrow \infty$, and has no physical interpretation.

Comparison of \cref{eq:chemicalpotentialalpha} with the second Euler-Lagrange equation in \cref{eq:eulerlagrange} suggests that the right-hand side of \cref{eq:euler_orderparameterprototype} vanishes and reduces to the conservation equation
\begin{align*}
\ckorteweg^\alphakorteweg_t + \gradient \cdot \l( \ckorteweg^\alphakorteweg \vel^\alphakorteweg \r) = 0.
\end{align*}
Furthermore, if the initial distribution of the order parameter is defined as $\ckorteweg^\alphakorteweg \l( \dimvec , 0 \r) \coloneqq \dens^\alphakorteweg \l( \dimvec , 0 \r)$ then it holds that $\ckorteweg^\alphakorteweg \l( \dimvec , \tend \r) = \dens^\alphakorteweg \l( \dimvec , \tend \r)$ for all time $\tend$.
However, from a numerical point of view this would require a mixed discretization, where similar to \cite{neusser2015}, the second Euler-Lagrange equation in \cref{eq:eulerlagrange} has to be solved as an additional elliptic constraint.
An alternative approach would be to directly solve \cref{eq:euler_orderparameterprototype}, where $\betakorteweg$ and $\upsilonkorteweg$ are then additional relaxation parameters.
However, the Cahn-Hillard-type diffusion term in \cref{eq:euler_orderparameterprototype} would require the evaluation of a fourth-order term which is computationally even more expensive than the third-order term in the original \nsk ~model.
Therefore, we propose to neglect the Cahn-Hillard-type diffusion term, i.e. $\jc = 0$.
This assumption is valid as the order parameter is only a relaxation variable.
Hence, the diffusion term has no physical interpretation and can be neglected as long as the order parameter is still guaranteed to asymptotically converge to the density.
This can be ensured by the second contribution in \cref{eq:euler_orderparameterprototype}, the Allen-Cahn-type source term.
Hence, the evolution equation for the order parameter in the Eulerian form is given as
\begin{align*}
\ckorteweg^\alphakorteweg_t + \gradient \cdot \l( \ckorteweg^\alphakorteweg \vel^\alphakorteweg \r) = \betakorteweg \l[ \gammakorteweg \laplace \ckorteweg^\alphakorteweg + \alphakorteweg \l( \dens^\alphakorteweg - \ckorteweg^\alphakorteweg \r)  \r].
\end{align*}
The structure of this equation is very similar to the relaxation equation for the order parameter proposed by \citeauthor{hitz2020}.
The only difference is the additional convection term for the order parameter, which ensures that $\ckorteweg^\alphakorteweg$ asymptotically converges to the density for all $\tend$ as long as $\alphakorteweg \rightarrow \infty$ and $\betakorteweg \rightarrow \infty$.
In the formulation of \citeauthor{hitz2020} this is only true for $\tend \rightarrow \infty$.

Consequently, the complete relaxation system for the \nsk ~equations is given as
\begin{subequations}
	\label{eq:navierstokeskorteweg_alpha}
	\begin{align}
	\label{eq:euler_mass_alpha}
	\dens_t^\alphakorteweg = \gradient \cdot \l( \dens^\alphakorteweg \vel^\alphakorteweg \r) &= 0, \\
	\label{eq:euler_linearmomentum_alpha}
	\l( \dens^\alphakorteweg \vel^\alphakorteweg \r)_t + \gradient \cdot \l( \dens^\alphakorteweg \vel^\alphakorteweg \otimes \vel^\alphakorteweg + \pres^\alphakorteweg \Ibb \r) & = \gradient \cdot \l[ \viscousstressalpha +\kortewegstressalpha \r], \\
	\label{eq:euler_energy_alpha}
	\l( \dens^\alphakorteweg \ener^\alphakorteweg \r)_t + \gradient \cdot \l( \l( \dens^\alphakorteweg \ener^\alphakorteweg + \pres^\alphakorteweg \r) \vel^\alphakorteweg \r) &= \gradient \cdot \l[ \l( \viscousstressalpha + \kortewegstressalpha \r) \cdot \vel^\alphakorteweg \r] + \gradient \cdot \je^\alphakorteweg, \\
	\label{eq:euler_orderparameter_alpha}
	\ckorteweg^\alphakorteweg_t + \gradient \cdot \l( \ckorteweg^\alphakorteweg \vel^\alphakorteweg \r) &= \betakorteweg \l[ \gammakorteweg \laplace \ckorteweg^\alphakorteweg + \alphakorteweg \l( \dens^\alphakorteweg - \ckorteweg^\alphakorteweg \r)  \r]
	\end{align}
\end{subequations}
with the simplified diffusive energy flux
\begin{align*}
\je^\alphakorteweg &= \jfourier^\alphakorteweg - \gammakorteweg \ckorteweg^\alphakorteweg \gradient \ckorteweg^\alphakorteweg \gradient \cdot \vel^\alphakorteweg + \gammakorteweg \gradient \ckorteweg^\alphakorteweg \src^\alphakorteweg.
\end{align*}
In the following, we refer to the relaxation model as given in \cref{eq:navierstokeskorteweg_alpha} as \relaxationmodelone.
In terms of the function vector of the affinities, the rate of entropy production for the relaxation model of the \nsk ~equations is given by
\begin{align*}
\srs \l( \affinitiesvec^\alphakorteweg \r) = \frac{1}{\temp} \l\lbrace \frac{4}{3} \munewton \l| \gradient \cdot \vel^\alphakorteweg \r|^2 + \munewton \l| \Dbbdevfreealpha \r|^2 + \betakorteweg \l| \chempotc^\alphakorteweg \r|^2 + \heatcoef \l| \frac{\gradient \temp^\alphakorteweg}{\temp^\alphakorteweg} \r|^2 \r\rbrace \geq 0,
\end{align*} 
which underlines the consistency with the Second Law of Thermodynamics.

Similar to \cite{rohde2010}, the divergence of the Korteweg tensor in the momentum equation \eqref{eq:euler_linearmomentum_alpha} can be simplified to
\begin{align}
\label{eq:divkortewegstress_linearmomentum}
\gradient \cdot \kortewegstressalpha = \alphakorteweg \dens^\alphakorteweg \gradient \l( \ckorteweg^\alphakorteweg - \dens^\alphakorteweg  \r).
\end{align}
The same reformulation of the expression in the energy equation yields
\begin{align}
\label{eq:divkortewegstress_energy}
\gradient \cdot \l( \kortewegstressalpha \cdot \vel^\alphakorteweg  \r) =~& \alphakorteweg \dens^\alphakorteweg \gradient \l( \ckorteweg^\alphakorteweg - \dens^\alphakorteweg \r) \cdot \vel^\alphakorteweg + \frac{\alphakorteweg}{2} \l( \l( \ckorteweg^\alphakorteweg \r)^2 - \l( \dens^\alphakorteweg \r)^2 \r) \gradient \cdot \vel^\alphakorteweg \\
&+ \gammakorteweg \l[ \frac{1}{2} \l| \gradient \ckorteweg^\alphakorteweg \r|^2 \gradient \cdot \vel^\alphakorteweg - \gradient \ckorteweg^\alphakorteweg \otimes \gradient \ckorteweg^\alphakorteweg : \gradient \otimes \vel^\alphakorteweg \r]. \nonumber
\end{align}
In contrast to the original \nsk ~model, both expression in \cref{eq:divkortewegstress_linearmomentum,eq:divkortewegstress_energy} contain only first-order derivatives.
Note that, for the sake of simplicity the upper index $\alphakorteweg$ is omitted from now on if the meaning is clear from the context.

The \relaxationmodelone ~still exhibits mixed hyperbolic-elliptic first-order fluxes similar to the original NSK model.
However, according to \cite{rohde2010}, the expression $\alphakorteweg \dens \gradient \dens = \gradient \cdot \l( \frac{\alphakorteweg}{2} \dens^2 \r)$ in \cref{eq:divkortewegstress_linearmomentum} as well as the term $\alphakorteweg \dens \gradient \dens + \frac{\alphakorteweg}{2} \l( \dens \r)^2 \gradient \cdot \vel = \gradient \cdot \l( \frac{\alphakorteweg}{2} \dens^2 \vel \r) $ in \cref{eq:divkortewegstress_energy} can be used to define a modified pressure function
\begin{align}
\label{eq:modified_pressure}
\pres_\alphakorteweg = \pres + \frac{\alphakorteweg}{2} \dens^2
\end{align}
in the linear momentum equations \eqref{eq:euler_linearmomentum_alpha} as well as in the energy equation \eqref{eq:euler_energy_alpha}.
The modified pressure function results from comprising these non-conservative products into the convective flux.
This is a purely mathematical re-ordering which does not alter the physically effective pressure.
We refer to this formulation as \relaxationmodeltwo.

Both relaxation models can be written in the form
\begin{align*}
\UPDE_t + \gradient \cdot \FPDE \l( \UPDE , \gradient \UPDE , \laplace \UPDE \r) = \SPDE \l( \UPDE , \gradient \UPDE \r)
\end{align*}
with the vector of unknowns $\UPDE = \transpose{\l( \dens , \dens \vel , \dense , \ckorteweg \r)}$, the source $\SPDE \l( \UPDE , \gradient \UPDE \r)$ and the flux vector $\FPDE \l( \UPDE , \gradient \UPDE , \laplace \UPDE \r) = \FcPDE \l( \UPDE \r) - \FvPDE \l( \UPDE, \gradient \UPDE , \laplace \UPDE \r)$ composed of the convective fluxes $\FcPDE = \l( \FPDE^{\mathrm{C},1} , ... , \FPDE^{\mathrm{C},\dimensions} \r)$ and the viscous fluxes $\FvPDE = \l( \FPDE^{\mathrm{V},1} , ... , \FPDE^{\mathrm{V},\dimensions} \r)$.
For $\dimensions = 3$ and for the \relaxationmodelone, they are explicitly given as
\begin{align}
\label{eq:navierstokeskorteweg_p}
\FPDE^{\mathrm{C},i} &= \l( \begin{array}{c} \dens \velx_i \\ \dens \velx_i \velx_1 + \delta_{1i} \pres \\ \dens \velx_i \velx_2 + \delta_{2i} \pres \\ \dens \velx_i \velx_3 + \delta_{3i} \pres \\ \l( \dense + \pres \r) \velx_i \\ \ckorteweg \velx_i \end{array} \r) , \quad 
\FPDE^{\mathrm{V},i} = \l( \begin{array}{c} 0 \\ \viscousstress_{1i} \\ \viscousstress_{2i} \\ \viscousstress_{3i} \\ \viscousstress_{i1} \velx_1 + \viscousstress_{i2} \velx_2 + \viscousstress_{i3} \velx_3 + \physflux_{\ener,i}\\ \gammakorteweg \betakorteweg \ckorteweg_{x_i} \end{array} \r) , \\
\SPDE &= \l( \begin{array}{c} 0 \\ \alphakorteweg \dens \l( \ckorteweg_{x} - \dens_{x} \r) \\ \alphakorteweg \dens \l( \ckorteweg_{y} - \dens_{y} \r) \\ \alphakorteweg \dens \l( \ckorteweg_{z} - \dens_{z} \r) \\ \alphakorteweg \dens \gradient \l( \ckorteweg - \dens \r) \cdot \vel + \frac{\alphakorteweg}{2} \l( \ckorteweg^2 - \dens^2 \r) \gradient \cdot \vel + \gammakorteweg \l[ \frac{1}{2} \l| \gradient \ckorteweg \r|^2 \gradient \cdot \vel - \gradient \ckorteweg \otimes \gradient \ckorteweg : \gradient \otimes \vel \r] \\ \alphakorteweg \betakorteweg \l( \dens - \ckorteweg \r) \end{array} \r). \nonumber
\end{align}
For the \relaxationmodeltwo, they are given by
\begin{align}
\label{eq:navierstokeskorteweg_palpha}
\FPDE^{\mathrm{C},i} &= \l( \begin{array}{c} \dens \velx_i \\ \dens \velx_i \velx_1 + \delta_{1i} \pres_\alphakorteweg \\ \dens \velx_i \velx_2 + \delta_{2i} \pres_\alphakorteweg \\ \dens \velx_i \velx_3 + \delta_{3i} \pres_\alphakorteweg \\ \l( \dense + \pres_\alphakorteweg \r) \velx_i \\ \ckorteweg \velx_i \end{array} \r) , \quad 
\FPDE^{\mathrm{V},i} = \l( \begin{array}{c} 0 \\ \viscousstress_{1i} \\ \viscousstress_{2i} \\ \viscousstress_{3i} \\ \viscousstress_{i1} \velx_1 + \viscousstress_{i2} \velx_2 + \viscousstress_{i3} \velx_3 + \physflux_{\ener,i}\\ \gammakorteweg \betakorteweg \ckorteweg_{x_i} \end{array} \r) , \\
\SPDE &= \l( \begin{array}{c} 0 \\ \alphakorteweg \dens \ckorteweg_{x1} \\ \alphakorteweg \dens \ckorteweg_{x2} \\ \alphakorteweg \dens \ckorteweg_{x2} \\ \alphakorteweg \dens \gradient \ckorteweg \cdot \vel + \frac{\alphakorteweg}{2} \ckorteweg^2 \gradient \cdot \vel + \gammakorteweg \l[ \frac{1}{2} \l| \gradient \ckorteweg \r|^2 \gradient \cdot \vel - \gradient \ckorteweg \otimes \gradient \ckorteweg : \gradient \otimes \vel \r] \\ \alphakorteweg \betakorteweg \l( \dens - \ckorteweg \r) \end{array} \r). \nonumber
\end{align}
Hence, the eigenvalues of the Jacobian of the first-order fluxes of \relaxationmodeltwo ~are given by
\begin{align*}
\eigenvalue_1 = \vel \cdot \normvec - \sqrt{ \l( \fracp{\pres_\alphakorteweg}{\dens} \r)_\entropymass }, \quad
\eigenvalue_{2,...,\dimensions+1} = \vel \cdot \normvec, \quad
\eigenvalue_{\dimensions+2} = \vel \cdot \normvec + \sqrt{ \l( \fracp{\pres_\alphakorteweg}{\dens} \r)_\entropymass }.
\end{align*}
Thus, this formulation of the relaxation system remains the strict hyperbolicity of the flux $\FcPDE$ for large enough values of the Korteweg parameter
\begin{align*}
\alphakorteweg > \alphakorteweg_* = \l| \text{min} \l( \frac{1}{r} \l(  \fracp{\pres \l( r , \temp \r) }{\dens} \r)_\entropymass : r \in ( \dens^\spin_\vap \l( \temp \r) , \dens^\spin_\liq \l( \temp \r) ) \wedge \temp \in \l( 0 , 1 \r) \r) \r|
\end{align*}
even if the Helmholtz free energy per unit volume in \cref{eq:helmholtzfreeenergyvol} becomes non-convex.
\subsection{Thermodynamic Consistent Boundary Conditions}
\label{sec:boundary_conditions}
Following \citeauthor{soucek2020} \cite{soucek2020}, the concept of \cit~ introduced in \cref{sec:cit} can also be used to derive constitutive relations at the wall boundary, which are compatible with the Second Law of Thermodynamics.
For this purpose, additional balance equations at the surface have to be defined.
According to the framework of \citeauthor{soucek2020}, see also \cite{slattery1980}, the balance law of an arbitrary quantity $\surf{\arbvolquant}$ at the wall boundary surface $\wallsurface$ is given by
\begin{align}
\surf{\material{\arbvolquant}} + \surf{\arbvolquant} \l( \surf{\gradient} \cdot \surftang{\vel} - \l( \dimensions - 1 \r) \curvature \surfnorm{\vel} \cdot \surfnormvec \r) = \surf{\srarb}^\arbvolquant - \surf{\gradient} \cdot \surf{\arbvolflux}^\arbvolquant - \jumpbig{\arbvolflux^\arbvolquant + \arbvolquant \l( \vel - \surf{\vel} \r)} \cdot \surfnormvec
\end{align}
if the corresponding balance law in the bulk is given as
\begin{align*}
\arbvolquant_\tend + \gradient \cdot \l( \arbvolquant \vel + \arbvolflux^{\arbvolquant} \r) = \srarb^\arbvolquant.
\end{align*}
For simplicity, external supplies have been neglected.
The symbols $\arbvolflux^\arbvolquant$ and $\surf{\arbvolflux}^\arbvolquant$ indicate the diffusive bulk and surface fluxes, respectively, $\srarb^\arbvolquant$ is a production term and $\curvature$ is the mean curvature.
The outward pointing normal vector of the surface is denoted by $\surfnormvec$, and the subscripts $\surftang{\l( \cdot \r)}$ and $\surfnorm{\l( \cdot \r)}$ indicate the tangential and the normal component of a vectorial surface quantity in the full space, respectively.
Finally, $\jump{\cdot} = \l( \cdot \r)^+ - \l( \cdot \r)^-$ is the jump operator across the wall boundary surface $\wallsurface$, where the superscript $\l( \cdot \r)^-$ corresponds to a quantity at the surface in the domain $\domain^- = \domain$ and $\l( \cdot \r)^+$ to a quantity at the surface in the solid domain $\domain^+$.

We assume that the solid domain, and hence the wall boundary surface is non-moving.
Moreover, the surface $\wallsurface$ should be impermeable, in-surface heat conduction is neglected and the surface and the solid are assumed to have the same temperature.
In terms of equations, this assumptions can be expressed as
\begin{align}
\label{eq:bc_assumptions}
\vel^+ = 0, \quad \surf{\vel} = 0, \quad \vel^- \cdot \surfnormvec = 0, \quad \jesurf = 0, \quad \surf{\temp} = \temp^+. 
\end{align}
Thus, the balance equations at the surface for the mass, the linear momentum, the total energy and the entropy are given by
\begin{subequations}
	\begin{align}
	\label{eq:mass_surface}
	\l( \surf{\dens} \r)_\tend &= 0, \\
	\label{eq:momentum_surface}
	- \surf{\gradient} \cdot \surf{\Tbb} &= \jump{\Tbb} \cdot \surfnormvec, \\
	\label{eq:energy_surface}
	\l( \surf{\dens} \surf{\ener} \r)_\tend &= \jump{\Tbb \cdot \vel} \cdot \surfnormvec - \je^- \cdot \surfnormvec, \\
	\label{eq:entropy_surface}
	\l( \surf{\dens} \surf{\entropymass} \r)_\tend - \surf{\gradient} \cdot \l( \jssurf \r) &= \srssurf + \jumpBig{\js} \cdot \surfnormvec.
	\end{align}
\end{subequations}
They are valid for the original \nsk ~model in \cref{eq:navierstokeskorteweg} as well as the relaxation models in \cref{eq:navierstokeskorteweg_p,eq:navierstokeskorteweg_palpha}.
However, with the relaxation models we obtain an additional equation
\begin{align}
\l( \surf{\ckorteweg} \r)_\tend &= \srcsurf
\end{align}
for the evolution of the order parameter on the surface.
Following \citeauthor{soucek2020}, the total Helmholtz free energy per unit area at the surface $\helmfreevolsurftot \coloneqq \helmfreevol_{\Gamma,\mathrm{tot}}$ can be expressed via its Legendre transform as
\begin{align}
\label{eq:helmfreesurftot}
\helmfreevolsurftot \l( \dummyorderparameter^- , \surf{\temp} \r) = \surf{\dens} \surf{\ener} - \surf{\temp} \surf{\dens} \surf{\entropymass} = \enervolsurf - \surf{\temp} \entropyvolsurf.
\end{align}
Based on statistical physics \cite{rowlinson1989}, the Helmholtz free energy per unit area is assumed as a function of the temperature at the surface $\surf{\temp}$ and $\dummyorderparameter^- \coloneqq \dens^-$ the bulk density in case of the original \nsk ~model or $\dummyorderparameter^- \coloneqq \ckorteweg^-$ the bulk order parameter in case of the relaxation models.
The material time derivative at the surface of \cref{eq:helmfreesurftot} combined with the assumptions in \cref{eq:bc_assumptions} yields
\begin{align}
\label{eq:entropy_surface_two}
\surf{\temp} \l( \entropyvolsurf \r)_t = \l( \enervolsurf \r)_t - \l( \fracp{\helmfreevolsurftot}{\dummyorderparameter^-} \r)_{\surf{\temp}} \l( \dummyorderparameter^- \r)_t.
\end{align}
Substitution of the surface energy balances, \cref{eq:energy_surface}, and the respective governing equation for $\dummyorderparameter$ in the bulk, \eqref{eq:euler_mass} or \eqref{eq:euler_orderparameter_alpha}, into \cref{eq:entropy_surface_two} results in
\begin{align}
\label{eq:entropy_surface_three}
\surf{\temp} \l( \entropyvolsurf \r)_t = \jump{\Tbb \cdot \vel} \cdot \surfnormvec -\je^- \cdot \surfnormvec - \l( \fracp{\helmfreevolsurftot}{\dummyorderparameter^-} \r)_{\surf{\temp}} \l( -\dummyorderparameter^- \gradient \cdot \vel^- - \vel^- \gradient \dummyorderparameter^- + \srdummy^- \r)
\end{align}
for the entropy balance equation with $\srx{\dummyorderparameter=\dens} = 0$ and $\srx{\dummyorderparameter=\ckorteweg} = \src$.
Similar to \cite{soucek2020}, a membrane model is considered, where the surface stress tensor
\begin{align*}
\surf{\Tbb} = \surftenscoeff_{\mathrm{vl}} \l( \Ibb - \surfnormvec \otimes \surfnormvec \r)
\end{align*}
only depends on the equilibrium surface tension coefficient $\surftenscoeff_{\mathrm{vl}}$.
This enables the simplification
\begin{align*}
\jump{\Tbb \cdot \vel} \cdot \surfnormvec = - \tang{\vel}^- \cdot \tang{\l( \Tbb \cdot \surfnormvec \r)}^{\pm}
\end{align*}
in \cref{eq:entropy_surface_three}, where additionally \cref{eq:momentum_surface} and the assumptions in \cref{eq:bc_assumptions} have been used.
The specification of appropriate thermodynamic driving forces at the surface $\fkappavec \l( \surf{\affinitiesvec} \r)$ and a comparison of \cref{eq:entropy_surface_three} with \cref{eq:entropy_surface} permits the derivation of constitutive relations at the boundary, similar to the procedure in the bulk.
The interested reader is referred to \cite{heida2013,soucek2020} for a more detailed insight into the derivation of boundary conditions of Korteweg-type fluids.

The choice of the function vector of the affinities as
\begin{align*}
\fkappavec \l( \affinitiesvec \r) = \l( \tang{\vel}^- , \gradient \cdot \vel^- , \frac{1}{\surf{\temp}} - \frac{1}{\temp^-} \r)
\end{align*}
yields the following constitutive conditions for the original \nsk ~model
\begin{subequations}
	\label{eq:bc_nsk}
	\begin{align}
	\label{eq:bc_navierstokes}
	2 \munewton \tang{\l( \gradvel \cdot \surfnormvec \r)}^- &= \affinitycoef{1} \gradient \cdot \vel^- \surf{\gradient} \dens^- - \affinitycoef{2} \tang{\vel}^-, \\
	\label{eq:bc_contactangle}
	\gradient \dens^- \cdot \surfnormvec &= \frac{1}{\gammakorteweg} \l[ \affinitycoef{1} \gradient \cdot \vel^- - \l( \fracp{\helmfreevolsurftot}{\dens^-} \r)_{\surf{\temp}} \r], \\
	\label{eq:bc_energy}
	\gradient \temp^- \cdot \surfnormvec &= - \frac{\affinitycoef{3}}{\heatcoef} \l( \frac{1}{\surf{\temp}} - \frac{1}{\temp^-} \r).
	\end{align}
\end{subequations}
According to \citeauthor{soucek2020}, they are called the generalized Navier-slip conditions, where \cref{eq:bc_navierstokes} governs the slip condition due to the viscous forces at the wall surface, \cref{eq:bc_contactangle} prescribes the contact angle between the vapor-liquid phase interface and the wall boundary, and \cref{eq:bc_energy} specifies the diffusive energy flux between the solid and the domain $\domain$.
The symbols $\affinitycoef{1}$, $\affinitycoef{2}$ and $\affinitycoef{3}$ are positive parameters as introduced in \cref{eq:linearfluxforcerelation}.
In \cref{eq:bc_navierstokes}, the parameter $\affinitycoef{2}$ corresponds to the Navier-slip condition and is related to a non-dimensional slip-length.
The parameter $\affinitycoef{3}$ in \cref{eq:bc_energy} is a non-dimensional heat transfer coefficient.
The additional terms related to $\affinitycoef{1}$ enable the consideration of contact angle hysteresis effects.
Following \cite{soucek2020}, these dynamic contact angle effects, similar to the Navier-slip condition, are caused by unresolved surface roughnesses which induce a slip-stick behavior of the vapor-liquid interface on the wall.

The same procedure combined with the function vector of the thermodynamic affinities
\begin{align*}
\fkappavec \l( \affinitiesvec \r) = \l( \tang{\vel}^- , \gradient \cdot \vel^- , \chempotc^- , \frac{1}{\surf{\temp}} - \frac{1}{\temp^-} \r)
\end{align*}
yields the boundary conditions
\begin{subequations}
	\label{eq:bc_nsk_alpha}
	\begin{align}
	\label{eq:bc_navierstokes_alpha}
	\tang{\l( 2 \munewton \gradvel \cdot \surfnormvec \r)}^- &= \affinitycoef{1} \gradient \cdot \vel^- \surf{\gradient} \ckorteweg^- - \affinitycoef{2} \tang{\vel}^-, \\
	\label{eq:bc_contactangle_alpha}
	\gradient \ckorteweg^- \cdot \surfnormvec &= \frac{1}{\gammakorteweg} \l[ \affinitycoef{1} \gradient \cdot \vel^- - \l( \fracp{\helmfreevolsurftot}{\ckorteweg^-} \r)_{\surf{\temp}} \r], \\
	\label{eq:bc_relaxation_alpha}
	\gammakorteweg \laplace \ckorteweg^- + \alphakorteweg \l( \dens^- - \ckorteweg^- \r) &= - \frac{\affinitycoef{1}\affinitycoef{4}}{\betakorteweg} \gradient \cdot \vel^-, \\
	\label{eq:bc_energy_alpha}
	\gradient \temp^- \cdot \surfnormvec &= - \frac{\affinitycoef{3}}{\heatcoef} \l( \frac{1}{\surf{\temp}} - \frac{1}{\temp^-} \r)
	\end{align}
\end{subequations}
for the relaxation models.
The boundary conditions in \cref{eq:bc_nsk_alpha,eq:bc_nsk} have a similar structure, with the difference that in \cref{eq:bc_nsk_alpha} the contact angle effects depend on the order parameter $\ckorteweg$.
Moreover, the relaxation procedure induces an additional boundary condition due to the Allen-Cahn like source term $\src$, which is scaled by the corresponding parameter $\affinitycoef{4}$.

\begin{figure}
\begin{center}
	\includegraphics[width=\linewidth,trim = 0.25cm 0 0 0, clip]{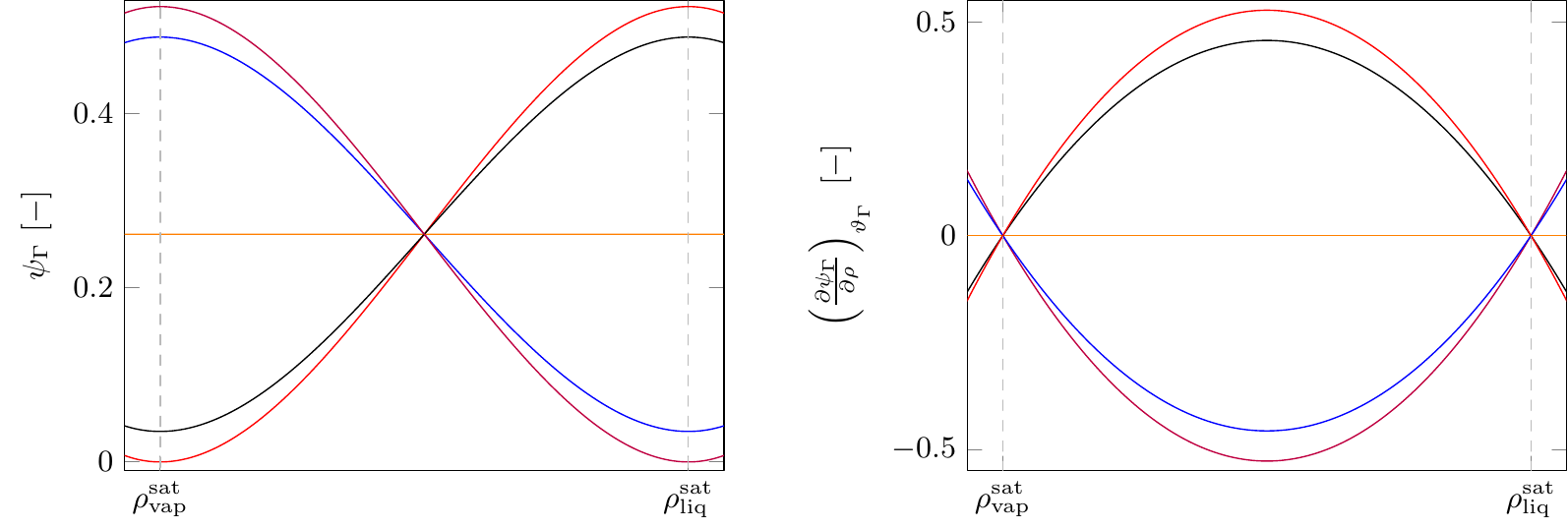}
\end{center}
\caption{\label{fig:helmfreesurfwall}Left: Helmholtz free energy per unit area at the surface. Right: The first derivative of the Helmholtz free energy per unit area at the surface with respect to the order parameter. The different curves correspond to the contact angles $\contactangle = \l\lbrace {\color{red}{0 \degree},\color{black}{30 \degree},\color{orange}{90 \degree},\color{blue}{150 \degree},\color{purple}{180 \degree}} \r\rbrace$ at a constant temperature of $\temp = 0.85$.}
\end{figure}

It remains to specify the total surface Helmholtz free energy per unit area.
In this work, we consider a cubic polynomial as proposed in \cite{sibley2013a,sibley2013b} and used in \cite{desmarais2016,soucek2020,gelissen2020}.
This enables to prescribe the wall interaction potentials $\surftenscoeff_{\mathrm{sv}}$ and $\surftenscoeff_{\mathrm{sl}}$ which act between the solid and the pure vapor bulk phase or the pure liquid bulk phase, respectively, and additionally ensures a smooth blending over the phase interface.
Furthermore, any contact angle effects away from the phase interface can be avoided.
In terms of equations these conditions are give as
\begin{align*}
\helmfreevolsurftot \l( \dens^\sat_\vap , \surf{\temp} \r) = \surftenscoeff_{\mathrm{sv}}, \quad \helmfreevolsurftot \l( \dens^\sat_\liq , \surf{\temp} \r) = \surftenscoeff_{\mathrm{sl}}, \quad 
\l( \fracp{\helmfreevolsurftot}{\dummyorderparameter^-} \r)_{\surf{\temp}} \Bigg|_{\dummyorderparameter^- = \dens^\sat_\vap} = \quad \l( \fracp{\helmfreevolsurftot}{\dummyorderparameter^-} \r)_{\surf{\temp}} \Bigg|_{\dummyorderparameter^- = \dens^\sat_\liq} = 0.
\end{align*}
Hence, the total Helmholtz free energy per unit area on the surface and its first derivative with respect to the order parameter are given as
\begin{align*}
\helmfreevolsurftot \l( \dummyorderparameter^- , \surf{\temp} \r) = \frac{\l( \dummyorderparameter^- - \dens^\sat_\liq \r)^2 \l( 2 \dummyorderparameter^- + \dens^\sat_\liq - 3 \dens^\sat_\vap \r) \surftenscoeff_{\mathrm{sv}} - \l( \dummyorderparameter^- - \dens^\sat_\vap \r)^2 \l( 2 \dummyorderparameter^- + \dens^\sat_\vap - 3 \dens^\sat_\liq \r) \surftenscoeff_{\mathrm{sl}}}{\l( \dens^\sat_\liq - \dens^\sat_\vap \r)^3}
\end{align*}
and
\begin{align}
\label{eq:helmfreesurftot_derivative}
\l( \fracp{\helmfreevolsurftot}{\dummyorderparameter^-} \r)_{\surf{\temp}} = \frac{6 \l( \dummyorderparameter - \dens^\sat_\liq \r) \l( \dummyorderparameter - \dens^\sat_\vap \r)}{ \l( \dens^\sat_\liq - \dens^\sat_\vap \r)^3 } \surftenscoeff_{\mathrm{vl}} \cos \l( \contactangle \r),
\end{align}
where the explicit dependence of the Maxwellian states $( \dens_\vap^\sat,\dens_\liq^\sat )$ and the equilibrium surface tension coefficient $\surftenscoeff_{\mathrm{vl}}$ on the surface temperature is omitted, for the sake of clarity.
Additionally, the Young-Laplace law
\begin{align*}
\surftenscoeff_{\mathrm{sv}} - \surftenscoeff_{\mathrm{sl}} = \surftenscoeff_{\mathrm{vl}} \cos \l( \contactangle \r)
\end{align*}
has been used in \cref{eq:helmfreesurftot_derivative} to enable a formulation in terms of the contact angle $\contactangle$.
The surface Helmholtz free energy per unit area and its derivative are depicted in \cref{fig:helmfreesurfwall} as a function of the order parameter for fixed contact angles and at a constant temperature.
The equilibrium surface tension coefficient in \cref{eq:helmfreesurftot_derivative} is a fluid specific and temperature dependent property.
Following \cite{cahn1958}, see also \cite{pismen2000,diehl2007,gelissen2020}, for a planar diffuse interface which connects two equilibrium states, it is given by
\begin{align}
\label{eq:surftenscoef}
\surftenscoeff_{\mathrm{vl}} \l( \surf{\temp} \r) = \gammakorteweg \int_{- \infty}^{+\infty} \l( \fracp{\dens}{\xdir} \r)^2 \dimensions \xdir = \sqrt{\frac{2}{\gammakorteweg}} \int_{\dens^\sat_\vap \l( \surf{\temp} \r)}^{\dens^\sat_\liq \l( \surf{\temp} \r)} \sqrt{\helmfreevol \l( \dens , \surf{\temp} \r) - \bitangent \l( \dens , \surf{\temp} \r)} \dimensions \dens,
\end{align}
where the second term under the square root is the so called bi-tangent
\begin{align*}
\bitangent \l( \dens , \surf{\temp} \r) &= \l( \dens - \dens^\sat_\vap \r) \l( \fracp{\helmfreevol}{\dens} \r)_{\surf{\temp}} \Bigg|_{\dens^\sat_\vap} + \helmfreevol \l( \dens^\sat_\vap , \surf{\temp} \r) = \dens \l( \fracp{\helmfreevol}{\dens} \r)_{\surf{\temp}} \Bigg|_{\dens^\sat_\vap} - \pres \l( \dens^\sat_\vap , \surf{\temp} \r),
\end{align*}
which connects the Maxwellian states as depicted in \cref{fig:thermodynamics} on the right.
\section{Numerical Methods}
\label{sec:numerics_dgsem}
\subsection{Flow Solver}
\label{sec:numerics_bulk}
An extension of the open source code \flexi ~to multiphase flow is used for the discretization of the original Navier-Stokes-Korteweg equations \eqref{eq:navierstokeskorteweg} and the relaxation systems, \cref{eq:navierstokeskorteweg_p,eq:navierstokeskorteweg_palpha}.
\flexi ~is based on the discontinuous Galerkin spectral element method (\dgsem) \cite{kopriva2009}.
A detailed description of the implementation is given in \cite{krais2021}, and we will only give a short overview of the major building blocks.

The domain $\domain$ is tessellated into non-overlapping hexahedral elements, which may be organized fully unstructured and feature hanging nodes, so called mortar cells.
Each physical grid element is mapped onto the reference cube $\refelem = \l[ -1 , 1 \r]^3$ to transform the balance law from the physical space $\dimvec = \transpose{\l( \xdir_1 = \xdir , \xdir_2 = \ydir , \xdir_3 = \zdir \r)}$ into the reference space $\refspace = \transpose{\l( \xiref , \etaref , \zetaref \r)}$.
Multiplication of the balance law with a test function and integration by parts yields the weak formulation.
In each grid cell, Lagrangian polynomials of degree $\Npoly$ defined by Legendre-Gauss points are used to approximate the solution and the fluxes.
This allows discontinuities across the surface of an element.
The volume and the surface integrals are approximated by a Gaussian quadrature rule, with the interpolation points as integration points (collocation).
The HLLC and the Rusanov \cite{toro2009} Riemann solvers are used to enable the coupling between the elements.

To handle the viscous fluxes, we follow the method of \citeauthor{bassi1997} \cite{bassi1997} known as the \brone ~scheme or the \textit{lifting} procedure.
Similar to \cite{hitz2020}, the second gradients of the density required in the Korteweg stress of the original NSK model are evaluated by applying the procedure of \citeauthor{bassi1997} twice.
The source terms of the relaxation model include non-conservative products which depend on the gradient of the solution.
In this work, they are approximated as point-wise source terms with the lifted gradients, which proved to be stable.
By the use of the method of lines, the solution is advanced with an explicit fourth-order low storage Runge-Kutta (\rk) method as presented in \cite{kennedy2000}.

The used van der Waals \eos ~is implemented in the framework of \citeauthor{foell2019} \cite{foell2019}, which enables a direct evaluation of the \eos ~or a tabulation approach.
\subsection{Numerical Treatment of the Contact Angle Boundary Condition}
\label{sec:numerics_bc}
Boundary conditions in \dg ~methods are typically enforced weakly through the definition of the corresponding interface flux.
Therefore, the implementation of the boundary conditions \cref{eq:bc_navierstokes,eq:bc_energy,eq:bc_nsk_alpha} is straightforward.
However, this is not the case for the contact angle boundary condition, \cref{eq:bc_contactangle}, if the original \nsk ~model is solved in the conservative form.
As outlined by \citeauthor{desmarais2016} \cite{desmarais2016}, the density as well as the first and the second gradient of the density in the wall normal direction have to be consistent.
However, the boundary conditions in \cref{eq:bc_nsk} do not provide any second gradient of the density.
Moreover, the approach of \citeauthor{bassi1997} cannot be applied directly for the computation of the gradients as no numerical flux for the computation of the first density gradient is provided.
However, this is mandatory to determine a consistent second-order gradient of the density.

Hence, in this work we transfer the iterative scheme used in \cite{desmarais2016} for the discretization of the original \nsk ~equations with a second-order accurate \fv ~scheme on Cartesian meshes to an unstructured \dg ~discretization of arbitrary order.
The key idea in \cite{desmarais2016} is to define two ghost cells with the indices $\l(\idir,\jdir-1,\kdir\r)$ and $\l(\idir,\jdir-2,\kdir\r)$ for the \fv ~cell $\l(\ijk\r)$ adjacent to a wall which itself is oriented normal to the \ydir-direction.
The indices $\idir$, $\jdir$ and $\kdir$  correspond to the $\xdir$, $\ydir$, $\zdir$ directions, respectively. 
The first ghost cell $\l(\idir,\jdir-1,\kdir\r)$ provides ghost states for $\l( \dens, \vel, \temp \r)$ as common in \fv~ schemes and the second $\l(\idir,\jdir-2,\kdir\r)$ only a ghost state for $\dens$.
In a first step, a consistent ghost state for the density $\dens_{\idir,\jdir-1,\kdir}$ is iterated with the boundary condition, \cref{eq:bc_contactangle}, the wall temperature $\temp_{\idir,\jdir-\frac{1}{2},\kdir}$ and the density at the wall $\dens_{\idir,\jdir-\frac{1}{2},\kdir} = \frac{1}{2} \l( \dens_{\idir,\jdir,\kdir} + \dens_{\idir,\jdir-1,\kdir} \r)$, see Eq. (4.59) in \cite{desmarais2016}.
In the second step, the remaining ghost state $\dens_{i,j-2,k}$ is determined such that
\begin{align*}
\l( \fracp{\dens}{\ydir} \r)_{\idir,\jdir-\frac{1}{2},\kdir}^{\Ocal \l( 2 \r)} = \l( \fracp{\dens}{\ydir} \r)_{\idir,\jdir-\frac{1}{2},\kdir}^{\Ocal \l( 4 \r)}
\end{align*}
holds, where the derivatives are approximated with finite differences.
This idea applied to a \dg ~scheme with the methodology of \citeauthor{bassi1997} yields exemplary for a wall surface normal to the $\etaref$ direction an iterative procedure for the interface numerical flux $\numflux{\dens}_{\oq} ( \etaref = 1 )$, where $\odir$ and $\qdir$ correspond to the $\xiref$ and $\zetaref$ directions in the reference space, respectively.
Hence, the objective function to be minimized is given by
\begin{align}
\label{eq:objectivefunction_contactangle}
\l( \gradient \dens^-_{\oq} \cdot {\surf{\normvec}}_{\oq} \r)_{\bc} - \l( \gradient \dens^-_{\oq} \cdot {\surf{\normvec}}_{\oq} \r)_{\brone} = 0, \qquad o,q = 0,\ldots,N,
\end{align}
where according to \cref{eq:bc_contactangle} the first term $\l( \gradient \dens^-_{\oq} \cdot {\surf{\normvec}}_{\oq} \r)_{\bc} = \function{\dens_{\oq} ( \etaref = 1 ), {\surf{\temp}}_{\oq}}$ is a function of the density prolongated to the surface and the temperature at the wall.
The second term $\l( \gradient \dens^-_{\oq} \cdot {\surf{\normvec}}_{\oq} \r)_{\brone} = \function{\dens_{\opq} , \numflux{\dens}_{\oq} ( \etaref = 1 ),{\surf{\normvec}}_{\oq}}$ is a function of the density in the volume (and its direct Voronoi neighbors), the specified numerical flux at the wall boundary and the surface normal vector ${\surfnormvec}_{\oq}$.
\cref{eq:objectivefunction_contactangle} is solved for $\numflux{\dens}_{\oq}$ numerically by a Newton scheme.
This approach is an element local procedure and showed to converge with in at most 10 iteration steps.
For the evaluation of the second gradients, the numerical flux $\numflux{\l( \gradient \dens_\oq \r)} \coloneqq \gradient \dens_{\oq} ( \etaref = 1 )$ is defined as the density gradient in the volume prolongated to the surface.

Moreover, for the evaluation of the boundary conditions, \cref{eq:bc_contactangle,eq:bc_contactangle_alpha}, the integral in \cref{eq:surftenscoef} has to be determined.
Since, this integral cannot be evaluated analytically, a Gaussian quadrature with 15 support points is used in this work.
To avoid the computation of the integral at each boundary degree of freedom (\dof) in each \rk-stage, the equilibrium surface tension is initially tabulated with $10000$ equidistant grid points in the interval $\surf{\temp} \in \l[ 0.3, 1 \r]$.
During runtime a linear interpolation of the tabulated data is used.
\section{Results}
\label{sec:results}
In this section, numerical experiments of the relaxation model for the \nsk ~equations are presented.
First, simulations in the bulk are discussed.
We start with the validation in 1D with benchmark problems where the reference solutions have been computed with the original NSK model.
Turning to 2D, results of a merging droplet event are presented and the resolution requirement of the two relaxation formulations are compared.
Then, 3D results of colliding droplets are shown.
Secondly, we turn to confined domains.
In 3D, we show that the boundary conditions used are able to reproduce the Young-Laplace law once the fluid reached its equilibrium.
This is followed by the simulation of a moving droplet on a solid surface, which considers contact angle hysteresis effects.
Finally, we prove that our framework is capable to deal with complex domains.
For this, we simulate a spinodal decomposition in a 3D porous medium.
\begin{table}[ht!]
	\begin{center}
	\begin{tabular}{l ? c c c c c c }
		\hline
		\hline
		\rule{0pt}{14pt} Quantity & $\gammakorteweg$ & $\munewton$ & $\heatcoef$ & $\cvdw$ & $\surftenscoeff_{\mathrm{vl}}$ & $\interfacewidth$ \\[1ex]
		\Xhline{2\arrayrulewidth}
		\rule{0pt}{15pt} Value & $10^{-4}$ & $0.01$ & $\frac{1}{150}$ & $5$ & $0.0052$ & $0.0598$ \\[1ex]
		\hline
		\hline
	\end{tabular}
	\caption{\label{table:fluidproperties}The fluid properties used over all test cases, if not stated otherwise.}
	\end{center}
\end{table}

In all examples the initial configuration for the relaxation variable is identical to the initial density field.
Hence, only initial density, velocity and temperature fields are given in the sequel.
Moreover, the fluid properties have been kept constant over all test cases, if not stated otherwise and are specified in \cref{table:fluidproperties}, where the symbol $\interfacewidth$ indicates the initial interface thickness.
According to \cite{gelissen2018}, it is approximated as
\begin{align*}
\interfacewidth \l( \temp \r) = \sqrt{2} \gammakorteweg \frac{\l( \dens_\liq^\sat \l( \temp \r) - \dens_\vap^\sat \l( \temp \r) \r)^2}{\surftenscoeff_{\mathrm{vl}} \l( \temp \r)}.
\end{align*}
The definition of the equilibrium surface tension coefficient $\surftenscoeff_{\mathrm{vl}}$ is given in \cref{eq:surftenscoef}.

The following setup is chosen for all test cases, if not stated otherwise:
As a numerical flux function, the HLLC method was used for the \relaxationmodeltwo ~and the Rusanov flux for the \originalmodel ~and the \relaxationmodelone.
Time integration was performed explicitly with a fourth-order low storage \rk ~method and $CFL=0.9$.
\subsection{1D Test Cases}
\label{sec:1D}
\subsubsection{Static Solution}
\label{sec:staticsolution}
According to \cite{gelissen2018}, the density of a quiescent bubble at a constant temperature $\temp_0$ with a radius $\radius_{0}$ and its center at $\xdir_{0}$ can be approximated by
\begin{align}
\label{eq:staticbubble1D}
\dens \l( \xdir , \tend_0 \r) &= \frac{\dens_\liq^\sat \l( \temp_0 \r) + \dens_\vap^\sat \l( \temp_0 \r)}{2} + \frac{\dens_\liq^\sat \l( \temp_0 \r) - \dens_\vap^\sat \l( \temp_0 \r)}{2} \tanh \l( 4 \frac{\l| \xdir - \xdir^0 \r| - \radius}{\interfacewidth \l( \temp_0 \r)} \r).
\end{align}
To determine the sensitivity of the relaxation model on the relaxation variable $\alphakorteweg$, a static bubble was initialized based on \cref{eq:staticbubble1D} and relaxed to an equilibrium state.
For this, the Korteweg parameter was varied in between $\alphakorteweg \in \lbrace 5 , 10 , 100 , 1000 \rbrace$ whereas the relaxation parameter was fixed to $\betakorteweg = 1000$.
The physical parameters used are summarized in \cref{table:staticbubble1D}.
\begin{table}[ht!]
	\begin{center}
	\begin{tabular}{l ? c c c c c c }
		\hline
		\hline
		\rule{0pt}{14pt} Quantity & $\temp_0$ & $\dens_\vap^\sat$ & $\dens_\liq^\sat$ & $\velx_1$ & $\xdir^0$ & $\radius$ \\[1ex]
		\Xhline{2\arrayrulewidth}
		\rule{0pt}{15pt} Value & $0.85$ & $0.3197$ & $1.8071$ & $0$ & $0.5$ & $0.1$ \\[1ex]
		\hline
		\hline
	\end{tabular}
	\caption{\label{table:staticbubble1D}The values used for the initialization of the 1D static bubble test case.}
	\end{center}
\end{table}
The domain $\domain \in \l( 0 , 1 \r)$ was discretized by $110$ elements with a minimum grid spacing of $\Delta \xdir = 0.005$ at the interface positions $\xdir = 0.4$ and $\xdir = 0.6$.
A linear stretching of the mesh was used with $40$ elements in each part of the liquid and $30$ elements inside the bubble.
The polynomial degree was fixed to $\Npoly = 4$ for all simulations which corresponds to fifth-order of accuracy.

\begin{figure}[ht!]
\begin{center}
	\includegraphics[width=\linewidth,trim = 2.25cm 0 0 0, clip]{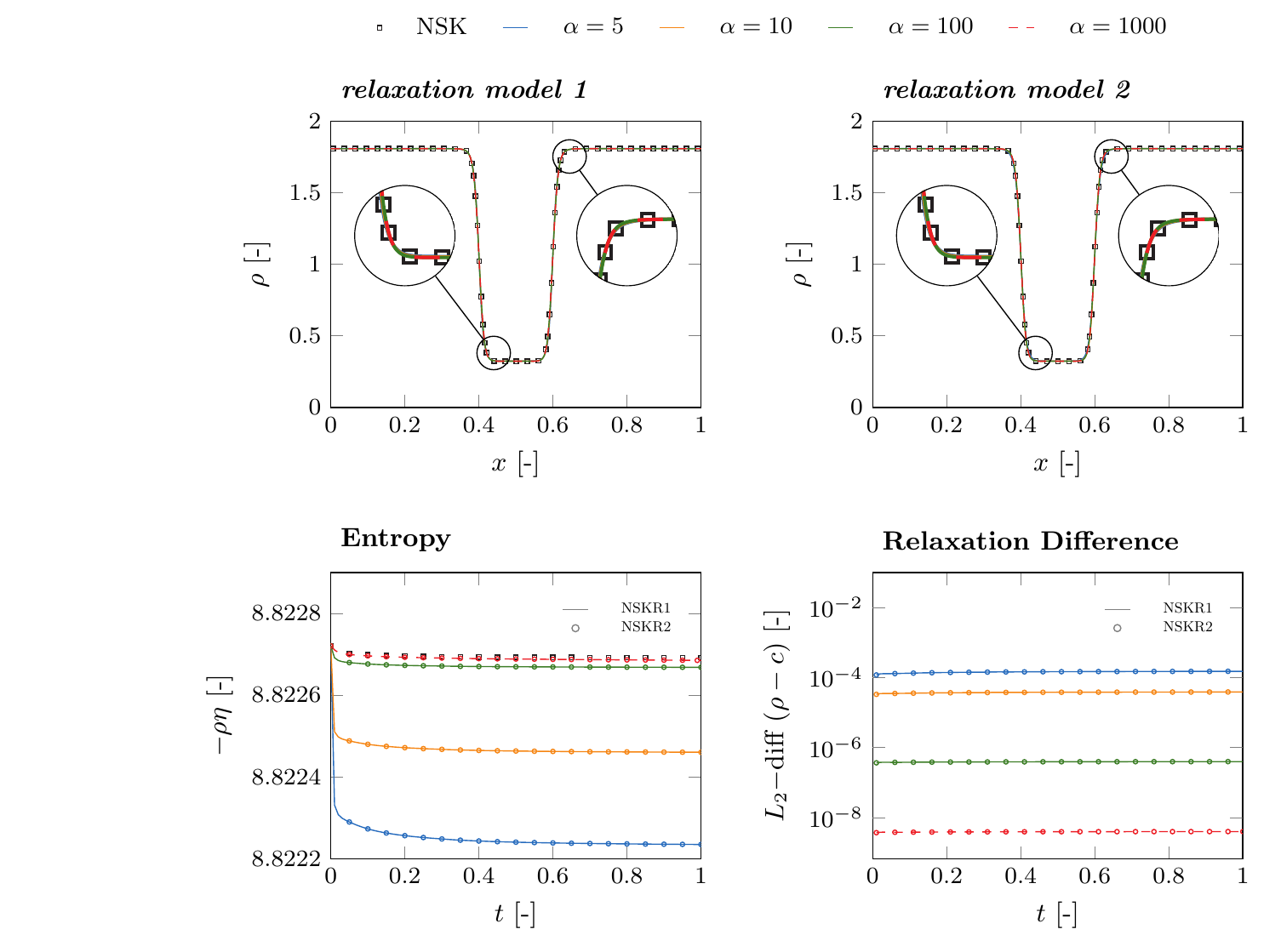}
\end{center}
\caption{\label{fig:staticbubble1D}Top: Comparison of the density field achieved with the \relaxationmodelone ~(left) and the \relaxationmodeltwo ~(right) to the \originalmodel ~at $\tend = 1$. Bottom: Evolution of the entropy (left) and the relaxation difference (right) for both relaxation formulations over the time; \relaxationmodelone ({\color{gray}{\textbf{--}}}), \relaxationmodeltwo ({\color{gray}{$\boldsymbol{\circ}$}}).}
\end{figure}
A comparison of the density profiles at equilibrium ($t = 1$) computed with the \relaxationmodelone ~and the \relaxationmodeltwo ~to the \originalmodel ~are shown in the top left and top right views in \cref{fig:staticbubble1D}, respectively.
Both formulations provide excellent results.
At the edges of the phase interface a minor deviation of the density profile to the \originalmodel ~is observed for small Korteweg parameters.
However, this deviation vanishes with an increasing Korteweg parameter.
A closer look at the temporal evolution of the integral entropy per unit volume depicted on the bottom left in \cref{fig:staticbubble1D} reveals that both relaxation formulations are consistent to the Second Law of Thermodynamics.
This is indicated by the monotonously decreasing mathematical entropy $\overbar{\dens \entropymass} = - \dens \entropymass$.
An increase of the Korteweg parameter causes the entropy production to converge to the one of the \originalmodel.
The temporal evolution of the $L_2$-difference between $\dens - \ckorteweg$ is given in the plot on the bottom right in \cref{fig:staticbubble1D}.
The $L_2$-difference decreases with an increasing Korteweg parameter and converges to a constant value for all simulations.
\subsubsection{Traveling Wave Solution}
\label{sec:travelingwavesolution}
To determine the influence of the relaxation parameter $\betakorteweg$, a moving test case with a traveling wave solution was investigated.
For this, a single phase interface was advected over the time.
The initial density profile was approximated by
\begin{align*}
\dens \l( \xdir , \tend_0 \r) &= \frac{\dens_\liq^\sat \l( \temp_0 \r) + \dens_\vap^\sat \l( \temp_0 \r)}{2} + \frac{\dens_\liq^\sat \l( \temp_0 \r) - \dens_\vap^\sat \l( \temp_0 \r)}{2} \tanh \l( 4 \frac{ \xdir - \xdir^0 }{\interfacewidth \l( \temp_0 \r)} \r),
\end{align*}
where $\xdir_0$ specifies the initial position of the phase interface.
The physical parameters used are depicted in \cref{table:travelingwave1D}.
\begin{table}[ht!]
	\begin{center}
	\begin{tabular}{l ? c c c c c }
		\hline
		\hline
		\rule{0pt}{14pt} Quantity & $\temp_0$ & $\dens_\vap^\sat$ & $\dens_\liq^\sat$ & $\velx_1$ & $\xdir^0$ \\[1ex]
		\Xhline{2\arrayrulewidth}
		\rule{0pt}{15pt} Value & $0.85$ & $0.3197$ & $1.8071$ & $-0.5$ & $0.5$ \\[1ex]
		\hline
		\hline
	\end{tabular}
	\caption{\label{table:travelingwave1D}The values used for the initialization of the 1D traveling wave test case.}
	\end{center}
\end{table}
Based on the results of the static bubble test case, the Korteweg parameter was fixed to $\alphakorteweg = 100$ and the relaxation parameter was varied in between $\betakorteweg \in \lbrace 100 , 1000 \rbrace$.
The domain $\domain \in \l( -1 , 1 \r)$ was discretized by $200$ elements with a polynomial degree of $\Npoly = 5$.

\begin{figure}[ht!]
\begin{center}
	\includegraphics[width=\linewidth,trim = 2.25cm 0 0 0, clip]{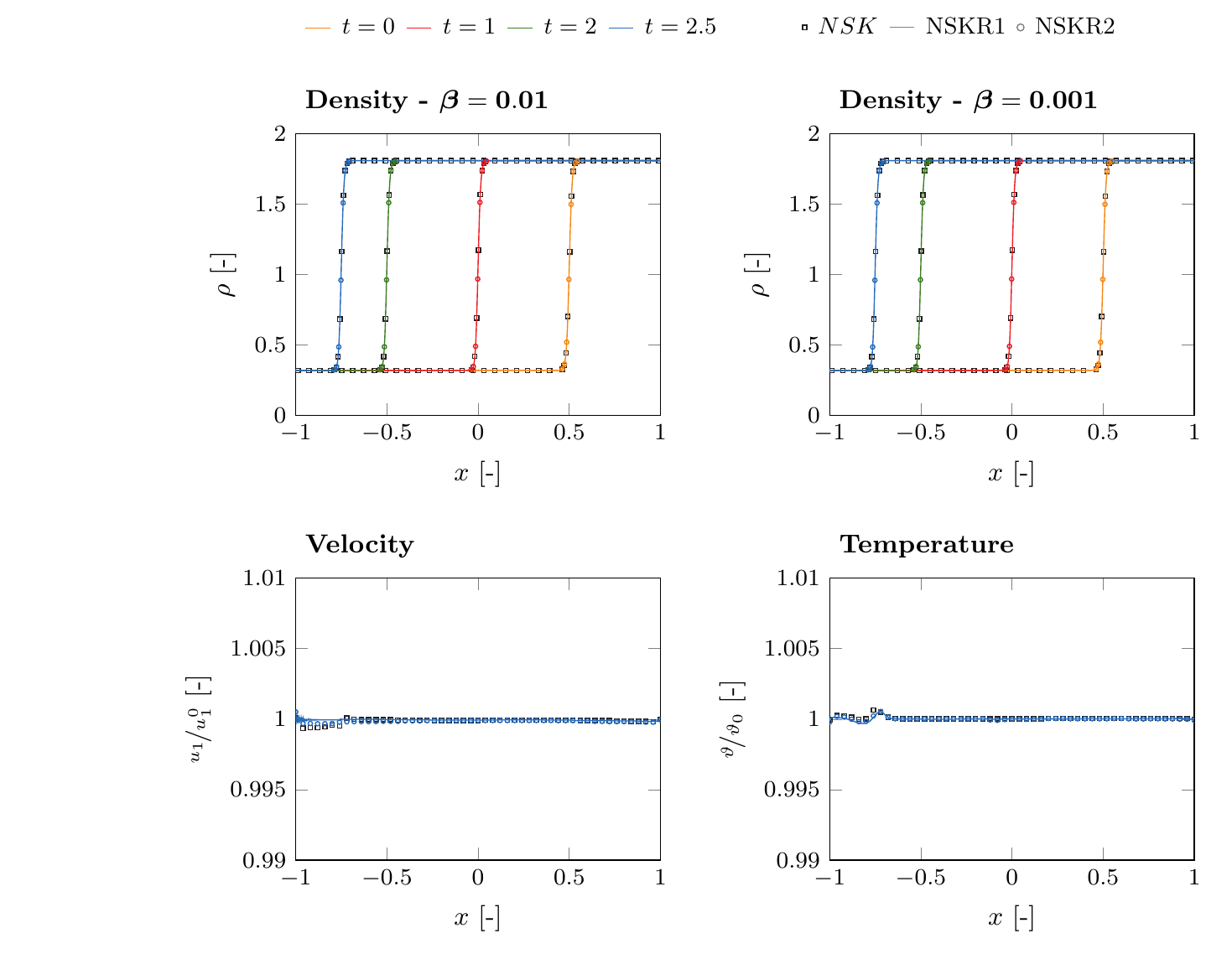}
\end{center}
\caption{\label{fig:travelingwave1D}Top: Evolution of the density profile for the \originalmodel ~and both relaxation model formulations. Bottom: Comparison of the velocity distributions (left) and the temperature distributions (right) at $\tend = 2.5$. The abbreviations NSK, NSKR1 and NSKR2 indicate the \originalmodel, the \relaxationmodelone ~and the \relaxationmodeltwo, respectively.}
\end{figure}

In the upper part of \cref{fig:travelingwave1D}, the density profiles at the time levels $\tend \in \lbrace 0 , 1 , 2 , 2.5 \rbrace$ are depicted for the \originalmodel ~and both relaxation model formulations.
An excellent agreement with the \originalmodel ~is achieved for both specified relaxation parameters $\betakorteweg = 100$ (left) and $\betakorteweg = 1000$ (right).
In the bottom view on the left, the velocity distributions are depicted at the time $\tend = 2.5$ for all model formulations.
A minor perturbation in the velocity can be observed for all models.
This is due to the imperfect initialization of an equilibrium density profile and the interaction of these perturbations with the boundaries.
However, these perturbations are in a similar order of magnitude for all model formulations and not an intricate feature of the relaxation models.
An additional view of the temperature at the time $\tend = 2.5$ is given in the bottom right of \cref{fig:travelingwave1D}.
The temperature field reflects the results already discussed for the velocity.

The authors want to highlight that the new relaxation equation for the order parameter $\ckorteweg$, which additionally considers a convection term, is a significant improvement in comparison to the results given in \cite{hitz2020}.
In \cite{hitz2020}, even with a relaxation parameter of $\betakorteweg = 0.001$, which corresponds to $\betakorteweg = 1000$ in the present work, the relaxation model was not able to fit the velocity profile of the original NSK model.
\subsection{2D Test Case}
\label{sec:2D}
Turning to multi-dimensional test cases, a merging event of two droplets in 2D was simulated.
The density field was initially defined as
\begin{align}
\dens \l( \dimvec , \tend_0 \r) &= \dens_\liq^\sat + \sum_{j = 1}^2 \frac{\dens_\vap^\sat - \dens_\liq^\sat}{2} \tanh \l( 4 \frac{ | \dimvec - \dimvec^0_j | - \radius_j}{\interfacewidth} \r).
\end{align}
The physical parameters used are summarized in \cref{table:mergingdroplets2D}.
\begin{table}[ht!]
	\begin{center}
	\begin{tabular}{l ? c c c c c c c c }
		\hline
		\hline
		\rule{0pt}{14pt} Quantity & $\temp_0$ & $\dens_\vap^\sat$ & $\dens_\liq^\sat$ & $\vel$ & $\dimvec_1^0$ & $\dimvec_2^0$ & $\radius_1$ & $\radius_2$ \\[1ex]
		\Xhline{2\arrayrulewidth}
		\rule{0pt}{15pt} Value & $0.85$ & $0.3197$ & $1.8071$ & $\transpose{\l( 0 , 0 , 0 \r)}$ & $\transpose{\l( 0.4 , 0.5 , 0 \r)}$ & $\transpose{\l( 0.7 , 0.5 , 0 \r)}$ & $0.2$ & $0.1$ \\[1ex]
		\hline
		\hline
	\end{tabular}
	\caption{\label{table:mergingdroplets2D}The values used for the initialization of the 2D merging droplets test case.}
	\end{center}
\end{table}
Based on the results of the 1D test cases presented in \cref{sec:1D}, the Korteweg parameter was fixed to $\alphakorteweg = 100$ and the relaxation parameter to $\betakorteweg = 100$.
The domain $\domain \in \l( 0 , 1 \r)^2$ was discretized by $100^2$ elements.
In case of the \originalmodel ~and the \relaxationmodelone ~a polynomial degree of $\Npoly = 3$ was used, and for the \relaxationmodeltwo ~the polynomial degree was varied in between $\Npoly \in \lbrace 4 , 5 , 6 \rbrace$.
The reasoning for the variation of the polynomial degree for the \relaxationmodeltwo ~was to investigate the influence of the mixed discretization of the Korteweg tensor.
As one part of the Korteweg tensor is approximated via the solution of a Riemann problem and the other part is treated as a point-wise source term, see also \cref{sec:numerics_bulk}, the \relaxationmodeltwo ~might be more sensitive to the chosen grid resolution in comparison to the \relaxationmodelone.

\begin{figure}[ht!]
\begin{center}
	\includegraphics[width=0.9\linewidth]{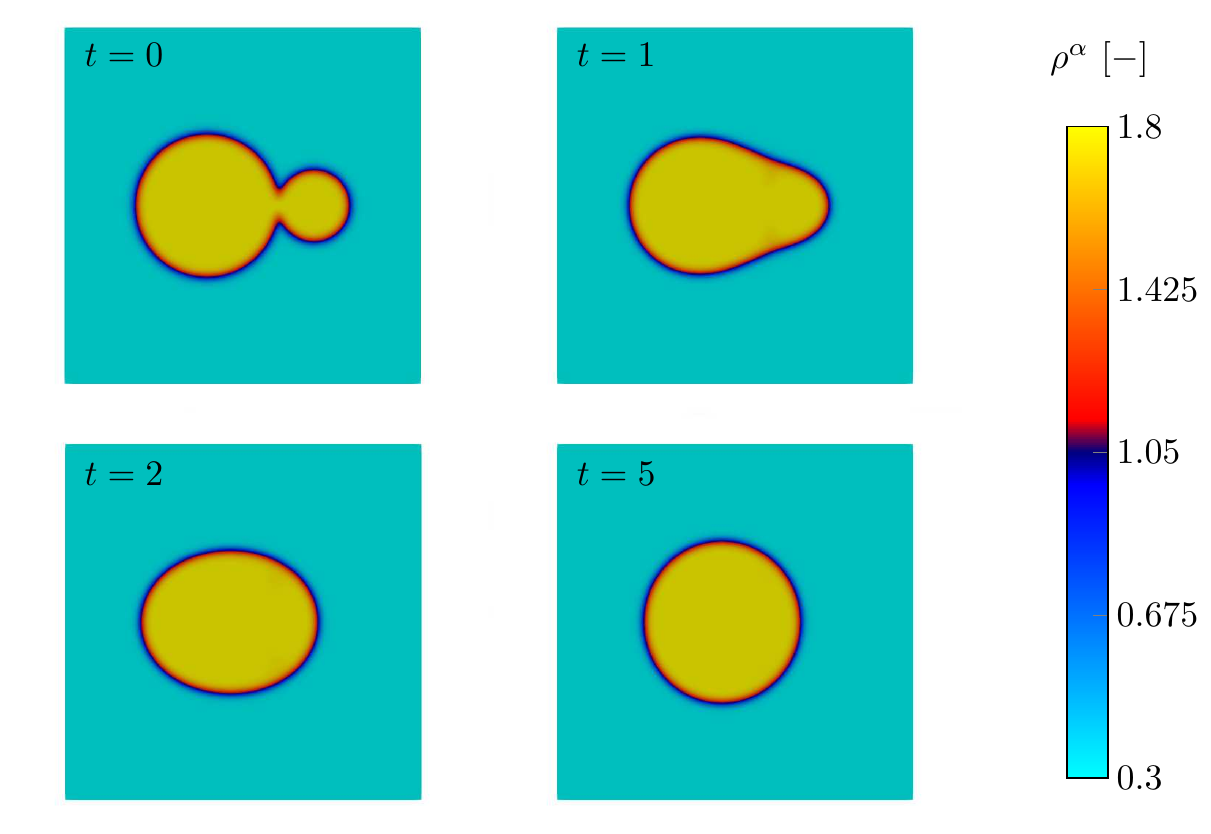}
\end{center}
\caption{\label{fig:mergingdroplets2D}Temporal evolution of the density field computed with the \relaxationmodelone.}
\end{figure}

\begin{figure}[ht!]
\begin{center}
	\includegraphics[width=0.9\linewidth]{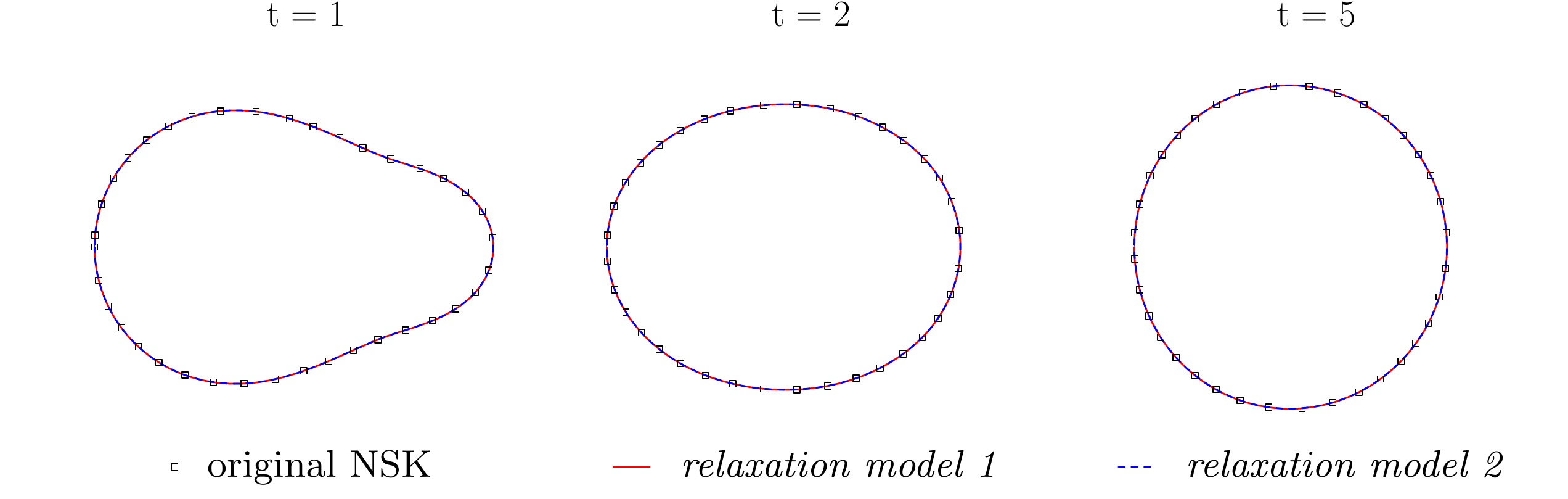}
\end{center}
\caption{\label{fig:mergingdroplets2Dbeta}Comparison of the phase interface position visualized by the density contour $\tilde{\dens} = 1.05$ at different time instances. The results of the \relaxationmodeltwo ~depicted in the figure were achieved with a polynomial degree $\Npoly = 4$.}
\end{figure}

\begin{figure}[ht!]
\begin{center}
	\includegraphics[width=\linewidth,,trim = 1.0cm 0 0 0, clip]{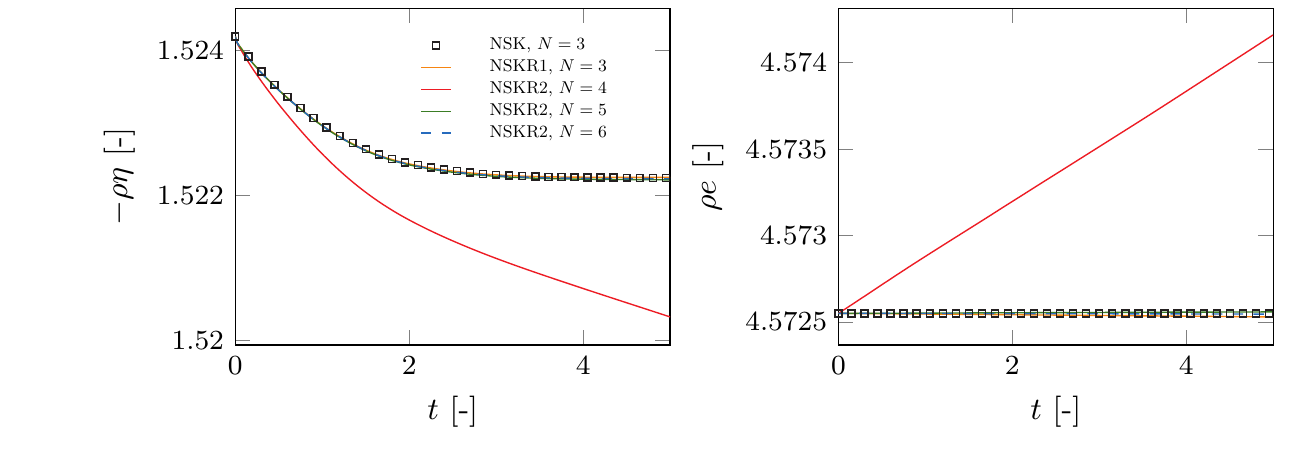}
\end{center}
\caption{\label{fig:mergingdroplets2Dentropy}The entropy per unit volume (left) and the total energy per unit volume (right) over the time for the merging droplet test case. The abbreviations NSK, NSKR1 and NSKR2 indicate the \originalmodel, the \relaxationmodelone ~and the \relaxationmodeltwo, respectively.}
\end{figure}

In \cref{fig:mergingdroplets2D}, the temporal evolution of the density field computed with the \relaxationmodelone ~is depicted.
Driven by the surface tension forces, both droplets eventually merge to a single spherical droplet.
Both relaxation model formulations are able to approximate the temporal evolution of the phase interface very well, as illustrated in \cref{fig:mergingdroplets2Dbeta}, where contours of the density at $\tilde{\dens} = 1.05$ for the original and both relaxation model formulations are presented.
Moreover, both relaxation model formulations fulfilled the Second Law of Thermodynamics, as depicted in \cref{fig:mergingdroplets2Dentropy}.
However, the entropy production predicted by the \relaxationmodeltwo ~with $N=4$ is larger compared to the results of the \relaxationmodelone, which reproduces the temporal evolution of the entropy per unit volume of the \originalmodel ~in \cref{fig:mergingdroplets2Dentropy} very well.
An increase of the spatial resolution. e.g. through an increase of the polynomial degree $\Npoly \in \lbrace 5, 6 \rbrace$, resolved this issue for the \relaxationmodeltwo.
The reason for the incorrect entropy production rate in case of the \relaxationmodeltwo ~is the splitting of the Korteweg tensor into a hyperbolic part and a point-wise source term which violates the well-balanced property without a sufficient spatial resolution.
This in turn violates the conservation of the total energy, as depicted in \cref{fig:mergingdroplets2Dentropy} on the right, and eventually causes the wrong entropy production rate.

Based on the results of the merging droplet event, the \relaxationmodelone ~is chosen as the standard formulation in the following due to the lower resolution requirements.
\subsection{3D Test Cases}
\label{sec:3D}
Binary head on collisions, as presented in \cite{gelissen2018,rajkotwala2020}, where used to validate the \relaxationmodelone ~in 3D.
In \cite{gelissen2018,rajkotwala2020}, the authors defined different numerical test cases based on collision Weber and Reynolds numbers which lie in different collision regimes.
However, during the comparison of the initialization given in \cite{gelissen2018} and \cite{rajkotwala2020} it turned out that the Reynolds numbers specified in \cite{rajkotwala2020} (Table 3) have been mixed up in between the test cases.
Moreover, the velocities specified in \cite{gelissen2018} (Table 2) proved to be inconsistent to the specified collision Weber numbers given in \cite{gelissen2018} (Table 2) and \cite{rajkotwala2020} (Table 3).
Therefore, in the present work, the initializations have been adjusted to fit the initial internal, kinetic and surface energy depicted in \cite{gelissen2018} (Figure 15).

In this work, two head-on droplet collisions, one in the \textit{reflexive separation} regime and one in the \textit{toroidal droplet breakup} regime have been simulated.
The fluid properties and initial data common to both setups are summarized in \cref{table:collidingdroplets3D}.
\begin{table}[ht!]
	\begin{center}
	\begin{tabular}{l ? c c c c c c c c c }
		\hline
		\hline
		\rule{0pt}{14pt} Quantity & $\temp_0$ & $\dens_\vap$ & $\dens_\liq $ & $\dimvec_1^0$ & $\dimvec_2^0$ & $\radius_d$ & $\gammakorteweg$ & $\surftenscoeff_{\mathrm{vl}}$ & $\interfacewidth$\\[1ex]
		\Xhline{2\arrayrulewidth}
		\rule{0pt}{15pt} Value & $0.85$ & $0.3537$ & $1.8493$ & $\transpose{\l( 0.3 , 0.5 , 0.5 \r)}$ & $\transpose{\l( 0.7 , 0.5 , 0 \r)}$ & $0.1$ & $\frac{1}{6000}$ & $0.0068$ & $0.0772$ \\[1ex]
		\hline
		\hline
	\end{tabular}
	\caption{\label{table:collidingdroplets3D}The values used for the initialization of the 3D droplet collision test cases.}
	\end{center}
\end{table}
The vapor and liquid densities specified in \cref{table:collidingdroplets3D} were chosen to fulfill the Young-Laplace law for the given temperature $\temp_0$ and droplet radius $\radius_d$.
For this, the pressure in the liquid and the vapor are given as
\begin{align*}
\pres_\vap = \pres^\sat \l( \temp_0 \r) + \frac{\dens_\vap^\sat}{\dens_\liq^\sat - \dens_\vap^\sat} \frac{2 \surftenscoeff_{\mathrm{vl}}}{\radius_d} , \quad \pres_\liq = \pres^\sat \l( \temp_0 \r) + \frac{\dens_\liq^\sat}{\dens_\liq^\sat - \dens_\vap^\sat} \frac{2 \surftenscoeff_{\mathrm{vl}}}{\radius_d}
\end{align*}
and the densities were evaluated as $\dens_\vap = \dens (\temp_0 , \pres_\vap)$ and $\dens_\liq =  \dens (\temp_0 , \pres_\liq)$ in the vapor and the liquid, respectively.
The remaining parameters as well as the resulting similarity parameters are summarized in \cref{table:collidingdropletssetup3D}.

\begin{table}[ht!]
	\begin{center}
	\begin{tabular}{l ? c c c c c c c }
		\hline
		\hline
		\rule{0pt}{14pt} Regime & $\vel_{rel}$ & $\munewton$ & $\heatcoef$ & $\cvdw$ & $\Reynolds$ & $\Weber_{\text{col}}$ & $\Prandtl$ \\[1ex]
		\Xhline{2\arrayrulewidth}
		\rule{0pt}{15pt} Reflexive separation & $\transpose{\l( 2.2 ,0 , 0 \r)}$ & $6.944 \cdot 10^{-4}$ & $\frac{1}{150}$ & $5$ & $1440$ & $265$ & $ 20 $ \\[1ex]
		\rule{0pt}{15pt} Toroidal breakup & $\transpose{\l( 4   ,0 , 0 \r)}$ & $1.010 \cdot 10^{-3}$ & $\frac{1}{150}$ & $5$ & $990$  & $876$ & $ 20 $ \\[1ex]
		\hline
		\hline
	\end{tabular}
	\caption{\label{table:collidingdropletssetup3D}The values used for the initialization of the 3D droplet collision test cases.}
	\end{center}
\end{table}
The initial density and velocity fields were given by
\begin{align*}
\dens \l( \dimvec , \tend_0 \r) &= \dens_\liq + \sum_{j = 1}^2 \frac{\dens_\vap - \dens_\liq}{2} \tanh \l( 4 \frac{ | \dimvec - \dimvec^0_j | - \radius_d}{\interfacewidth} \r), \\
\vel \l( \dimvec , \tend \r) &= 
\begin{cases}
	{\color{white}{-}} \frac{ \vel_{rel}}{2} \l( 1 - \tanh \l( 4 \frac{ | \dimvec - \dimvec^0_j | - \radius_j}{\interfacewidth} \r) \r) &: \xdir < 0.5, \\
	               -   \frac{ \vel_{rel}}{2} \l( 1 - \tanh \l( 4 \frac{ | \dimvec - \dimvec^0_j | - \radius_j}{\interfacewidth} \r) \r) &: \xdir > 0.5
\end{cases}
\end{align*}
for both simulations.
The Korteweg parameter was fixed to $\alphakorteweg = 200$ as a value of $\alphakorteweg = 100$ turned out to be insufficient to reproduce the separation of the two droplets in the \textit{reflexive separation} test case.
The relaxation parameter was specified as $\betakorteweg = 1000$ to adequately capture the strong dynamics of the test cases.
The domain $\domain \in \l( 0 , 1 \r)^3$ was discretized by $50^3$ elements with $\Npoly = 3$.
\begin{figure}[ht!]
\vspace{0.5cm}
\begin{center}
	\includegraphics[width=\linewidth]{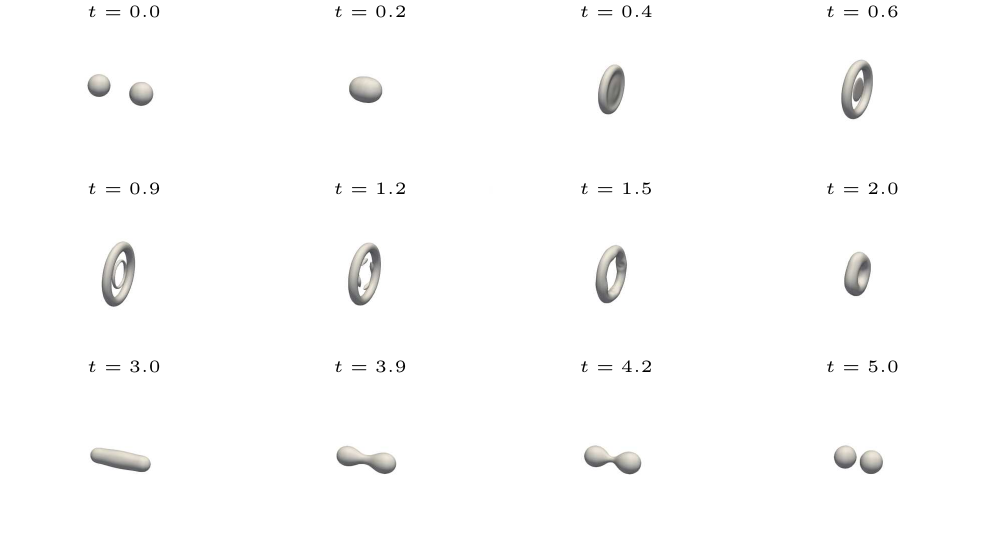}
\end{center}
\caption{\label{fig:reflexive3D}Temporal evolution of the \textit{reflexive separation} test case visualized by the iso-contour of the mean density $\tilde{\dens}$. The depicted results were achieved with the \relaxationmodelone.}
\end{figure}

\begin{figure}[ht!]
\begin{center}
	\includegraphics[width=\linewidth,trim = 1.0cm 0 0 0, clip]{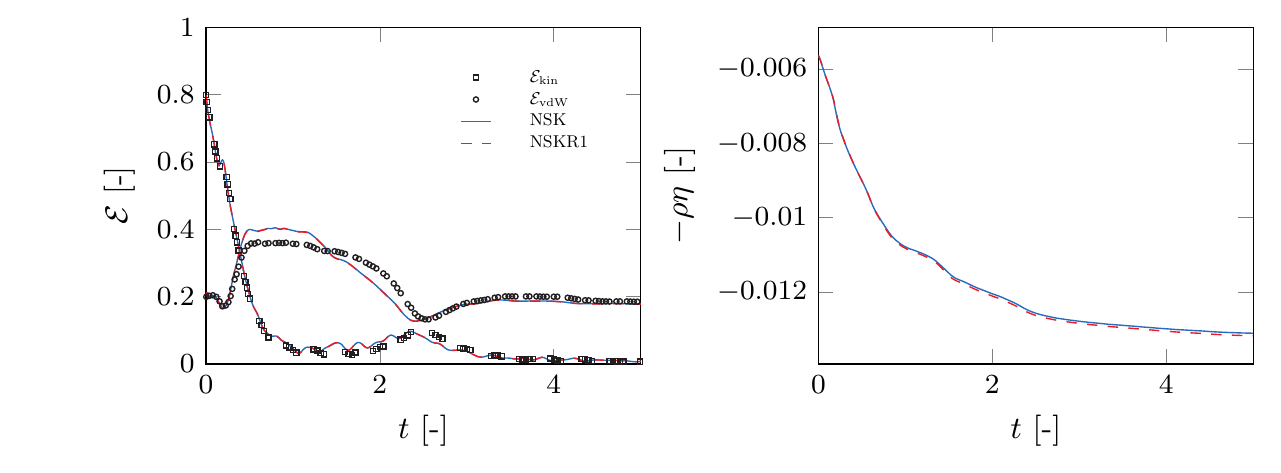}
\end{center}
\caption{\label{fig:reflexiveenergy}Left: The temporal evolution of the integral kinetic and surface energy computed with the \originalmodel ~and the \relaxationmodelone ~for the \textit{reflexive separation} test case. Moreover, the results of \cite{rajkotwala2020} depicted as symbols are plotted as a reference. Right: The temporal evolution of the integral entropy per unit volume. The abbreviations NSK and NSKR1 indicate the \originalmodel ~and the \relaxationmodelone, respectively.}
\end{figure}

The time evolution of the \textit{reflexive separation} test case is depicted in \cref{fig:reflexive3D}.
The results are visualized by the iso-contour of the mean density $\tilde{\dens} = \frac{1}{2} \l( \dens_\liq^\sat\l( \temp_0 \r) + \dens_\vap^\sat \l( \temp_0 \r) \r)$.
Both droplets move towards each other, and at a critical distance they merge to a single droplet.
In the further evolution, this single droplet elongates and forms a disc.
The inner disc separates from the outer ring of the elongated droplet and forms an inner ring which then breaks up into four droplets.
Due to the counteracting surface tension forces the outer ring contracts, merges with the four inner droplets and forms again a single droplet which now starts to elongate in the $\xdir$-direction.
In the further evolution, this elongated droplet starts to form two droplets moving away from each other, but they are still connected by a liquid bridge.
Eventually, the liquid bridge breaks, and two separated droplets remain.
The overall temporal evolution of the test case is in qualitative agreement with the results in \cite{gelissen2018,rajkotwala2020}.
A quantitative comparison of the results is given in \cref{fig:reflexiveenergy}, where on the left the integral kinetic and surface energy of the \originalmodel ~and the \relaxationmodelone ~are drawn over the time and compared to the results in \cite{rajkotwala2020} which are highlighted by the symbols.
The results of the \relaxationmodelone ~match the one of the \originalmodel ~perfectly.
However, both models overpredict the integral surface energy compared with the results in \cite{rajkotwala2020}.
Moreover, the temporal evolution seems to be faster in the present work.
The same behavior is also mirrored in the integral kinetic energy.
Hence, the results in the present work seem to be less dissipative compared with the results in \cite{rajkotwala2020}.
The reason for this might be twofold.
First, in the present study the same number of degrees of freedom were used as in \cite{gelissen2018} and \cite{rajkotwala2020}, however, we employed a fourth-order \dgsem ~whereas in \cite{rajkotwala2020} a second-order \fv ~scheme was used.
The authors in \cite{rajkotwala2020} did not discuss if their presented results already achieved grid convergence.
In the present work, a further grid refinement did not alter the overall dynamics in \cref{fig:reflexive3D} or the results in \cref{fig:reflexiveenergy}.
A second reason for the discrepancies might evolve from the ambiguously specified Reynolds numbers in \cite{rajkotwala2020}, which even deviate form the Reynolds numbers specified in \cite{gelissen2018}.
Nevertheless, the authors want to highlight that the deviations of the \relaxationmodelone ~to the reference are not an issue of the model itself, as it perfectly fits the results of the \originalmodel ~for the setup specified in the \cref{table:collidingdroplets3D,table:collidingdropletssetup3D}.
Furthermore, in the right panel of \cref{fig:reflexiveenergy}, the temporal evolution of the integral entropy per unit volume is drawn over the time.
Also for this highly dynamic test case, the \relaxationmodelone ~is consistent to the Second Law of Thermodynamics.

\begin{figure}[ht!]
\begin{center}
	\includegraphics[width=\linewidth]{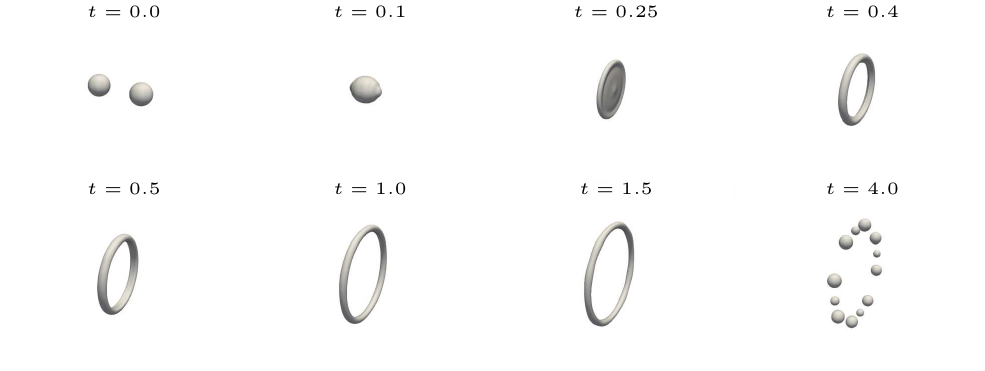}
\end{center}
\caption{\label{fig:toroidal3D}Temporal evolution of the \textit{toroidal breakup} test case visualized by the iso-contour of the mean density $\tilde{\dens}$. The depicted results were achieved with the \relaxationmodelone.}
\end{figure}

\begin{figure}[ht!]
\begin{center}
	\includegraphics[width=\linewidth,trim = 1.0cm 0 0 0, clip]{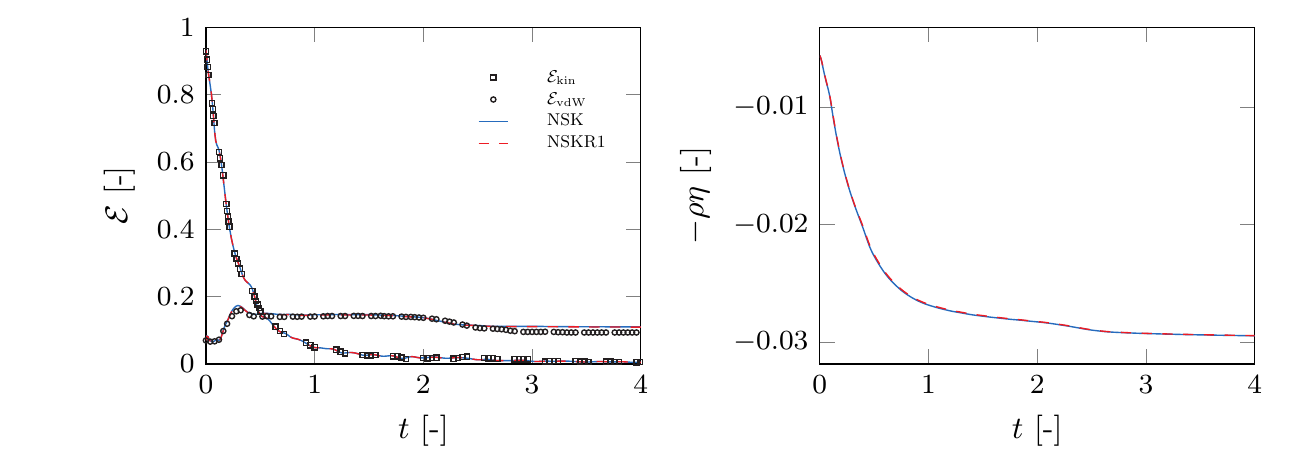}
\end{center}
\caption{\label{fig:toroidalenergy}Left: The temporal evolution of the integral kinetic and surface energies computed with the \originalmodel ~and the \relaxationmodelone ~for the \textit{toroidal breakup} test case. Moreover, the results of \cite{rajkotwala2020} depicted as symbols are plotted as a reference. Right: The temporal evolution of the integral entropy per unit volume. The abbreviations NSK and NSKR1 indicate the \originalmodel ~and the \relaxationmodelone, respectively.}
\end{figure}

Turning to the second test case, the \textit{toroidal droplet breakup}, the temporal evolution is again visualized by the iso-contour of the mean density $\tilde{\dens}$, as depicted in \cref{fig:toroidal3D}.
Also in this test case, the two droplets move towards each other and merge to a single droplet.
This new droplet elongates and forms a disc with a thin film connected to the outer ring.
Over the time, the inner film breaks and the remaining ring further expands and starts to oscillate.
These oscillations eventually cause the ring to break into twelve single droplets.
Also the results of this test case are in qualitative agreement with the one presented in \cite{gelissen2018,rajkotwala2020}.
A quantitative comparison for the \textit{toroidal breakup} test case is given in \cref{fig:toroidalenergy}.
The \relaxationmodelone ~fits the results of the \originalmodel ~perfectly, as depicted on the left.
Additionally, a very good agreement with the reference given in \cite{rajkotwala2020} is achieved.
However, the surface energies in the present work slightly overestimate the ones given in \cite{rajkotwala2020} at approximately $\tend \approx 2.5$ .
The deviation can be traced back to the different amount of droplets, eight in \cite{rajkotwala2020} and twelve in the present study, at the final time.
This is surprising in the sense that we would expect a symmetrical result from a symmetrical test case, where surface tension and viscous forces are still dominating.
However, the results presented in \cite{rajkotwala2020} show different mirror-symmetrical patterns to the $\xdir \ydir$-plane as to the $\xdir \zdir$-plane.
Furthermore, consistency with the Second Law of Thermodynamics is achieved, as depicted in \cref{fig:toroidalenergy} on the right.
\subsection{Confined Domains}
\label{sec:CD}
\subsubsection{Static Contact Angles}
\label{sec:staticcontactangles}
Turning to wall bounded flow, we first validated the equilibrium part of the contact angle boundary conditions presented in \cref{eq:bc_contactangle,eq:bc_contactangle_alpha}.
For this reason, a liquid pillar
\begin{align*}
\dens \l( \dimvec , \tend_0 \r) &= \frac{\dens_\vap + \dens_\liq}{2} + \frac{\dens_\vap - \dens_\liq}{2} \tanh \l( 4 \frac{ \dist - \radius_d}{\interfacewidth} \r)
\end{align*}
with $\dist = \sqrt{\l( \xdir - \xdir_0 \r)^2 + \l( \ydir - \ydir_0 \r)^2}$ was initialized in the domain $\domain \in (0,1) \times (0,1) \times (0,0.2)$.
A summary of the parameters used is given in \cref{table:staticcontactangle3D}.
\begin{table}[ht!]
	\begin{center}
	\begin{tabular}{l ? c c c c c c c c }
		\hline
		\hline
		\rule{0pt}{14pt} Quantity & $\temp_0$ & $\dens_\vap^\sat$ & $\dens_\liq^\sat$ & $\vel$ & $\dimvec^0$ & $\radius_d$ & $\surf{\temp}$ & $\contactangle$ \\[1ex]
		\Xhline{2\arrayrulewidth}
		\rule{0pt}{15pt} Value & $0.85$ & $0.3197$ & $1.8071$ & $\transpose{\l( 0 , 0 , 0 \r)}$ & $\transpose{\l( 0.5 , 0.5 , 0 \r)}$ & $0.2$ & $0.85$ & $\lbrace 30 , 60 , 90 , 120 , 150 \rbrace$ \\[1ex]
		\hline
		\hline
	\end{tabular}
	\caption{\label{table:staticcontactangle3D}The values used for the initialization of the test cases to validate the 3D static contact angle boundary conditions.}
	\end{center}
\end{table}
The Korteweg parameter was fixed to $\alphakorteweg = 100$, and the relaxation parameter was specified as $\betakorteweg = 100$.
At the top and the bottom of the domain, isothermal no-slip walls were set as boundary conditions.
The remaining boundary conditions have been chosen as periodic.
In the subdomain $\domain_I \in (0.25,0.75) \times (0.25,0.75) \times (0,0.2)$, a grid spacing of $\Delta \xdir = \Delta \ydir =  \Delta \zdir = 0.01$ has been used.
In the remaining domain $\domain_{II} = \domain \backslash \domain_I$ a mesh coarsening by the use of mortar interfaces was used, which results in a total number of $64000$ elements each with $\Npoly = 3$.
For each specified contact angle the simulation has been run until an equilibrium state had been reached.

\begin{figure}[ht!]
\begin{center}
	\includegraphics[width=\linewidth,trim = 0 0 0 0, clip]{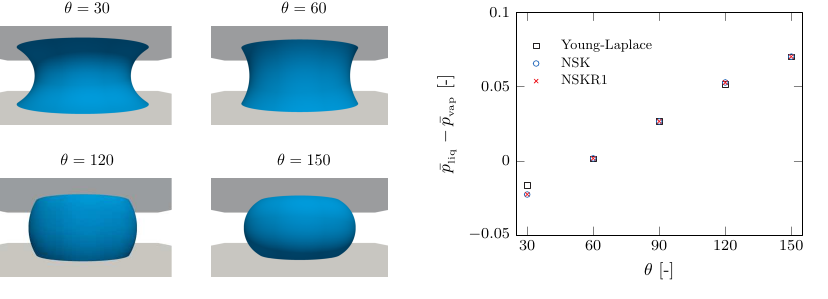}
\end{center}
\caption{\label{fig:staticcontactangle3D}Left: Liquid bridges for different contact angles visualized by the iso-contour of the mean density $\tilde{\dens}$ at $t = 15$. All presented results have been computed with the \relaxationmodelone. Right: Comparison of the pressure jump across the phase interface for the \originalmodel, the \relaxationmodelone ~and the theoretical value predicted by the Young-Laplace law. The abbreviations NSK and NSKR1 indicate the \originalmodel ~and the \relaxationmodelone, respectively.}
\end{figure}

On the left in \cref{fig:staticcontactangle3D}, the liquid bridges formed and visualized by the iso-contour of the mean density $\tilde{\dens} = 0.5 \l( \dens_\vap^\sat \l( \temp_0 \r) + \dens_\liq^\sat \l( \temp_0 \r) \r)$ are depicted for the static contact angles $\contactangle \in \lbrace 30 , 60 , 120 , 150 \rbrace$.
A quantitative validation of the contact angle boundary conditions is given in \cref{fig:staticcontactangle3D} on the right, where the pressure jump across the phase interface is compared to the theoretical value predicted by the Young-Laplace law
\begin{align}
\label{eq:younglaplacelaw}
\Delta \pres = \pres_\liq - \pres_\vap = \surftenscoeff_{\mathrm{vl}} \l( \frac{1}{\radius_1} + \frac{1}{\radius_2} \r)
\end{align}
with the surface tension coefficient $\surftenscoeff$ specified in \cref{table:fluidproperties}.
In \cref{eq:younglaplacelaw}, the radii $\radius_1$ and $\radius_2$ are chosen according to \cite{doermann2015,mastrangeli2015}.
Hence, the radius $\radius_1$ is defined as the theoretical value prescribed by the boundary condition
\begin{align*}
\radius_1 = -\frac{\height}{2 \cos \l( \contactangle \r)}
\end{align*}
with the channel height $\height = 0.2$.
The radius $\radius_2$ is defined as the distance between the point $\dimvec_\liq = \transpose{\l( 0.5 , 0.5 , 0.1 \r)}$ and the phase interface measured in the $\xdir \ydir$-plane.
In the simulations, the pressure jump across the phase interface was determined as the difference of the averaged pressures $\bar{\pres}_\liq$ and $\bar{\pres}_\vap$ defined as
\begin{align*}
\bar{\pres}_\liq \coloneqq \frac{1}{\height} \int_0^\height \pres \l( 0.5 , 0.5 , z \r) \dist \zdir \quad \mathrm{and} \quad \bar{\pres}_\vap \coloneqq \frac{1}{\height} \int_0^\height \pres \l( 0 , 0.5 , z \r) \dist \zdir.
\end{align*}
An excellent agreement between the \originalmodel ~and the \relaxationmodelone ~is achieved as depicted in \cref{fig:staticcontactangle3D} on the right.
Moreover, both model formulations coincide very well with the theoretical values predicted by the Young-Laplace law for nearly all prescribed contact angles.
A minor deviation can be observed for the $\contactangle = 30 \degree$ case.
The reason for this is a non-constant pressure profile in the $\zdir$-direction inside the droplet with strong deviations at the wall from the mean value, which is induced by the contact angle boundary condition.
Due to the nano sized droplet, this near-wall region is not negligible and alters the averaged pressure.
Hence, the use of the Young-Laplace as a reference for the $\contactangle = 30 \degree$ case and such nano sized channels might be invalid.

This is also supported by simulations performed by the authors, but not presented in this work, where a initial radius $\radius_d = 0.1$ was used.
Although the phase interfaces did not get in contact with each other, they still effected each other, which caused significant deviations from the Young-Laplace law for all specified contact angles.
However, these deviations are expected to vanish for larger droplets and channel heights.
\subsubsection{Dynamic Contact Angles}
\label{sec:dynamiccontactangles}
To validate the dynamic contact angle effects, a droplet which slides over a wall was simulated.
For this, a droplet was initialized as a half-sphere in the domain $\domain \in (0,1) \times (0,0.5) \times (0,0.25)$ with
\begin{align*}
\dens \l( \dimvec , \tend_0 \r) &= \frac{\dens_\vap + \dens_\liq}{2} + \frac{\dens_\vap - \dens_\liq}{2} \tanh \l( 4 \frac{ \dist - \radius_d}{\interfacewidth} \r),
\end{align*}
where $\dist = \sqrt{\l( \xdir - \xdir_0 \r)^2 + \l( \ydir - \ydir_0 \r)^2  + \l( \zdir - \zdir_0 \r)^2}$.
A summary of the parameters used for the setup is given in \cref{table:dynamiccontactangle3D}.
\begin{table}[ht!]
	\begin{center}
	\begin{tabular}{l ? c c c c c c c c c c c }
		\hline
		\hline
		\rule{0pt}{14pt} Quantity & $\temp_0$ & $\dens_\vap^\sat$ & $\dens_\liq^\sat$ & $\dimvec^0$ & $\radius_d$ & $\surf{\temp}$ & $\surf{\vel}$ & $\contactangle$ & $\affinitycoef{1}$ & $\affinitycoef{2}$ & $\affinitycoef{4}$ \\[1ex]
		\Xhline{2\arrayrulewidth}
		\rule{0pt}{15pt} Value & $0.85$ & $0.3197$ & $1.8071$ & $\transpose{\l( 0.5 , 0.5 , 0 \r)}$ & $0.1$ & $0.85$ & $\transpose{\l( 0.5 , 0 , 0 \r)}$ & $90$ & $20$ & $8 \cdot 10^{-4}$ & $0$ \\[1ex]
		\hline
		\hline
	\end{tabular}
	\caption{\label{table:dynamiccontactangle3D}The values used for the initialization of the test cases to validate the 3D dynamic contact angle boundary conditions.}
	\end{center}
\end{table}
The Korteweg parameter was fixed to $\alphakorteweg = 100$ and the relaxation parameter was specified as $\betakorteweg = 100$.
At the top of the domain, an isothermal no-slip wall was imposed, which moves with the velocity $\surf{\vel}$.
The bottom of the domain was set as an isothermal generalized Navier-slip boundary condition by the use of \cref{eq:bc_nsk,eq:bc_nsk_alpha}.
The remaining boundary conditions have been chosen as periodic.
In the subdomain $\domain_I \in (0,1) \times (0.1,0.4) \times (0,0.15)$, a grid spacing of $\Delta \xdir = \Delta \ydir =  \Delta \zdir = 0.01$ was used.
In the remaining domain $\domain_{II} = \domain \backslash \domain_I$, a mesh coarsening by the use of mortar interfaces was applied, which results in a total number of $58750 $ elements each with $\Npoly = 3$.
The specified boundary conditions induce the formation of a Couette flow, which then causes the droplet to slide over the bottom wall and form a contact angle hysteresis.

\begin{figure}[ht!]
\begin{center}
	\includegraphics[width=0.9\linewidth,trim = 0.25cm 7.0cm 0.5cm 15.0cm, clip]{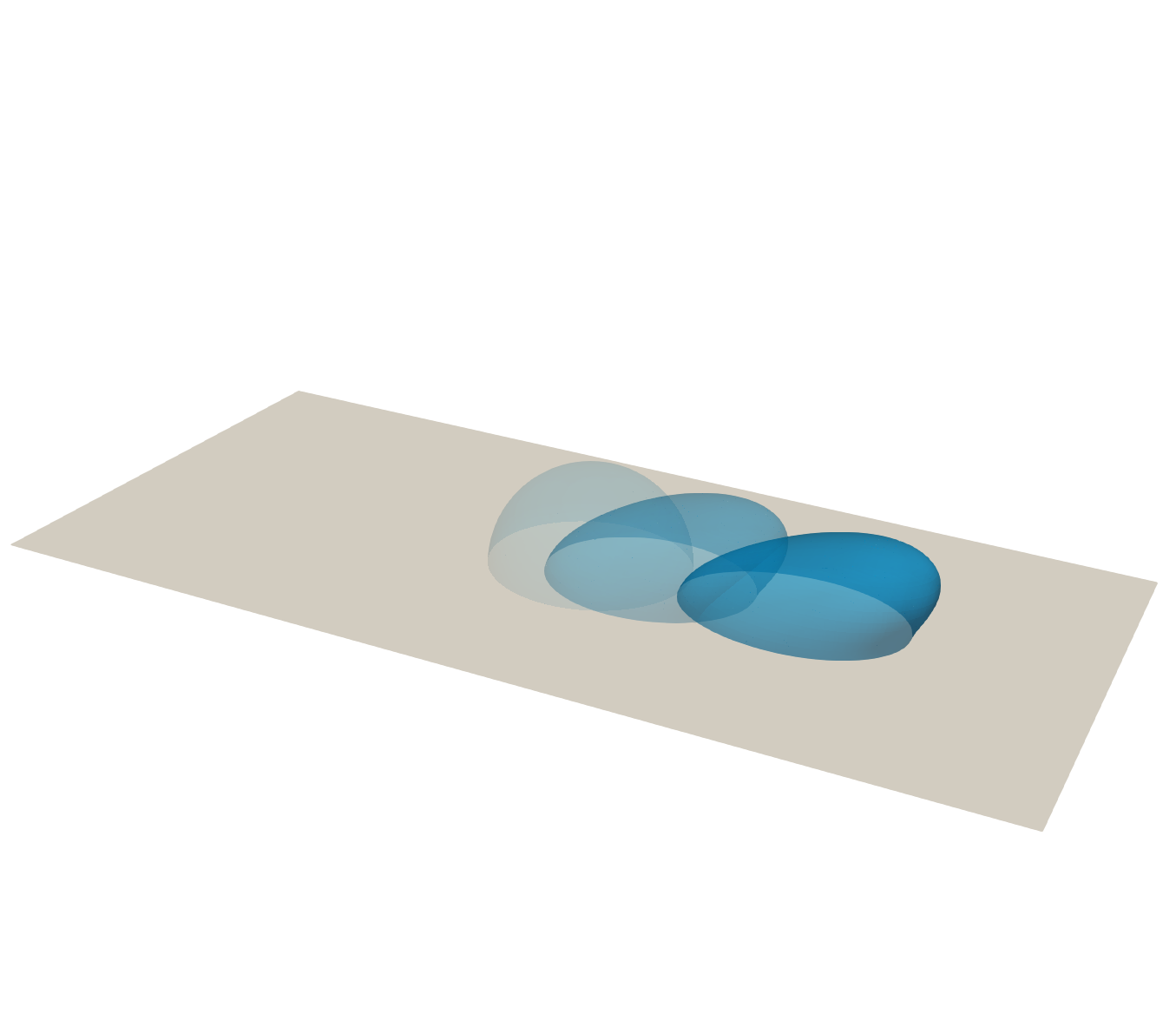}
\end{center}
\caption{\label{fig:dynamiccontactangle3D}Sliding droplet at the time instances $\tend \in \lbrace 0 , 2 , 5 \rbrace$. The droplet surface is visualized by the iso-contour of the mean density $\tilde{\dens}$.}
\end{figure}

\begin{figure}[ht!]
\begin{center}
	\includegraphics[width=0.9\linewidth,trim = 2.0cm 0 6.5cm 0cm, clip]{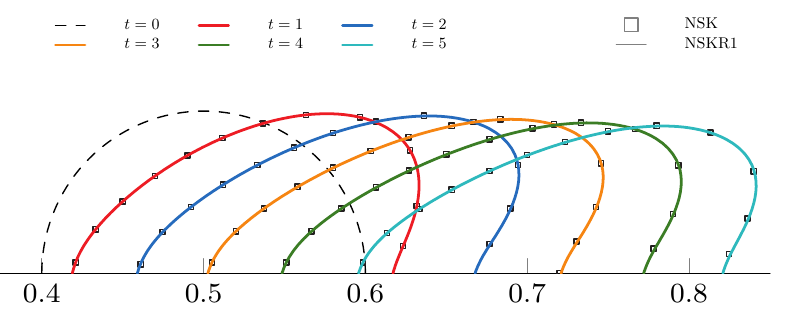}
\end{center}
\caption{\label{fig:dynamiccontactangle_comparison}Phase interface of the sliding droplet visualized by the iso-contour of the mean density $\tilde{\dens}$ at different time instances. The abbreviations NSK and NSKR1 indicate the \originalmodel ~and the \relaxationmodelone, respectively.}
\end{figure}

The evolution of the sliding droplet is depicted for the time instances $\tend \in \lbrace 0 , 2 , 5 \rbrace$ in \cref{fig:dynamiccontactangle3D}, where the surface of the droplets is visualized by the iso-contour of the mean density $\tilde{\dens} = 0.5 \l( \dens_\vap^\sat \l( \temp_0 \r) + \dens_\liq^\sat \l( \temp_0 \r) \r)$.
A comparison of the \relaxationmodelone ~to the \originalmodel ~is given in \cref{fig:dynamiccontactangle_comparison}, where the positions of the phase interfaces in the $\xdir \zdir$-plane with its origin at $\transpose{\l( 0.5 , 0.25 , 0.125 \r)}$ are presented.
The contact angle hysteresis induced by the boundary conditions prescribed is clearly visible.
Moreover, the results of the \relaxationmodelone ~are in very good agreement to the \originalmodel ~at all time levels.
\subsection{A Porous Media Example}
\label{sec:PM}
Finally, to illustrate that the new relaxation model is also capable to cope with complex three-dimensional domains and even curved surfaces, a spinodal decomposion was simulated in a structured porous media.
For this purpose, a domain $\domain \in (0,0.3)^3$ with 27 solid spheres, each with a radius $\radius$, was considered.
The positions of the centers $\dimvec_i^0 = \transpose{( \xdir_i^0 , \ydir_i^0 , \zdir_i^0 )}$ of the spheres were defined as permutations of $\xdir^0 \in \lbrace 0.05 , 0.15 , 0.25 \rbrace$, $\ydir^0 \in \lbrace 0.05 , 0.15 , 0.25 \rbrace$ and $\zdir^0 \in \lbrace 0.05 , 0.15 , 0.25 \rbrace$.
Initially, the porous domain was filled with a pure liquid at a constant density $\dens_\liq$ and a temperature $\temp_0$.
They were chosen such that the initial state is outside of the saturation dome.
The solid spheres have been specified as isothermal no-slip walls with a constant temperature $\surf{\temp}$ which is below the initial temperature $\temp_0$.
To trigger the spinodal decomposition, the wall temperature $\surf{\temp}$ was fixed to a value such that the density $\dens_\liq$ at the temperature $\surf{\temp}$ corresponds to an unstable state in the spinodal region.
All the remaining boundaries were chosen as periodic boundary conditions.
A summary of the parameters used to initialize the test case is given in \cref{table:porousdomain3D}.
\begin{table}[ht!]
	\begin{center}
	\begin{tabular}{l ? c c c c c }
		\hline
		\hline
		\rule{0pt}{14pt} Quantity & $\temp_0$ & $\dens_\liq^\sat$ & $\radius$ & $\surf{\temp}$ & $\contactangle$ \\[1ex]
		\Xhline{2\arrayrulewidth}
		\rule{0pt}{15pt} Value & $0.95$ & $1.5$ & $0.03$ & $0.75$ & $150$ \\[1ex]
		\hline
		\hline
	\end{tabular}
	\caption{\label{table:porousdomain3D}The values used for the initialization of the 3D spinodal decomposition in a structured porous media.}
	\end{center}
\end{table}
The Korteweg parameter was fixed to $\alphakorteweg = 100$ and the relaxation parameter was specified as $\betakorteweg = 100$.
To approximate the spherical shape of the solids a curved mesh with a polynomial degree of $\Ngeo = 2$ was used.
The interested reader is referred to \cite{krais2021} for more details on the used mesh curving technique.
An overall amount of $648000$ elements was used for the discretization of the domain each with $\Npoly = 3$.
\begin{figure}[ht!]
\begin{center}
	\includegraphics[width=\linewidth,trim = 0 0 0 0, clip]{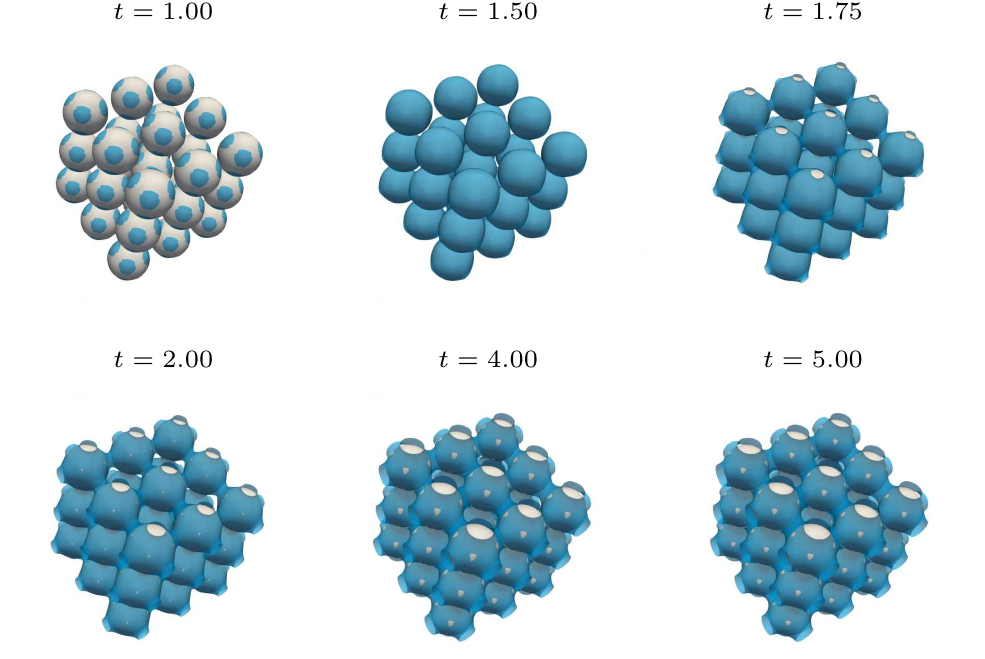}
\end{center}
\caption{\label{fig:spinodaldecomposition3D}Spinodal decomposition in a structured porous media. The phase interface is visualized by the iso-contour of the mean density $\tilde{\dens}$. Regions enclosed by the blue iso-surfaces indicate vapor. The domain was initially filled with pure liquid.}
\end{figure}

The temporal evolution of the spinodal decomposition is depicted in \cref{fig:spinodaldecomposition3D}, where the iso-contour of the mean density $\tilde{\dens} = 0.5 \l( \dens_\vap^\sat \l( \surf{\temp} \r) + \dens_\liq^\sat \l( \surf{\temp} \r) \r)$ is used to visualize the phase interface.
At the time $\tend = 1$, a small amount of vapor has formed which begins to enclose the solid spheres.
Complete entrapment of the solid is achieved at $\tend = 1.5$.
At approximately $\tend = 1.75$, the vapor regions encapsulating the solid start to connect with their direct neighbors.
As depicted in the bottom left of \cref{fig:spinodaldecomposition3D}, the liquid begins to re-wet the solid spheres ($\tend = 2$), while the vapor bridges between the solids continue to grow ($\tend = 4$).
Finally, at $\tend = 5$, an equilibrium state is reached with a co-existing vapor and liquid phase.
\section{Conclusion}
\label{sec:conclusion}
Two relaxation model formulations for the non-isothermal Navier-Stokes-Korteweg equations have been presented.
The models are an extension of the recently presented parabolic relaxation models of \cite{hitz2020} for the isothermal \nsk ~equations.
In contrast to \cite{hitz2020}, the models are extended by a convection-diffusion-reaction equation for the relaxation variable.
This guarantees the thermodynamic consistency of the relaxation model even for the non-isothermal formulation.
By the use of a modified pressure function, fully hyperbolic first-order fluxes can be constructed for the \relaxationmodeltwo.
In addition, the new bulk models were complemented by thermodynamically consistent contact angle boundary conditions, which consider static as well as dynamic contact angle effects.

Both model formulations were validated against solutions of the original \nsk ~model in one and two dimensions.
This includes static droplets and traveling wave solutions as well as a merging bubble event.
To fulfill the Second Law of Thermodynamics on a discrete level, the relaxation formulation referred to as \relaxationmodeltwo ~has proven to be much more sensitive with respect to the grid resolution than the \relaxationmodelone ~and the \originalmodel.
Nevertheless, this formulation provides an access for analytical methods which enables the derivation of homogenized models at the macro-scale, as presented in \cite{rohde2020}.
The capability of the \relaxationmodelone ~to handle complex binary head-on droplet collisions in 3D was proven by a comparison with the literature.

Turning to wall bounded flow, the introduced boundary conditions showed consistency with the Young-Laplace law.
Moreover, in a setup with dynamic contact angle effects, the relaxation model generated comparable results to the original \nsk ~model.
Finally, to prove the capability of the new relaxation model to handle complex domains, a spinodal decomposition in an artificial porous media with curved boundaries was simulated.

Future work aims to investigate wall bounded and porous media flow, e.g. nucleation processes of droplets and bubbles on a wall as well as evaporation and condensation processes in porous media.
This will take geometry variations into account, and the influence will by quantified by the use of uncertainty quantification methods, as presented in \cite{kuhn2019}.
Following \cite{rohde2020}, the derivation of a homogenized model of Darcy-type for the \nsk ~equations will be a further field of research.
\section*{Acknowledgements}
Funded by Deutsche Forschungsgemeinschaft (DFG, German Research Foundation) under Germany's Excellence Strategy - EXC 2075 – 390740016. We acknowledge the support by the Stuttgart Center for Simulation Science (SimTech).
The simulations were performed on the national supercomputer HPE Apollo System Hawk at the High Performance Computing Center Stuttgart (HLRS) under the grant number \textit{hpcmphas/44084}.
  \bibliographystyle{elsarticle-num-names} 
  \bibliography{literature}

\begin{thebibliography}{73}
\expandafter\ifx\csname natexlab\endcsname\relax\def\natexlab#1{#1}\fi
\providecommand{\url}[1]{\texttt{#1}}
\providecommand{\href}[2]{#2}
\providecommand{\path}[1]{#1}
\providecommand{\DOIprefix}{doi:}
\providecommand{\ArXivprefix}{arXiv:}
\providecommand{\URLprefix}{URL: }
\providecommand{\Pubmedprefix}{pmid:}
\providecommand{\doi}[1]{\href{http://dx.doi.org/#1}{\path{#1}}}
\providecommand{\Pubmed}[1]{\href{pmid:#1}{\path{#1}}}
\providecommand{\bibinfo}[2]{#2}
\ifx\xfnm\relax \def\xfnm[#1]{\unskip,\space#1}\fi
\bibitem[{Guo and Liu(2007)}]{guo2007}
\bibinfo{author}{Z.~Guo}, \bibinfo{author}{W.~Liu},
\newblock \bibinfo{title}{{Biomimic from the superhydrophobic plant leaves in
  nature: Binary structure and unitary structure}},
\newblock \bibinfo{journal}{Plant Science} \bibinfo{volume}{172}
  (\bibinfo{year}{2007}) \bibinfo{pages}{1103--1112}.
\bibitem[{Class et~al.(2009)Class, Ebigbo, Helmig, Dahle, Nordbotten, Celia,
  Audigane, Darcis, Ennis-King, Fan, Flemisch, Gasda, Jin, Krug, Labregere,
  {Naderi Beni}, Pawar, Sbai, Thomas, Trenty, and Wei}]{class2009}
\bibinfo{author}{H.~Class}, \bibinfo{author}{A.~Ebigbo},
  \bibinfo{author}{R.~Helmig}, \bibinfo{author}{H.~K. Dahle},
  \bibinfo{author}{J.~M. Nordbotten}, \bibinfo{author}{M.~A. Celia},
  \bibinfo{author}{P.~Audigane}, \bibinfo{author}{M.~Darcis},
  \bibinfo{author}{J.~Ennis-King}, \bibinfo{author}{Y.~Fan},
  \bibinfo{author}{B.~Flemisch}, \bibinfo{author}{S.~E. Gasda},
  \bibinfo{author}{M.~Jin}, \bibinfo{author}{S.~Krug},
  \bibinfo{author}{D.~Labregere}, \bibinfo{author}{A.~{Naderi Beni}},
  \bibinfo{author}{R.~J. Pawar}, \bibinfo{author}{A.~Sbai},
  \bibinfo{author}{S.~G. Thomas}, \bibinfo{author}{L.~Trenty},
  \bibinfo{author}{L.~Wei},
\newblock \bibinfo{title}{{A benchmark study on problems related to CO2 storage
  in geologic formations}},
\newblock \bibinfo{journal}{Computational Geosciences} \bibinfo{volume}{13}
  (\bibinfo{year}{2009}) \bibinfo{pages}{409--434}.
\bibitem[{Weinstein and Ruschak(2004)}]{weinstein2004}
\bibinfo{author}{S.~J. Weinstein}, \bibinfo{author}{K.~J. Ruschak},
\newblock \bibinfo{title}{{COATING} {FLOWS}},
\newblock \bibinfo{journal}{Annual Review of Fluid Mechanics}
  \bibinfo{volume}{36} (\bibinfo{year}{2004}) \bibinfo{pages}{29--53}.
\bibitem[{de~Gans et~al.(2004)de~Gans, Duineveld, and Schubert}]{degans2004}
\bibinfo{author}{B.-J. de~Gans}, \bibinfo{author}{P.~C. Duineveld},
  \bibinfo{author}{U.~S. Schubert},
\newblock \bibinfo{title}{Inkjet {P}rinting of {P}olymers: {S}tate of the {A}rt
  and {F}uture {D}evelopments},
\newblock \bibinfo{journal}{Advanced Materials} \bibinfo{volume}{16}
  (\bibinfo{year}{2004}) \bibinfo{pages}{203--213}.
\bibitem[{Blunt et~al.(2013)Blunt, Bijeljic, Dong, Gharbi, Iglauer, Mostaghimi,
  Paluszny, and Pentland}]{blunt2013}
\bibinfo{author}{M.~J. Blunt}, \bibinfo{author}{B.~Bijeljic},
  \bibinfo{author}{H.~Dong}, \bibinfo{author}{O.~Gharbi},
  \bibinfo{author}{S.~Iglauer}, \bibinfo{author}{P.~Mostaghimi},
  \bibinfo{author}{A.~Paluszny}, \bibinfo{author}{C.~Pentland},
\newblock \bibinfo{title}{{Pore-scale imaging and modelling}},
\newblock \bibinfo{journal}{Advances in Water Resources} \bibinfo{volume}{51}
  (\bibinfo{year}{2013}) \bibinfo{pages}{197--216}.
\bibitem[{Ishii and Hibiki(2011)}]{ishii2011}
\bibinfo{author}{M.~Ishii}, \bibinfo{author}{T.~Hibiki},
  \bibinfo{title}{{Thermo-Fluid Dynamics of Two-Phase Flow}},
  \bibinfo{edition}{2} ed., \bibinfo{publisher}{Springer New York},
  \bibinfo{address}{New York, NY}, \bibinfo{year}{2011}.
\bibitem[{Anderson et~al.(1998)Anderson, McFadden, and Wheeler}]{anderson1998}
\bibinfo{author}{D.~M. Anderson}, \bibinfo{author}{G.~B. McFadden},
  \bibinfo{author}{A.~A. Wheeler},
\newblock \bibinfo{title}{{DIFFUSE}-{INTERFACE} {METHODS} {IN} {FLUID}
  {MECHANICS}},
\newblock \bibinfo{journal}{Annual Review of Fluid Mechanics}
  \bibinfo{volume}{30} (\bibinfo{year}{1998}) \bibinfo{pages}{139--165}.
\bibitem[{Hitz et~al.(2021)Hitz, J{\"{o}}ns, Heinen, Vrabec, and
  Munz}]{hitz2021}
\bibinfo{author}{T.~Hitz}, \bibinfo{author}{S.~J{\"{o}}ns},
  \bibinfo{author}{M.~Heinen}, \bibinfo{author}{J.~Vrabec},
  \bibinfo{author}{C.-D. Munz},
\newblock \bibinfo{title}{{Comparison of macro- and microscopic solutions of
  the {R}iemann problem {II}. {T}wo-phase shock tube}},
\newblock \bibinfo{journal}{Journal of Computational Physics}
  \bibinfo{volume}{429} (\bibinfo{year}{2021}) \bibinfo{pages}{110027}.
\bibitem[{J{\"{o}}ns and Munz(2022)}]{joens2022}
\bibinfo{author}{S.~J{\"{o}}ns}, \bibinfo{author}{C.-D. Munz},
\newblock \bibinfo{title}{{Riemann Solvers for Phase Transition in a
  Compressible Sharp-Interface Method}},
\newblock \bibinfo{journal}{Submitted to Applied Mathematics and Computation}
  (\bibinfo{year}{2022}).
\bibitem[{Hirt and Nichols(1981)}]{hirt1981}
\bibinfo{author}{C.~W. Hirt}, \bibinfo{author}{B.~D. Nichols},
\newblock \bibinfo{title}{Volume of fluid ({VOF}) method for the dynamics of
  free boundaries},
\newblock \bibinfo{journal}{Journal of Computational Physics}
  \bibinfo{volume}{39} (\bibinfo{year}{1981}) \bibinfo{pages}{201 -- 225}.
\bibitem[{Fedkiw et~al.(1999)Fedkiw, A., Merriman, and Osher}]{fedkiw1999}
\bibinfo{author}{R.~P. Fedkiw}, \bibinfo{author}{T.~A.},
  \bibinfo{author}{B.~Merriman}, \bibinfo{author}{S.~Osher},
\newblock \bibinfo{title}{A {N}on-oscillatory {E}ulerian {A}pproach to
  {I}nterfaces in {M}ultimaterial {F}lows (the {G}host {F}luid {M}ethod)},
\newblock \bibinfo{journal}{Journal of Computational Physics}
  \bibinfo{volume}{152} (\bibinfo{year}{1999}) \bibinfo{pages}{457--492}.
\bibitem[{Cueto-Felgueroso et~al.(2018)Cueto-Felgueroso, Fu, and
  Juanes}]{cueto2017}
\bibinfo{author}{L.~Cueto-Felgueroso}, \bibinfo{author}{X.~Fu},
  \bibinfo{author}{R.~Juanes},
\newblock \bibinfo{title}{Pore-scale modeling of phase change in porous media},
\newblock \bibinfo{journal}{Phys. Rev. Fluids} \bibinfo{volume}{3}
  (\bibinfo{year}{2018}) \bibinfo{pages}{084302}.
\bibitem[{Van~der Waals(1894)}]{vanderwaals1894}
\bibinfo{author}{J.~Van~der Waals},
\newblock \bibinfo{title}{Thermodynamische {Theorie} der {Kapillarität} unter
  {Voraussetzung} stetiger {Dichteänderung}},
\newblock \bibinfo{journal}{Zeitschrift für Physikalische Chemie}
  \bibinfo{volume}{13} (\bibinfo{year}{1894}) \bibinfo{pages}{657--725}.
\bibitem[{Korteweg(1901)}]{korteweg1901}
\bibinfo{author}{D.~J. Korteweg},
\newblock \bibinfo{title}{Sur la forme que prennent les équations du
  mouvements des fluides si l'on tient compte des forces capillaires causées
  par des variations de densité considérables mais connues et sur la théorie
  de la capillarité dans l'hypothèse d'une variation continue de la
  densité},
\newblock \bibinfo{journal}{Archives Néerlandaises des Sciences exactes et
  naturelles} \bibinfo{volume}{6} (\bibinfo{year}{1901})
  \bibinfo{pages}{1--24}.
\bibitem[{Dunn and Serrin(1985)}]{dunn1985}
\bibinfo{author}{J.~E. Dunn}, \bibinfo{author}{J.~Serrin},
\newblock \bibinfo{title}{{On the thermomechanics of interstitial working}},
\newblock \bibinfo{journal}{Archive for Rational Mechanics and Analysis}
  \bibinfo{volume}{88} (\bibinfo{year}{1985}) \bibinfo{pages}{95--133}.
\bibitem[{Sou{\v{c}}ek et~al.(2020)Sou{\v{c}}ek, Heida, and
  M{\'{a}}lek}]{soucek2020}
\bibinfo{author}{O.~Sou{\v{c}}ek}, \bibinfo{author}{M.~Heida},
  \bibinfo{author}{J.~M{\'{a}}lek},
\newblock \bibinfo{title}{{On a thermodynamic framework for developing boundary
  conditions for Korteweg-type fluids}},
\newblock \bibinfo{journal}{International Journal of Engineering Science}
  \bibinfo{volume}{154} (\bibinfo{year}{2020}) \bibinfo{pages}{103316}.
\bibitem[{Rohde(2018)}]{rohde2018}
\bibinfo{author}{C.~Rohde},
\newblock \bibinfo{title}{Fully {Resolved} {Compressible} {Two}-{Phase} {Flow}:
  {Modelling}, {Analytical} and {Numerical} {Issues}},
\newblock in: \bibinfo{editor}{M.~Bulicek}, \bibinfo{editor}{E.~Feireisl},
  \bibinfo{editor}{M.~Pokorny} (Eds.), \bibinfo{booktitle}{New {Trends} and
  {Results} in {Mathematical} {Description} of {Fluid} {Flows}},
  \bibinfo{publisher}{Springer International Publishing},
  \bibinfo{address}{Cham}, \bibinfo{year}{2018}, pp. \bibinfo{pages}{115--181}.
\bibitem[{Hattori and Li(1994)}]{hattori1994}
\bibinfo{author}{H.~Hattori}, \bibinfo{author}{D.~Li},
\newblock \bibinfo{title}{Solutions for {Two}-{Dimensional} {System} for
  {Materials} of {Korteweg} {Type}},
\newblock \bibinfo{journal}{SIAM Journal on Mathematical Analysis}
  \bibinfo{volume}{25} (\bibinfo{year}{1994}) \bibinfo{pages}{85--98}.
\bibitem[{Bresch et~al.(2003)Bresch, Desjardins, and Lin}]{bresch2003}
\bibinfo{author}{D.~Bresch}, \bibinfo{author}{B.~Desjardins},
  \bibinfo{author}{C.-K. Lin},
\newblock \bibinfo{title}{On {Some} {Compressible} {Fluid} {Models}:
  {Korteweg}, {Lubrication}, and {Shallow} {Water} {Systems}},
\newblock \bibinfo{journal}{Communications in Partial Differential Equations}
  \bibinfo{volume}{28} (\bibinfo{year}{2003}) \bibinfo{pages}{843--868}.
\bibitem[{Rohde(2005)}]{rohde2005}
\bibinfo{author}{C.~Rohde},
\newblock \bibinfo{title}{On local and non-local {Navier}-{Stokes}-{Korteweg}
  systems for liquid-vapour phase transitions},
\newblock \bibinfo{journal}{ZAMM - Journal of Applied Mathematics and Mechanics
  / Zeitschrift für Angewandte Mathematik und Mechanik} \bibinfo{volume}{85}
  (\bibinfo{year}{2005}) \bibinfo{pages}{839--857}.
\bibitem[{Kotschote(2008)}]{kotschote2008}
\bibinfo{author}{M.~Kotschote},
\newblock \bibinfo{title}{Strong solutions for a compressible fluid model of
  {Korteweg} type},
\newblock \bibinfo{journal}{Annales de l'Institut Henri Poincare (C) Non Linear
  Analysis} \bibinfo{volume}{25} (\bibinfo{year}{2008}) \bibinfo{pages}{679 --
  696}.
\bibitem[{Jamet et~al.(2001)Jamet, Lebaigue, Coutris, and Delhaye}]{jamet2001}
\bibinfo{author}{D.~Jamet}, \bibinfo{author}{O.~Lebaigue},
  \bibinfo{author}{N.~Coutris}, \bibinfo{author}{J.~M. Delhaye},
\newblock \bibinfo{title}{The {Second} {Gradient} {Method} for the {Direct}
  {Numerical} {Simulation} of {Liquid}–{Vapor} {Flows} with {Phase}
  {Change}},
\newblock \bibinfo{journal}{Journal of Computational Physics}
  \bibinfo{volume}{169} (\bibinfo{year}{2001}) \bibinfo{pages}{624 -- 651}.
\bibitem[{Coquel et~al.(2005)Coquel, Diehl, Merkle, and Rohde}]{coquel2005}
\bibinfo{author}{F.~Coquel}, \bibinfo{author}{D.~Diehl},
  \bibinfo{author}{C.~Merkle}, \bibinfo{author}{C.~Rohde},
\newblock \bibinfo{title}{Sharp and {Diffuse} {Interface} {Methods} for {Phase}
  {Transition} {Problems} in {Liquid}-{Vapour} {Flows}},
\newblock \bibinfo{journal}{Numerical methods for hyperbolic and kinetic
  problems} \bibinfo{volume}{7} (\bibinfo{year}{2005})
  \bibinfo{pages}{239--270}.
\bibitem[{Diehl(2007)}]{diehl2007}
\bibinfo{author}{D.~Diehl}, \bibinfo{title}{Higher {Order} {Schemes} for
  {Simulation} of {Compressible} {Liquid}-{Vapour} {Flows} with {Phase}
  {Change}}, \bibinfo{type}{{PhD} thesis}, Albert-Ludwigs-Universität,
  \bibinfo{address}{Freiburg}, \bibinfo{year}{2007}.
\bibitem[{Haink and Rohde(2008)}]{haink2008}
\bibinfo{author}{J.~Haink}, \bibinfo{author}{C.~Rohde},
\newblock \bibinfo{title}{Local {Discontinuous}-{Galerkin} {Schemes} for
  {Model} {Problems} in {Phase} {Transition} {Theory}},
\newblock \bibinfo{journal}{Commun. Comput. Phys} \bibinfo{volume}{4}
  (\bibinfo{year}{2008}) \bibinfo{pages}{860--893}.
\bibitem[{Gomez et~al.(2010)Gomez, Hughes, Nogueira, and Calo}]{gomez2010}
\bibinfo{author}{H.~Gomez}, \bibinfo{author}{T.~J.~R. Hughes},
  \bibinfo{author}{X.~Nogueira}, \bibinfo{author}{V.~M. Calo},
\newblock \bibinfo{title}{Isogeometric analysis of the isothermal
  {Navier}–{Stokes}–{Korteweg} equations},
\newblock \bibinfo{journal}{Computer Methods in Applied Mechanics and
  Engineering} \bibinfo{volume}{199} (\bibinfo{year}{2010})
  \bibinfo{pages}{1828 -- 1840}.
\bibitem[{Braack and Prohl(2013)}]{braack2013}
\bibinfo{author}{M.~Braack}, \bibinfo{author}{A.~Prohl},
\newblock \bibinfo{title}{Stable discretization of a diffuse interface model
  for liquid-vapor flows with surface tension},
\newblock \bibinfo{journal}{ESAIM: Mathematical Modelling and Numerical
  Analysis} \bibinfo{volume}{47} (\bibinfo{year}{2013})
  \bibinfo{pages}{401--420}.
\bibitem[{Giesselmann and Pryer(2015)}]{giesselmann2015}
\bibinfo{author}{J.~Giesselmann}, \bibinfo{author}{T.~Pryer},
\newblock \bibinfo{title}{{Energy consistent discontinuous Galerkin methods for
  a quasi-incompressible diffuse two phase flow model}},
\newblock \bibinfo{journal}{ESAIM: Mathematical Modelling and Numerical
  Analysis} \bibinfo{volume}{49} (\bibinfo{year}{2015})
  \bibinfo{pages}{275--301}.
\bibitem[{Tian et~al.(2015)Tian, Xu, Kuerten, and Vegt}]{tian2015}
\bibinfo{author}{L.~Tian}, \bibinfo{author}{Y.~Xu}, \bibinfo{author}{J.~G.~M.
  Kuerten}, \bibinfo{author}{J.~J. W. v.~d. Vegt},
\newblock \bibinfo{title}{A local discontinuous {Galerkin} method for the
  (non)-isothermal {Navier}–{Stokes}–{Korteweg} equations},
\newblock \bibinfo{journal}{Journal of Computational Physics}
  \bibinfo{volume}{295} (\bibinfo{year}{2015}) \bibinfo{pages}{685 -- 714}.
\bibitem[{Diehl et~al.(2016)Diehl, Kremser, Kröner, and Rohde}]{diehl2016}
\bibinfo{author}{D.~Diehl}, \bibinfo{author}{J.~Kremser},
  \bibinfo{author}{D.~Kröner}, \bibinfo{author}{C.~Rohde},
\newblock \bibinfo{title}{Numerical solution of
  {Navier}–{Stokes}–{Korteweg} systems by {Local} {Discontinuous}
  {Galerkin} methods in multiple space dimensions},
\newblock \bibinfo{journal}{Applied Mathematics and Computation}
  \bibinfo{volume}{272} (\bibinfo{year}{2016}) \bibinfo{pages}{309 -- 335}.
\bibitem[{Gelissen et~al.(2018)Gelissen, Geld, Kuipers, and
  Kuerten}]{gelissen2018}
\bibinfo{author}{E.~J. Gelissen}, \bibinfo{author}{C.~W. M. v.~d. Geld},
  \bibinfo{author}{J.~A.~M. Kuipers}, \bibinfo{author}{J.~G.~M. Kuerten},
\newblock \bibinfo{title}{Simulations of droplet collisions with a {Diffuse}
  {Interface} {Model} near the critical point},
\newblock \bibinfo{journal}{International Journal of Multiphase Flow}
  \bibinfo{volume}{107} (\bibinfo{year}{2018}) \bibinfo{pages}{208 -- 220}.
\bibitem[{Martínez et~al.(2019)Martínez, Ramírez, Nogueira, Khelladi, and
  Navarrina}]{martinez2019}
\bibinfo{author}{A.~Martínez}, \bibinfo{author}{L.~Ramírez},
  \bibinfo{author}{X.~Nogueira}, \bibinfo{author}{S.~Khelladi},
  \bibinfo{author}{F.~Navarrina},
\newblock \bibinfo{title}{A high-order finite volume method with improved
  isotherms reconstruction for the computation of multiphase flows using the
  {Navier}-{Stokes}-{Korteweg} equations},
\newblock \bibinfo{journal}{Computers \& Mathematics with Applications}
  (\bibinfo{year}{2019}).
\bibitem[{Gelissen et~al.(2020)Gelissen, {van der Geld}, Baltussen, and
  Kuerten}]{gelissen2020}
\bibinfo{author}{E.~Gelissen}, \bibinfo{author}{C.~{van der Geld}},
  \bibinfo{author}{M.~Baltussen}, \bibinfo{author}{J.~Kuerten},
\newblock \bibinfo{title}{Modeling of droplet impact on a heated solid surface
  with a diffuse interface model},
\newblock \bibinfo{journal}{International Journal of Multiphase Flow}
  \bibinfo{volume}{123} (\bibinfo{year}{2020}) \bibinfo{pages}{103173}.
\bibitem[{Rohde and von Wolff(2020)}]{rohde2020}
\bibinfo{author}{C.~Rohde}, \bibinfo{author}{L.~von Wolff},
\newblock \bibinfo{title}{{H}omogenization of {N}onlocal
  {N}avier-{S}tokes-{K}orteweg {E}quations for {C}ompressible {L}iquid-{V}apor
  {F}low in {P}orous {M}edia},
\newblock \bibinfo{journal}{SIAM Journal on Mathematical Analysis}
  \bibinfo{volume}{52} (\bibinfo{year}{2020}) \bibinfo{pages}{6155--6179}.
\bibitem[{Rohde(2010)}]{rohde2010}
\bibinfo{author}{C.~Rohde},
\newblock \bibinfo{title}{A local and low-order {Navier}-{Stokes}-{Korteweg}
  system},
\newblock in: \bibinfo{booktitle}{Nonlinear {Partial} {Differential}
  {Equations} and {Hyperbolic} {Wave} {Phenomena}}, volume
  \bibinfo{volume}{526}, \bibinfo{publisher}{American Mathematical Society},
  \bibinfo{address}{Providence, RI}, \bibinfo{year}{2010}, pp.
  \bibinfo{pages}{315--337}.
\bibitem[{Neusser et~al.(2015)Neusser, Rohde, and Schleper}]{neusser2015}
\bibinfo{author}{J.~Neusser}, \bibinfo{author}{C.~Rohde},
  \bibinfo{author}{V.~Schleper},
\newblock \bibinfo{title}{Relaxation of the {Navier}-{Stokes}-{Korteweg}
  {Equations} for {Compressible} {Two-Phase} {Flow} with {Phase} {Transition}},
\newblock \bibinfo{journal}{International Journal for Numerical Methods in
  Fluids} \bibinfo{volume}{79} (\bibinfo{year}{2015})
  \bibinfo{pages}{615--639}.
\bibitem[{Chertock et~al.(2017)Chertock, Degond, and Neusser}]{chertock2017}
\bibinfo{author}{A.~Chertock}, \bibinfo{author}{P.~Degond},
  \bibinfo{author}{J.~Neusser},
\newblock \bibinfo{title}{An {Asymptotic}-{Preserving} {Method} for a
  {Relaxation} of the {Navier}-{Stokes}-{Korteweg} {Equations}},
\newblock \bibinfo{journal}{Journal of Computational Physics}
  \bibinfo{volume}{335} (\bibinfo{year}{2017}) \bibinfo{pages}{387--403}.
\bibitem[{Corli et~al.(2014)Corli, Rohde, and Schleper}]{corli2014}
\bibinfo{author}{A.~Corli}, \bibinfo{author}{C.~Rohde},
  \bibinfo{author}{V.~Schleper},
\newblock \bibinfo{title}{Parabolic approximations of diffusive–dispersive
  equations},
\newblock \bibinfo{journal}{Journal of Mathematical Analysis and Applications}
  \bibinfo{volume}{414} (\bibinfo{year}{2014}) \bibinfo{pages}{773 -- 798}.
\bibitem[{Hitz et~al.(2020)Hitz, Keim, Munz, and Rohde}]{hitz2020}
\bibinfo{author}{T.~Hitz}, \bibinfo{author}{J.~Keim}, \bibinfo{author}{C.-D.
  Munz}, \bibinfo{author}{C.~Rohde},
\newblock \bibinfo{title}{A parabolic relaxation model for the
  {Navier}-{Stokes}-{Korteweg} equations},
\newblock \bibinfo{journal}{Journal of Computational Physics}
  \bibinfo{volume}{421} (\bibinfo{year}{2020}) \bibinfo{pages}{109714}.
\bibitem[{Desmarais(2016)}]{desmarais2016}
\bibinfo{author}{J.~Desmarais}, \bibinfo{title}{Towards numerical simulation of
  phase-transitional flows}, Ph.D. thesis, Mechanical Engineering,
  \bibinfo{year}{2016}. \bibinfo{note}{Proefschrift}.
\bibitem[{Jacqmin(2000)}]{jacqmin2000}
\bibinfo{author}{D.~Jacqmin},
\newblock \bibinfo{title}{{Contact-line dynamics of a diffuse fluid
  interface}},
\newblock \bibinfo{journal}{Journal of Fluid Mechanics} \bibinfo{volume}{402}
  (\bibinfo{year}{2000}) \bibinfo{pages}{57--88}.
\bibitem[{Sibley et~al.(2013{\natexlab{a}})Sibley, Nold, Savva, and
  Kalliadasis}]{sibley2013a}
\bibinfo{author}{D.~N. Sibley}, \bibinfo{author}{A.~Nold},
  \bibinfo{author}{N.~Savva}, \bibinfo{author}{S.~Kalliadasis},
\newblock \bibinfo{title}{{On the moving contact line singularity: Asymptotics
  of a diffuse-interface model}},
\newblock \bibinfo{journal}{The European Physical Journal E}
  \bibinfo{volume}{36} (\bibinfo{year}{2013}{\natexlab{a}})
  \bibinfo{pages}{26}.
\bibitem[{Sibley et~al.(2013{\natexlab{b}})Sibley, Nold, Savva, and
  Kalliadasis}]{sibley2013b}
\bibinfo{author}{D.~N. Sibley}, \bibinfo{author}{A.~Nold},
  \bibinfo{author}{N.~Savva}, \bibinfo{author}{S.~Kalliadasis},
\newblock \bibinfo{title}{{The contact line behaviour of solid-liquid-gas
  diffuse-interface models}},
\newblock \bibinfo{journal}{Physics of Fluids} \bibinfo{volume}{25}
  (\bibinfo{year}{2013}{\natexlab{b}}) \bibinfo{pages}{092111}.
\bibitem[{Heida(2013)}]{heida2013}
\bibinfo{author}{M.~Heida},
\newblock \bibinfo{title}{{On the derivation of thermodynamically consistent
  boundary conditions for the Cahn–Hilliard–Navier–Stokes system}},
\newblock \bibinfo{journal}{International Journal of Engineering Science}
  \bibinfo{volume}{62} (\bibinfo{year}{2013}) \bibinfo{pages}{126--156}.
\bibitem[{Dhaouadi et~al.(2019)Dhaouadi, Favrie, and Gavrilyuk}]{dhaouadi2019}
\bibinfo{author}{F.~Dhaouadi}, \bibinfo{author}{N.~Favrie},
  \bibinfo{author}{S.~Gavrilyuk},
\newblock \bibinfo{title}{{Extended Lagrangian approach for the defocusing
  nonlinear Schr{\"{o}}dinger equation}},
\newblock \bibinfo{journal}{Studies in Applied Mathematics}
  \bibinfo{volume}{142} (\bibinfo{year}{2019}) \bibinfo{pages}{336--358}.
\bibitem[{Rajagopal and Srinivasa(2004)}]{rajagopal2004}
\bibinfo{author}{K.~R. Rajagopal}, \bibinfo{author}{A.~R. Srinivasa},
\newblock \bibinfo{title}{On thermomechanical restrictions of continua},
\newblock \bibinfo{journal}{Proceedings of the Royal Society of London. Series
  A: Mathematical, Physical and Engineering Sciences} \bibinfo{volume}{460}
  (\bibinfo{year}{2004}) \bibinfo{pages}{631--651}.
\bibitem[{Heida et~al.(2012{\natexlab{a}})Heida, Málek, and
  Rajagopal}]{heida2012a}
\bibinfo{author}{M.~Heida}, \bibinfo{author}{J.~Málek}, \bibinfo{author}{K.~R.
  Rajagopal},
\newblock \bibinfo{title}{On the development and generalizations of
  {C}ahn–{H}illiard equations within a thermodynamic framework},
\newblock \bibinfo{journal}{{Z}eitschrift für angewandte {M}athematik und
  {P}hysik} \bibinfo{volume}{63} (\bibinfo{year}{2012}{\natexlab{a}})
  \bibinfo{pages}{145--169}.
\bibitem[{Heida et~al.(2012{\natexlab{b}})Heida, Málek, and
  Rajagopal}]{heida2012b}
\bibinfo{author}{M.~Heida}, \bibinfo{author}{J.~Málek}, \bibinfo{author}{K.~R.
  Rajagopal},
\newblock \bibinfo{title}{On the development and generalizations of
  {A}llen–{C}ahn and {S}tefan equations within a thermodynamic framework},
\newblock \bibinfo{journal}{{Z}eitschrift für angewandte {M}athematik und
  {P}hysik} \bibinfo{volume}{63} (\bibinfo{year}{2012}{\natexlab{b}})
  \bibinfo{pages}{759--776}.
\bibitem[{Carr et~al.(1984)Carr, Gurtin, and Slemrod}]{carr1984}
\bibinfo{author}{J.~Carr}, \bibinfo{author}{M.~E. Gurtin},
  \bibinfo{author}{M.~Slemrod},
\newblock \bibinfo{title}{Structured phase transitions on a finite interval},
\newblock \bibinfo{journal}{Archive for rational mechanics and analysis}
  \bibinfo{volume}{86} (\bibinfo{year}{1984}) \bibinfo{pages}{317--351}.
\bibitem[{Heinen et~al.(2022)Heinen, Hoffmann, Diewald, Seckler, Langenbach,
  and Vrabec}]{heinen2022}
\bibinfo{author}{M.~Heinen}, \bibinfo{author}{M.~Hoffmann},
  \bibinfo{author}{F.~Diewald}, \bibinfo{author}{S.~Seckler},
  \bibinfo{author}{K.~Langenbach}, \bibinfo{author}{J.~Vrabec},
\newblock \bibinfo{title}{{Droplet coalescence by molecular dynamics and
  phase-field modeling}},
\newblock \bibinfo{journal}{Physics of Fluids} \bibinfo{volume}{34}
  (\bibinfo{year}{2022}) \bibinfo{pages}{042006}.
\bibitem[{Jou et~al.(2010)Jou, Casas-Vazquez, and Lebon}]{jou2010}
\bibinfo{author}{D.~Jou}, \bibinfo{author}{J.~Casas-Vazquez},
  \bibinfo{author}{G.~Lebon}, \bibinfo{title}{Extended Irreversible
  Thermodynamics}, \bibinfo{publisher}{Springer}, \bibinfo{year}{2010}.
\bibitem[{Truesdell(1985)}]{truesdell1984}
\bibinfo{author}{C.~Truesdell}, \bibinfo{title}{Rational Thermodynamics},
  \bibinfo{publisher}{Springer}, \bibinfo{year}{1985}.
\bibitem[{Heida and M{\'{a}}lek(2010)}]{heida2010}
\bibinfo{author}{M.~Heida}, \bibinfo{author}{J.~M{\'{a}}lek},
\newblock \bibinfo{title}{{On compressible Korteweg fluid-like materials}},
\newblock \bibinfo{journal}{International Journal of Engineering Science}
  \bibinfo{volume}{48} (\bibinfo{year}{2010}) \bibinfo{pages}{1313--1324}.
\bibitem[{Freist{\"{u}}hler and Kotschote(2017)}]{freistuehler2017}
\bibinfo{author}{H.~Freist{\"{u}}hler}, \bibinfo{author}{M.~Kotschote},
\newblock \bibinfo{title}{{Phase-Field and Korteweg-Type Models for the
  Time-Dependent Flow of Compressible Two-Phase Fluids}},
\newblock \bibinfo{journal}{Archive for Rational Mechanics and Analysis}
  \bibinfo{volume}{224} (\bibinfo{year}{2017}) \bibinfo{pages}{1--20}.
\bibitem[{Giovangigli(2020)}]{giovangigli2020}
\bibinfo{author}{V.~Giovangigli},
\newblock \bibinfo{title}{{Kinetic derivation of diffuse-interface fluid
  models}},
\newblock \bibinfo{journal}{Physical Review E} \bibinfo{volume}{102}
  (\bibinfo{year}{2020}) \bibinfo{pages}{012110}.
\bibitem[{Gavrilyuk and Gouin(1996)}]{gavrilyuk1996}
\bibinfo{author}{S.~Gavrilyuk}, \bibinfo{author}{H.~Gouin},
\newblock \bibinfo{title}{Symmetric form of governing equations for capillary
  fluid},
\newblock in: \bibinfo{booktitle}{Trends in Applications of Mathematics to
  Mechanics}, volume \bibinfo{volume}{106} of \textit{\bibinfo{series}{Monogr.
  Surv. Pure Appl. Math.}}, \bibinfo{publisher}{Chapman \& Hall/CRC, Boca
  Raton, FL}, \bibinfo{year}{1996}, p. \bibinfo{pages}{306–311}.
\bibitem[{Dreyer et~al.(2012)Dreyer, Giesselmann, Kraus, and
  Rohde}]{dreyer2012}
\bibinfo{author}{W.~Dreyer}, \bibinfo{author}{J.~Giesselmann},
  \bibinfo{author}{C.~Kraus}, \bibinfo{author}{C.~Rohde},
\newblock \bibinfo{title}{{Asymptotic analysis for Korteweg models}},
\newblock \bibinfo{journal}{Interfaces and Free Boundaries}
  \bibinfo{volume}{14} (\bibinfo{year}{2012}) \bibinfo{pages}{105--143}.
\bibitem[{Desmarais and Kuerten(2014)}]{desmarais2014}
\bibinfo{author}{J.~Desmarais}, \bibinfo{author}{J.~Kuerten},
\newblock \bibinfo{title}{{Open boundary conditions for the Diffuse Interface
  Model in 1-D}},
\newblock \bibinfo{journal}{Journal of Computational Physics}
  \bibinfo{volume}{263} (\bibinfo{year}{2014}) \bibinfo{pages}{393--418}.
\bibitem[{Noether(1918)}]{noether1918}
\bibinfo{author}{E.~Noether},
\newblock \bibinfo{title}{{Invariante Variationsprobleme}},
\newblock \bibinfo{journal}{Nachr. Ges. Wiss. G\"{o}ttingen, Math.-Phys. Kl.}
  (\bibinfo{year}{1918}) \bibinfo{pages}{235--257}.
\bibitem[{Slattery(1980)}]{slattery1980}
\bibinfo{author}{J.~C. Slattery},
\newblock \bibinfo{title}{{INTERFACIAL TRANSPORT PHENOMENA}},
\newblock \bibinfo{journal}{Chemical Engineering Communications}
  \bibinfo{volume}{4} (\bibinfo{year}{1980}) \bibinfo{pages}{149--166}.
\bibitem[{Rowlinson and Widom(1984)}]{rowlinson1989}
\bibinfo{author}{J.~S. Rowlinson}, \bibinfo{author}{B.~Widom},
  \bibinfo{title}{Molecular Theory of Capillarity}, \bibinfo{publisher}{Dover
  Publications, New York}, \bibinfo{year}{1984}.
\bibitem[{Cahn and Hilliard(1958)}]{cahn1958}
\bibinfo{author}{J.~W. Cahn}, \bibinfo{author}{J.~E. Hilliard},
\newblock \bibinfo{title}{{Free energy of a nonuniform system. I. Interfacial
  free energy}},
\newblock \bibinfo{journal}{The Journal of Chemical Physics}
  \bibinfo{volume}{28} (\bibinfo{year}{1958}) \bibinfo{pages}{258--267}.
\bibitem[{Pismen and Pomeau(2000)}]{pismen2000}
\bibinfo{author}{L.~M. Pismen}, \bibinfo{author}{Y.~Pomeau},
\newblock \bibinfo{title}{{Disjoining potential and spreading of thin liquid
  layers in the diffuse-interface model coupled to hydrodynamics}},
\newblock \bibinfo{journal}{Physical Review E} \bibinfo{volume}{62}
  (\bibinfo{year}{2000}) \bibinfo{pages}{2480--2492}.
\bibitem[{Kopriva(2009)}]{kopriva2009}
\bibinfo{author}{D.~A. Kopriva}, \bibinfo{title}{Implementing spectral methods
  for partial differential equations: Algorithms for scientists and engineers},
  \bibinfo{publisher}{Springer Science \& Business Media},
  \bibinfo{year}{2009}.
\bibitem[{Krais et~al.(2021)Krais, Beck, Bolemann, Frank, Flad, Gassner,
  Hindenlang, Hoffmann, Kuhn, Sonntag, and Munz}]{krais2021}
\bibinfo{author}{N.~Krais}, \bibinfo{author}{A.~Beck},
  \bibinfo{author}{T.~Bolemann}, \bibinfo{author}{H.~Frank},
  \bibinfo{author}{D.~Flad}, \bibinfo{author}{G.~Gassner},
  \bibinfo{author}{F.~Hindenlang}, \bibinfo{author}{M.~Hoffmann},
  \bibinfo{author}{T.~Kuhn}, \bibinfo{author}{M.~Sonntag},
  \bibinfo{author}{C.-D. Munz},
\newblock \bibinfo{title}{{FLEXI: A high order discontinuous Galerkin framework
  for hyperbolic–parabolic conservation laws}},
\newblock \bibinfo{journal}{Computers \& Mathematics with Applications}
  \bibinfo{volume}{81} (\bibinfo{year}{2021}) \bibinfo{pages}{186--219}.
\bibitem[{Toro(2009)}]{toro2009}
\bibinfo{author}{E.~F. Toro}, \bibinfo{title}{{Riemann Solvers and Numerical
  Methods for Fluid Dynamics}}, \bibinfo{publisher}{Springer Berlin
  Heidelberg}, \bibinfo{address}{Berlin, Heidelberg}, \bibinfo{year}{2009}.
\bibitem[{Bassi and Rebay(1997)}]{bassi1997}
\bibinfo{author}{F.~Bassi}, \bibinfo{author}{S.~Rebay},
\newblock \bibinfo{title}{{A High-Order Accurate Discontinuous Finite Element
  Method for the Numerical Solution of the Compressible Navier–Stokes
  Equations}},
\newblock \bibinfo{journal}{Journal of Computational Physics}
  \bibinfo{volume}{131} (\bibinfo{year}{1997}) \bibinfo{pages}{267--279}.
\bibitem[{Kennedy et~al.(2000)Kennedy, Carpenter, and Lewis}]{kennedy2000}
\bibinfo{author}{C.~A. Kennedy}, \bibinfo{author}{M.~H. Carpenter},
  \bibinfo{author}{R.~Lewis},
\newblock \bibinfo{title}{{Low-storage, explicit Runge–Kutta schemes for the
  compressible Navier–Stokes equations}},
\newblock \bibinfo{journal}{Applied Numerical Mathematics} \bibinfo{volume}{35}
  (\bibinfo{year}{2000}) \bibinfo{pages}{177--219}.
\bibitem[{F{\"{o}}ll et~al.(2019)F{\"{o}}ll, Hitz, M{\"{u}}ller, Munz, and
  Dumbser}]{foell2019}
\bibinfo{author}{F.~F{\"{o}}ll}, \bibinfo{author}{T.~Hitz},
  \bibinfo{author}{C.~M{\"{u}}ller}, \bibinfo{author}{C.-D. Munz},
  \bibinfo{author}{M.~Dumbser},
\newblock \bibinfo{title}{{On the use of tabulated equations of state for
  multi-phase simulations in the homogeneous equilibrium limit}},
\newblock \bibinfo{journal}{Shock Waves} \bibinfo{volume}{29}
  (\bibinfo{year}{2019}) \bibinfo{pages}{769--793}.
\bibitem[{Rajkotwala et~al.(2020)Rajkotwala, Gelissen, Peters, Baltussen,
  van~der Geld, Kuerten, and Kuipers}]{rajkotwala2020}
\bibinfo{author}{A.~Rajkotwala}, \bibinfo{author}{E.~Gelissen},
  \bibinfo{author}{E.~Peters}, \bibinfo{author}{M.~Baltussen},
  \bibinfo{author}{C.~van~der Geld}, \bibinfo{author}{J.~Kuerten},
  \bibinfo{author}{J.~Kuipers},
\newblock \bibinfo{title}{{Comparison of the local front reconstruction method
  with a diffuse interface model for the modeling of droplet collisions}},
\newblock \bibinfo{journal}{Chemical Engineering Science: X}
  \bibinfo{volume}{7} (\bibinfo{year}{2020}) \bibinfo{pages}{100066}.
\bibitem[{D{\"{o}}rmann and Schmid(2015)}]{doermann2015}
\bibinfo{author}{M.~D{\"{o}}rmann}, \bibinfo{author}{H.-J. Schmid},
\newblock \bibinfo{title}{{Simulation of Capillary Bridges between Particles}},
\newblock \bibinfo{journal}{Procedia Engineering} \bibinfo{volume}{102}
  (\bibinfo{year}{2015}) \bibinfo{pages}{14--23}.
\bibitem[{Mastrangeli(2015)}]{mastrangeli2015}
\bibinfo{author}{M.~Mastrangeli},
\newblock \bibinfo{title}{{The Fluid Joint: The Soft Spot of Micro- and
  Nanosystems}},
\newblock \bibinfo{journal}{Advanced Materials} \bibinfo{volume}{27}
  (\bibinfo{year}{2015}) \bibinfo{pages}{4254--4272}.
\bibitem[{Kuhn et~al.(2019)Kuhn, D{\"{u}}rrw{\"{a}}chter, Meyer, Beck, Rohde,
  and Munz}]{kuhn2019}
\bibinfo{author}{T.~Kuhn}, \bibinfo{author}{J.~D{\"{u}}rrw{\"{a}}chter},
  \bibinfo{author}{F.~Meyer}, \bibinfo{author}{A.~Beck},
  \bibinfo{author}{C.~Rohde}, \bibinfo{author}{C.-D. Munz},
\newblock \bibinfo{title}{{Uncertainty Quantification for Direct Aeroacoustic
  Simulations of Cavity Flows}},
\newblock \bibinfo{journal}{Journal of Theoretical and Computational Acoustics}
  \bibinfo{volume}{27} (\bibinfo{year}{2019}) \bibinfo{pages}{1850044}.

\end{thebibliography}

\end{document}